\documentclass[11pt,a4paper]{article}
\usepackage[T1]{fontenc}
\usepackage{lmodern}
\usepackage[margin=27mm]{geometry}
\usepackage{microtype}
\usepackage{amsmath,amssymb,amsthm,mathtools}
\usepackage{booktabs,array,enumitem}
\usepackage{tikz}
\usepackage{authblk}
\usetikzlibrary{arrows.meta,calc,positioning}
\usepackage{caption,subcaption}
\usepackage[numbers,sort&compress]{natbib}
\usepackage{xcolor}
\usepackage{mathrsfs}
\definecolor{safe}{HTML}{247A73}
\definecolor{bad}{HTML}{B3423A}
\usepackage[hidelinks]{hyperref}
\usepackage[nameinlink,noabbrev]{cleveref}
\crefname{theorem}{Theorem}{Theorems}
\crefname{lemma}{Lemma}{Lemmas}
\crefname{proposition}{Proposition}{Propositions}
\crefname{corollary}{Corollary}{Corollaries}
\crefname{definition}{Definition}{Definitions}
\crefname{example}{Example}{Examples}
\crefname{remark}{Remark}{Remarks}
\crefname{section}{Section}{Sections}
\crefname{subsection}{Section}{Sections}
\crefname{appendix}{Appendix}{Appendices}
\crefname{figure}{Figure}{Figures}
\crefname{table}{Table}{Tables}
\crefname{equation}{Equation}{Equations}
\hypersetup{pdftitle={A Combinatorial Characterization of Immunity to the Strong Braess Paradox},pdfsubject={Fixed resource sharing in nonatomic congestion games},pdfauthor={Xujin Chen, Changjun Wang, Daorong Yu}}
\numberwithin{equation}{section}
\newtheorem{theorem}{Theorem}[section]
\newtheorem{lemma}[theorem]{Lemma}
\newtheorem{proposition}[theorem]{Proposition}
\newtheorem{corollary}[theorem]{Corollary}
\theoremstyle{definition}
\newtheorem{definition}[theorem]{Definition}
\newtheorem{example}[theorem]{Example}
\theoremstyle{remark}

\newcommand{\R}{\mathbb R}

\newcommand{\F}{\mathcal F}
\newcommand{\HH}{\mathcal H}

\newcommand{\one}{\boldsymbol{1}}
\newcommand{\ee}{\boldsymbol{e}}
\newcommand{\T}{^{\mathsf T}}
\newcommand{\ma}[1]{\textcolor{black}{#1}}
\definecolor{cyanaccent}{HTML}{1596A6}
\newcommand{\cy}[1]{\textcolor{cyanaccent}{#1}}

\DeclareMathOperator{\aff}{aff}
\DeclareMathOperator{\conv}{conv}
\DeclareMathOperator{\spanop}{span}
\DeclareMathOperator{\cone}{cone}
\DeclareMathOperator{\supp}{supp}
\DeclareMathOperator{\rank}{rank}
\DeclareMathOperator{\ri}{ri}
\DeclareMathOperator{\argmin}{arg\,min}
\setlist[enumerate]{itemsep=3pt,topsep=5pt}
\setlist[itemize]{itemsep=3pt,topsep=5pt}
\tikzset{resource/.style={line width=.85pt,draw=black},flow/.style={resource,-{Stealth[length=2mm]}},vertex/.style={circle,fill=black,inner sep=1.8pt}}
\title{A Combinatorial Characterization of Immunity\\ to the Strong Braess Paradox}

\author[1,2]{Xujin Chen}
\author[1]{Changjun Wang}
\author[1,2]{Daorong Yu}

\affil[1 ]{\footnotesize SKLMS, Academy of Mathematics and Systems Science, Chinese Academy of Sciences, Beijing, China}
\affil[2 ]{\footnotesize  School of Mathematical Sciences, University of Chinese Academy of Sciences, Beijing, China \\

\texttt{\{xchen, wcj, yudaorong\}@amss.ac.cn}}

\date{}
\begin{document}
\maketitle
\vspace{-1.7em}
\begin{abstract}
\ma{In a nonatomic congestion game, a \emph{strong Braess paradox}
occurs when reducing resource costs, population demands, or both
increases at least one population's equilibrium cost.
We characterize immunity to this paradox when each population's
available strategies form a fixed finite family of subsets of
a common resource set.
Immunity means that no such paradox can arise for any choice of
nonnegative continuous nondecreasing resource cost functions
common to all populations and any positive demand vector.}
%We characterize immunity to the strong Braess paradox in finite nonatomic congestion games. 
\ma{Our characterization} %with fixed resource sharing, common continuous nondecreasing resource costs, and positive population demands. This 
resolves the open question posed by Fujishige, Goemans, Harks, Peis, and Zenklusen (2017) of characterizing immunity without varying the resource embedding.

\ma{We call a nonempty subfamily of a population's available
strategies a \emph{menu}. A menu is \emph{positively exposed}
if it consists of all available strategies of minimum total
weight for some assignment of strictly positive weights
to the resources.} For a single population, %consider complete menus consisting of all strategies that minimize some strictly positive additive resource weights. We 
we prove that immunity holds if and only if the column matroid of the augmented incidence matrix is series-parallel for every \ma{positively
exposed menu.} %such menu with a nonempty resource set.  We characterize these menus combinatorially by trade closure.
For multiple populations, \ma{we consider one menu for each
population, all positively exposed by the same assignment
of strictly positive resource weights. The associated dual matroid includes one labeled element for
each population to keep track of its demand separately.
We show that immunity holds if and only if, for every such collection
of menus, each marked resource block of the dual matroid
is series-parallel and all labeled elements retained in that block are parallel.} %immunity holds if and only if, for every jointly trade-closed menu tuple, each marked resource block of the dual representation is series-parallel and its nonloop demand labels are parallel. This parallelism condition cannot be omitted, even when the entire marked  matroid is series-parallel. \ma{We characterize the menus and collections of menus used in these criteria combinatorially by trade closure.} 
We establish these characterizations through an affine response criterion and a blockwise ray lemma. Our results recover the immunity of matroid base systems under arbitrary resource sharing and the series-parallel routing criterion for complete acyclic path systems.

Furthermore, whenever a \ma{strategy} system is not immune, we construct a strong Braess \ma{paradox} % on its original resources 
with all demands equal to one \ma{and affine
resource costs having strictly positive slopes and intercepts,
by lowering just one intercept while keeping it positive. Thus immunity against all such interventions is equivalent to
immunity against arbitrary cost and demand reductions.} %, strictly positive affine costs, and a reduction in the intercept of a single resource cost function. Thus this restricted class of interventions suffices to test immunity to all cost and demand reductions.
When strategies are listed explicitly, nonimmunity has polynomial-size certificates. 
\end{abstract}

\section{Introduction}\label{sec:introduction}
Improving a congestible resource can \ma{leave} %make 
its users worse off at equilibrium.
Braess's paradox \citep{braess} is the classical routing example, \ma{in which} a new or
improved road changes route choices and raises equilibrium travel time.
We study the corresponding structural question for nonatomic congestion
games with arbitrary finite families of resource subsets. With the strategies
and their shared resources fixed, which systems exclude an increase in any
population's equilibrium cost \ma{when} resource costs or demands are reduced?
The requirement \ma{must hold uniformly across} %is uniform over 
all %common 
continuous nondecreasing resource
costs \ma{common to all populations} and all positive population demands.

\ma{A resource-sharing pattern specifies which resources are shared by different populations, so that their loads on each shared resource are aggregated.}
%The resource-sharing pattern specifies which resources used by different populations are actually the same resource. 
\citet{fujishige} characterize
immunity to the strong Braess paradox when that pattern may vary. Their
condition must exclude \ma{the} paradox under every embedding of the populations'
strategies into a common resource set. They leave open the characterization
for a fixed embedding. We resolve this open question in the finite
nonatomic model with common resource costs.

An improvement means a reduction of resource costs, population demands,
or both. It produces a \emph{strong Braess paradox} if at least one
population pays a higher equilibrium strategy cost after the improvement.
A fixed system is \emph{immune} when no admissible costs, demands, and
improvement can produce such \ma{a paradox}. %an increase. 
Thus the object characterized
here is the strategy system with its resource-sharing pattern, not one numerical
cost-and-demand instance.

\ma{When the sharing pattern is fixed, immune systems extend}
%Fixing the sharing pattern admits immune systems 
beyond matroid base families. For example, \ma{consider the single-population family $\F=\{ab,c\}$, where, throughout, we write $ab$ for $\{a,b\}$. This system} is
immune, although its two strategies have different cardinalities. %;throughout, $ab$ abbreviates $\{a,b\}$. 
\ma{However, immunity breaks down if we introduce two additional populations, one} %Now add one population 
forced
to use $a$ and \ma{the other} %another 
forced to use $b$. For suitable costs, improving
$a$ shifts the first population toward $ab$\ma{, which in turn increases} %and raises 
the cost paid by
the population forced to use $b$. The original population's \ma{strategy set is unchanged, but resource sharing} %choices have not changed, but the shared users 
make the improvement harmful. \ma{Our characterization must therefore capture}
%The characterization must therefore record 
both the dependencies among
strategies and the separate effects of the populations' demands.

\subsection{Main contributions}
\ma{We call a nonempty subfamily of a population's available
strategies a \emph{menu}. A menu is \emph{positively exposed}
if it consists of all available strategies of minimum total
weight under some assignment of strictly positive weights
to the resources. For multiple populations, we consider one
menu for each population, all exposed by the same resource
weights. At an equilibrium with strictly positive resource
costs, these costs provide the exposing weights. The menus
therefore include every minimum-cost strategy, including
those carrying zero flow. We use trade closure to characterize these collections of menus combinatorially through
resource counts, without reference to cost functions or
equilibria.}

\ma{Our results have three parts.
We reduce the requirement over all admissible costs, demands,
and improvements to finite algebraic checks, derive complete
matroidal characterizations of immunity for single and multiple
populations, and construct explicit paradoxes on the original
resources whenever immunity fails.}

%A relevant menu \ma{of a population} is obtained by assigning a strictly positive weight to each resource and retaining \emph{all} strategies of the minimum total weight.  We call such a menu \emph{positively exposed}. For \ma{multiple} populations, the same resource weights are used for every population. At an equilibrium with positive resource costs,  the cost vector provides these weights. The menu \ma{consists of} every minimizing strategy, even one carrying zero flow. Trade closure characterizes \ma{positively exposed menus} by resource counts, without referring to cost functions or an equilibrium. Our results have three parts.

First, we provide \ma{\emph{a finite algebraic criterion} for immunity.
The definition of immunity quantifies over infinitely many
admissible cost functions, demand vectors, and improvements.
By analyzing linear equations for local equilibrium changes,
we prove that this universal requirement is equivalent to
finitely many nonnegativity conditions for each relevant
choice of menus  (\cref{prop:single-analytic,prop:multi-analytic}).
These conditions depend only on which resources each strategy
uses and how resources are shared.} 

Second, we establish \emph{complete structural characterizations} \ma{for fixed strategy system to be immune. For a single population, this algebraic criterion has an
exact matroidal interpretation. For each positively exposed
menu, we append an all-ones column to its strategy--resource
incidence matrix. The strategy system is immune if and only
if the column matroid of every such augmented matrix is
series-parallel (\cref{thm:single}). Equivalently, none of these
matroids has a $U_{2,4}$ or $M(K_4)$ minor. For multiple populations, we augment the incidence matrix
with one indicator column for each population. These columns
give one matroid element (called demand element) per population, allowing
the populations' demands to be treated separately.
In the dual matroid, the connected components of the resource
restriction define the resource blocks. For each block,
contracting all other resources and deleting demand elements
that become loops gives its associated marked block.
We prove that the system is immune if and only if, for every collection
of menus exposed by common strictly positive resource weights,
each marked block is series-parallel and its retained demand
elements are all parallel (\cref{thm:multi}).
The latter} parallelism condition cannot be omitted, even when the entire
marked matroid is series-parallel. \ma{Our characterizations recover immunity for populations whose
strategy families are matroid bases, regardless of which
resources are shared across populations (\cref{cor:matroid-sharing}),
and the classical series-parallel routing criterion for
complete acyclic path systems (\cref{cor:routing-immunity}).}

Furthermore, we construct  \ma{\emph{nonimmunity witnesses} with unit demands and affine costs.} Whenever a strategy system is not immune, we construct a
strong Braess paradox \ma{for the game, where} all
demands equal to one, the resource costs are affine with
strictly positive slopes and intercepts, \ma{and the intervention that increases at least one population's equilibrium cost
lowers just one intercept while keeping it positive.
Thus, immunity against all such interventions} is
equivalent to immunity against arbitrary cost and demand
reductions (\cref{cor:affine-test}).
For explicitly listed strategies, nonimmunity has
polynomial-size certificates, placing immunity in $\mathrm{coNP}$
(\cref{prop:coNP}). These are menu-level and certificate bounds; enumerating
all relevant menus may be exponential.

\subsection{High-level ideas}
\ma{A separation argument first characterizes positive exposure
by trade closure, both for individual menus and for collections
of menus exposed by common resource weights.
We begin with one population and a fixed positively
exposed menu. At unit demand, the menu determines linear
equations satisfied by all induced resource-load vectors.
Setting selected load coordinates to zero, we solve for the others.
Whenever the remaining coordinates are uniquely determined,
we obtain a \emph{basic solution}. There are finitely many
such solutions. They need not be feasible loads, since the
linear equations alone can admit negative coordinates. The reason for considering these algebraic solutions comes
from equilibrium sensitivity.}

\ma{For a single population, we analyze small improvements under
affine resource costs with strictly positive slopes, while
the minimizing menu remains unchanged. Differentiating the equilibrium conditions, we express the rate
of change of equilibrium cost of the population as a linear combination of
the rate of change of demand and the rates of change of
resource cost functions evaluated at the current loads. We call the numbers multiplying these changes
\emph{response coefficients}.
Because an improvement makes these changes nonpositive,
nonnegative response coefficients ensure a nonpositive rate of change of equilibrium cost.
Since the coefficient multiplying the demand change (\emph{demand response coefficient}) is always
nonnegative, we focus on the coefficients multiplying
the resource cost changes (\emph{resource-cost response coefficients}). A standard averaging
identity expresses the vector of these coefficients as a
weighted average of the basic solutions, with positive weights
summing to one (\Cref{lem:cb}; see also \citet[Theorem~5]{derezinskiwarmuth}). The weights depend on the cost slopes,
but the basic solutions do not. Conversely, suitable slope
choices make the coefficient vector approach any prescribed
basic solution. Thus the resource response coefficients are nonnegative for
every choice of strictly positive cost slopes if and only if all basic solutions are nonnegative.}

\ma{We next identify when all basic solutions are nonnegative.
The hyperplanes on which individual resource loads vanish
divide the solution space of the load equations into regions
with fixed coordinate signs. The basic solutions are the
vertices of the closures of these regions. All strategies in the menu have the same total weight under the
strictly positive exposing weights.
At unit demand, the load equations describe the affine hull
of the incidence vectors of the strategies in the menu. Every solution of the load equations
is a linear combination of these vectors with coefficients
summing to one. 
By linearity, the weighted sum of load coordinates equals
the common strategy weight throughout this solution space,
even when some coefficients in the combination are negative.
When all coordinates are nonnegative, each is bounded above
by this fixed sum divided by its strictly positive resource
weight. Hence the region with positive coordinates is bounded. 
\Cref{lem:bridge} relates nonnegativity of every basic solution to
the existence of exactly one bounded region, and then to
series-parallel structure of the augmented incidence column
matroid. The proof combines classical counts of bounded
regions with an elimination argument that tracks signs
through series and parallel reductions.}

\ma{For multiple populations, we use the load equations obtained
from all menus together, while keeping each population's
demand contribution separate. For a fixed collection of menus, each population's contribution
to the right-hand sides of the load equations depends only
on its total demand, not on how its flow is distributed
among its menu strategies. For each population, all resource-cost response coefficients
must be nonnegative for every choice of positive resource
cost slopes. By the averaging formula (mentioned above in the single-population case), this is equivalent
to requiring all basic solutions associated with that
population to be nonnegative. Nonnegative resource-cost response coefficients also ensure that
demand reductions cannot locally increase equilibrium costs.
Indeed, the derivative of population $i$'s equilibrium cost
with respect to population $j$'s demand is a weighted sum
of products of their resource-cost response coefficients, with
the positive resource cost slopes as weights. If all resource-cost response coefficients are nonnegative,
the demand coefficients are therefore nonnegative as well. The additional structural step is a ray lemma applied to
each connected resource block (\Cref{lem:ray}).
Under the positivity supplied by exposure, if the combined
demand contribution satisfies the nonnegativity criterion,
then every contribution satisfying that criterion must be
a nonnegative multiple of the combined contribution within
the block. These proportionality relations are precisely
the parallelism of the retained demand elements in the matroid.
Together with \Cref{lem:bridge}, they establish both sufficiency and necessity
of the multipopulation characterization.}

\ma{The preceding analysis also shows how to construct a paradox
when the strategy system fails to satisfy the immunity conditions.
Such a failure yields a basic solution with a negative
coordinate for some population.
The averaging formula then supplies positive resource cost
slopes for which the corresponding resource-cost response
coefficient is negative. We realize this response at an equilibrium with every
population's demand equal to one, positive affine slopes
and intercepts, and positive flow on every menu strategy.
We then lower the corresponding resource cost intercept.
The new load is explicitly constructed by solving the affine
load equations together with the conditions that strategy
costs remain equal within each menu.
Because the initial flow assigns positive flow to every
menu strategy, a sufficiently small reduction keeps the
constructed load realizable by feasible menu flows.
The strict cost gaps to outside strategies and the positive
intercepts also persist.
The constructed load is therefore an equilibrium with the
same sets of minimum-cost strategies, and the negative
response coefficient ensures that the selected population's
equilibrium cost increases.
This yields a paradox on the original resource set.}

\ma{To compare equilibrium costs before and after an improvement,
we first consider strictly positive piecewise-affine resource
costs with positive slopes.
We vary each resource cost function and each population's
demand linearly from its initial to its improved value.
The regularity argument (see \Cref{lem:regularity-reduction} and Appendix \ref{app:regularity}) gives a finite subdivision
of the parameter interval such that, on each open subinterval,
every population's complete set of minimum-cost strategies
is fixed, each resource load remains in a fixed affine cost
segment, and equilibrium loads and costs vary smoothly.
The response formulas therefore apply on each subinterval
and, under the nonnegativity criterion, give nonpositive
derivatives of all populations' equilibrium costs.
Integrating on each subinterval and using continuity at
the subdivision points shows that no population's equilibrium
cost increases from the initial instance to the improved one.
Finally, we approximate arbitrary continuous nonnegative
nondecreasing costs uniformly on the feasible load range
by strictly positive piecewise-affine costs with positive
slopes. We choose the approximations so that, for each resource
and every load, the approximation to the improved cost
function is no greater than the approximation to the
initial cost function. Passing to the limit and using continuity of equilibrium costs,
we conclude that, under the original cost functions, no
population has a higher equilibrium cost in the improved
instance than in the initial instance.}

\subsection{Related work}
Our model follows Wardrop's equilibrium principle and \ma{admits} the separable convex
potential of Beckmann, McGuire, and Winsten \citep{wardrop,beckmann}.
Braess's example \citep{braess} \ma{shows that adding a road to a network
can increase equilibrium travel times.
Motivated by this phenomenon, we seek structural conditions
on fixed finite strategy families over a common resource set
that guarantee that reducing resource costs, population
demands, or both cannot increase any population's
equilibrium cost.} %asks how an equilibrium changes when resources improve. This comparison differs from price-of-anarchy analysis, which compares equilibrium and socially optimal outcomes within one instance \citep{roughgardentardos,roughgardenbook}. The present question is structural: which fixed strategy incidences make every admissible improvement harmless to every population?

\paragraph{Matroid characterizations of immunity.}
%\paragraph{Matroid strategy families and fixed resource sharing.} 
\ma{The work of \citet{fujishige} is most closely related
to ours.
Our characterization differs from theirs in both the
immunity requirement and the matroid construction. They first consider immunity to the weak Braess paradox, which means that
reductions in resource cost functions, population demands,
or both can never increase the equilibrium cost of any resource. They prove in}
%The closest set-system result is \citet{fujishige}.
their Theorem~3.1 \ma{that a strategy system is immune
to the weak Braess paradox if and only if, for each population,
its inclusion-minimal strategies are exactly the bases
of a matroid on the resource set. They also study immunity to the strong Braess paradox
when the pattern of resource sharing between populations
is allowed to vary. Each population's strategy family is preserved up to
resource relabelling, with distinct resources within
each population receiving distinct labels.
Resources in different populations may receive the same
label and are then treated as a single resource carrying
their combined load. For systems with}
%characterizes immunity to the \emph{weak} Braess paradox, which excludes an increase in any equilibrium resource cost, through matroid bases formed by each population's inclusion-minimal strategies. With 
at least two
populations and no empty strategies, their Theorem~4.3 \ma{shows that immunity to the strong paradox under
\emph{all} such sharing patterns holds if and only if the same
matroid bases condition is satisfied.} %gives the same matroid condition for immunity to the strong paradox under \emph{every} populationwise embedding.

\ma{In contrast, we fix each population's strategy family on a common
resource set, thereby fixing which resources are shared
by different populations. Requiring immunity only for the given pattern of resource
sharing enlarges the class of immune systems and leads
to a different characterization.
In particular, the matroid bases condition of \citet{fujishige} is no longer
necessary, as illustrated by the immune family}  %Our quantifier fixes that embedding. This admits additional immune families, such as 
$\{ab,c\}$
(\cref{ex:abc}). %, but makes the precise sharing pattern consequential.
\ma{Our characterization uses column matroids of augmented
strategy-resource incidence matrices, together with
duals and minors of these matroids.
In these matrices, rows correspond to strategies in the menus,
and resource columns record which strategies use each
resource.
The matroid condition in~\cite{fujishige}, by contrast, requires each
population's inclusion-minimal strategies to be exactly
the bases of a matroid on the resource set.
We illustrate the distinction between these
two types of matroidal characterizations in Appendix~\ref{apx:differentmatriod}. From our marked-block criterion, we recover the sufficient
condition for immunity to the strong Braess paradox
established by \citet{fujishige} (\Cref{cor:matroid-sharing}). Our matroid construction records each population's contribution
to the affine load equations through a separate demand
element.
The marked-block criterion distinguishes these individual
contributions, since their sum alone is insufficient
to determine immunity.}
%The matroids used here also differ: they represent dependencies among augmented incidence \emph{columns}, whereas a strategy matroid has the strategies as its \emph{bases}. \Cref{app:matroid-comparisons} makes this distinction explicit. \Cref{cor:matroid-sharing} derives immunity to the strong Braess paradox for arbitrary populations of matroid bases from our marked-block criterion. The fixed-sharing theorem also admits nonmatroidal strategy families. It retains each population's demand direction, which a test of aggregate demand alone would lose.

\paragraph{Immune network topology and its recognition.}
For single-commodity undirected routing, \citet{milchtaich} identify the
two-terminal series-parallel topology associated with immunity.
\citet{chen} give topology characterizations for multicommodity directed
and undirected networks. Our theorem applies to arbitrary resource-subset
families, including systems with no prescribed graph representation.
\ma{Given a directed acyclic network in which every vertex
and edge lies on a source-sink path, we consider all such
paths as strategies, and derive an undirected multigraph by adding an edge from
the sink back to the source and then forgetting all
edge orientations.} %For irredundant directed acyclic networks with their complete path families, 
\cref{prop:path-dictionary} proves \ma{that the column matroid of the
augmented path-edge incidence matrix of the network is dual to the cycle matroid of this multigraph. Using this duality and our single-population theorem,
we recover Milchtaich's~\cite{milchtaich} classical series-parallel
characterization of immunity in the directed acyclic
setting considered here (\Cref{cor:routing-immunity}).
The full path family is immune to the strong Braess
paradox if and only if the network is two-terminal
series-parallel. Matroid duality preserves the series-parallel property
while interchanging series and parallel extensions. Specifically, subdividing a network edge corresponds to a parallel
extension of the augmented column matroid, whereas
replacing an edge by two parallel edges corresponds
to a series extension (\Cref{cor:network-reversal}).}
%$N_{\HH}=M(G^+)^*$, where $G^+$ closes the terminals by a return edge. This recovers the single-commodity series-parallel criterion in that acyclic setting and explains the reversal of network series and parallel operations in the augmented column matroid.

Directed graph recognition \ma{for immunity} is studied by \citet{cenciarelli} and
\citet{matsubayashi}; the latter give an $O(m^2)$ algorithm for one
commodity and an $O(km^2)$ algorithm for $k$ commodities, where $m$ is the number of edges.
Those algorithms \ma{take the network and its source-sink
pairs as input, rather than an explicit list of all
available paths. In contrast, our complexity result assumes that each population's
available strategies are explicitly listed as
$0$-$1$ resource-incidence vectors.
Under this input model, nonimmunity has certificates
whose size and verification time are polynomial in
the total input length (\Cref{prop:coNP}). A network may contain exponentially many source-sink paths; so explicitly listing all available paths as strategies can require an input exponentially larger than the network description.} %use compact graph inputs. Our certificate result counts every explicitly listed strategy; a graph may encode exponentially many such paths. The two input models therefore give different recognition bounds.

\paragraph{Sensitivity and monotonicity.}
Classical sensitivity analysis studies how equilibrium loads and costs
respond to parameter changes \citep{hall,patriksson,josefssonpatriksson}.
More recently, \ma{for multicommodity routing games.} \citet{cominettiphase} prove smoothness along demand curves
when the complete minimum-cost path sets remain fixed and costs have
strictly positive derivatives. They also study \ma{the left and right derivatives of equilibrium social cost and the price of anarchy at demand values where these minimum-cost path sets change. We likewise derive a sensitivity formula for each population’s equilibrium cost on intervals where the complete minimum-cost strategy sets remain fixed. Our formula allows resource cost functions and population demands to vary simultaneously.} %one-sided derivatives at support changes in multicommodity routing. Our local calculations use the same fixed-menu perspective, with simultaneous changes of costs and demands. 
\ma{Our structural characterization requires that reductions in resource costs or population demands cannot locally increase any population’s equilibrium cost. This requirement must hold for \emph{every} choice of strictly positive slopes of the resource cost functions.} 
%The structural question here requires the response signs to be correct for \emph{every} positive slope choice. Basis averaging in \cref{lem:cb} turns that requirement into the finite basis test, while \cref{lem:regularity-reduction} and \cref{app:regularity} justify menu changes and the extension to general continuous nondecreasing costs.
\citet{verbreecherukuri} \ma{study how equilibria vary with demand in single-commodity routing games with fixed affine resource cost functions. They use this analysis to derive sufficient conditions for detecting Braess paradox.} 
%analyze varying demand in affine routing games and its relation to Braess's paradox.

\citet{cominetti} study monotonicity \ma{of equilibrium costs and resource loads under demand changes. For singleton congestion games, they show that equilibria can be selected so that each resource load is nondecreasing in every population’s demand. They extend this result to the broader class of constrained series-parallel congestion games. Their structural conditions ensure that equilibria can be chosen for all demand vectors so that each resource load is nondecreasing in every population’s demand, while resource cost functions remain unchanged.} 
%under demand changes and the existence of monotone equilibrium selections, including singleton and constrained series-parallel games. Their constructive sufficient classes illuminate when increasing demand increases equilibrium loads; our
Our characterization instead \ma{concerns fixed finite strategy families and ensures that reductions in resource cost functions, population demands, or both never increase any population’s equilibrium cost.} 
%tests every fixed finite family and allows both resource-cost and demand reductions. 
In particular, the term
``series-parallel'' \ma{refers to different structures in their paper and ours. In \cite{cominetti}, it describes the recursive construction of congestion games from singleton games. In our characterization, it is a property of the column matroids of augmented strategy-resource incidence matrices and of marked-block matroids derived from their duals.}
%in their strategy constructions is not a substitute for the augmented-column and marked-block conditions here.
% Our necessity result is uniform over costs and constructive in a different sense: any structural failure can be realized on the original resources by lowering one intercept of strictly positive affine costs at unit demands.

%\paragraph{Organization and proof dependencies.}
\subsection{\ma{Paper organization}}
\Cref{sec:model} defines the model and presents both main theorems.
\Cref{sec:menus} proves the exposure and recognition results.
\Cref{sec:single-proof,sec:multi-proof} give the single- and multipopulation
proofs. \Cref{sec:sp-prelim} establishes the network correspondence, and
\cref{sec:examples} develops separating examples and certificate consequences.
The appendices supply the continuous-cost extension, supplementary matroid
facts, numerical witnesses, and expanded auxiliary proofs. The main proof
passes from positive exposure to the affine sign criterion and its matroidal
characterization; the multipopulation proof adds the ray lemma. Core
sensitivity formulas and the witness construction are derived in the main
text. The network correspondence and applications then interpret the
characterization.

\section{Model and main results}\label{sec:model}
%\paragraph{Vector and geometric notation.}
All vector spaces and matrix representations are over $\R$ unless a
different field is stated. Vectors are columns and $\boldsymbol{v}\T$ is the transpose \ma{of vector $\boldsymbol{v}$}.
The vector $\ee_e$ has a one in coordinate $e$ and zeros elsewhere;
$\one$ is an all-ones vector of the indicated dimension. Inequalities are
coordinatewise: $\boldsymbol{v}\ge\boldsymbol{0}$ means $v_e\ge0$ for every \ma{coordinate} $e$, whereas $\boldsymbol{v}>\boldsymbol{0}$
means every coordinate is strictly positive. Accordingly, $\R_{+}^E$ and
$\R_{>0}^E$ denote the nonnegative and strictly positive orthants.

For a finite nonempty list of points (vectors) $\boldsymbol v_1,\ldots,\boldsymbol v_m$, \ma{their linear span, affine hull, convex hull, and generated cone are written as} %our conventions are
\[
\begin{aligned}
 \spanop\{\boldsymbol v_j:j=1,\ldots,m\}&\triangleq\{\textstyle\sum_{j=1}^m\alpha_j\boldsymbol v_j:\alpha_j\in\R,j=1,\ldots,m\},\\
 \aff\{\boldsymbol v_j:j=1,\ldots,m\}&\triangleq\{\textstyle\sum_{j=1}^m\alpha_j\boldsymbol v_j:
                      \alpha_j\in\R,j=1,\ldots,m,\ \sum_{j=1}^m\alpha_j=1\},\\
 \conv\{\boldsymbol v_j:j=1,\ldots,m\}&\triangleq\{\textstyle\sum_{j=1}^m\alpha_j\boldsymbol v_j:
                      \alpha_j\ge0,j=1,\ldots,m,\ \sum_{j=1}^m\alpha_j=1\},\\
 \cone\{\boldsymbol v_j:j=1,\ldots,m\}&\triangleq\{\textstyle\sum_{j=1}^m\alpha_j\boldsymbol v_j:\alpha_j\ge0,j=1,\ldots,m\}.
\end{aligned}
\]
%These are the linear span, affine hull, convex hull, and generated cone.
An affine combination may use negative coefficients\ma{, but a convex combination
can not.} An affine space is a translate $\boldsymbol a+L$ of a linear subspace $L$,
and the affine hull is the smallest affine space containing the points.
For a convex set $Q$, let $\ri Q$ denote its \emph{relative interior}, taken
within $\aff Q$ rather than in the ambient space. For a vector $\boldsymbol{v}$, its support is defined as
$\supp \boldsymbol{v}\triangleq\{e:v_e\ne0\}$.

For a matrix $A$, let $\operatorname{row}(A)$ denote its \emph{row space},
the linear span of its rows, \ma{instead of} the finite collection of rows itself.
Let $\rank A$ represent the dimension of $A$'s row
space, equivalently the dimension of its column space. Set $\ker A\triangleq\{\boldsymbol{v}:A\boldsymbol{v}=\boldsymbol{0}\}$. We identify vectors in row space with
their transposes when taking inner products. For a subspace $L$, let
$L^\perp\triangleq\{\boldsymbol q:\boldsymbol {\boldsymbol  q}^\top \boldsymbol v= 0\text{ for every }\boldsymbol{v}\in L\}$ denote the orthogonal complement of \(L\).
For an index set $X$, \ma{we use $\boldsymbol v|_X$ or $\boldsymbol v_X$ to denote} the vector restricted to those
coordinates. %; when indices are written as subscripts, $v_X$ means the same restriction.

\subsection{Fixed sharing and the strong Braess paradox}
Let $E$ be a finite resource set and $I=\{1,\ldots,k\}$ a population set.
Population $i\in I$ has a positive demand $d_i>0$ and a nonempty finite strategy
family $\F_i\subseteq2^E\setminus\{\varnothing\}$. Thus, {every strategy
available to every population contains at least one resource}; this \emph{nonempty-strategy
assumption} is maintained throughout the paper.
All populations face the same resource costs (cost functions)
$c_e:[0,\infty)\to[0,\infty)$, \ma{$e\in E$,} which are continuous and nondecreasing.
%Here $2^E$ denotes the family of all subsets of $E$.
\ma{For a strategy $S$, let $\boldsymbol \chi^S\in\{0,1\}^E$ denote its incidence vector, which satisfies}
$\chi^S_e=1$ if $e\in S$, and $\chi^S_e=0$ otherwise.
A feasible \emph{strategy flow} \ma{$f$} satisfies $f_{i,S}\ge0$ \ma{for each strategy $S$} and
$\sum_{S\in\F_i}f_{i,S}=d_i$ \ma{for all $i\in I$}. It induces aggregate (\emph{resource}) \emph{load}
\begin{equation}\label{eq:resource-load}
\boldsymbol  x=\sum_{i\in I}\sum_{S\in\F_i}f_{i,S}\boldsymbol \chi^S.
\end{equation}

\begin{definition}[Wardrop equilibrium]\label{def:wardrop}
A feasible flow is an equilibrium if\ma{, for each $i\in I$,} every strategy receiving positive flow
from population $i$ minimizes the strategy cost $\sum_{e\in S}c_e(x_e)$ over $S\in\F_i$.
Let $\lambda_i(c,\boldsymbol d)$ denote this minimum strategy cost. 
\cy{It will be shown below that $\lambda_i(c,d)$ is unique, independent of the equilibrium realized.}
\end{definition}
Define
\begin{equation}\label{eq:strategy-polytope}
 P_i\triangleq\conv\{\boldsymbol \chi^S:S\in\F_i\}\text{ for }i\in I,\qquad
 P(\boldsymbol d)\triangleq\sum_{i\in I} d_iP_i=\left\{\sum_{i\in I} d_i \boldsymbol y_i:\boldsymbol y_i\in P_i,i\in I\right\}.
\end{equation}
%The last sum is a Minkowski sum: $p(\boldsymbol d)=\{\sum_i d_i y_i:y_i\in P_i\}$.
A convex combination $\boldsymbol y_i\in P_i$ specifies a unit-demand strategy distribution;
multiplying its coefficients by $d_i$ gives population $i$'s feasible
flow. Thus $P(\boldsymbol d)$ is exactly the set of feasible aggregate resource loads.
The Beckmann formulation \citep{beckmann} identifies equilibrium loads with
minimizers of
\begin{equation}\label{eq:beckmann}
 \Phi_c(\boldsymbol x)=\sum_{e\in E}\int_0^{x_e}c_e(s)\,\mathrm{d}s\quad\text{over }\boldsymbol x\in P(\boldsymbol d).
\end{equation}
\ma{The minimizer exists due to the compactness of $P(\boldsymbol d)$ and continuity of the
potential $\Phi_c(\boldsymbol x)$}. Its gradient is $c(\boldsymbol x)=(c_e(x_e))_{e\in E}$; so a load $\boldsymbol x\in P(\boldsymbol{d})$ minimizes
the convex potential $\Phi_c(\boldsymbol x)$ exactly when $c(\boldsymbol x)\T(\boldsymbol y-\boldsymbol x)\ge0$ holds for every \ma{load
$\boldsymbol y\in P(\boldsymbol{d})$}. For two minimizers \ma{$\boldsymbol{x}$ and $\boldsymbol y$}, applying this inequality in both directions gives
\(
 \sum_{e\in E}(c_e(x_e)-c_e(y_e))(x_e-y_e)\le0\).
By the cost monotonicity, every summand on the left-hand side of this inequality is nonnegative, and hence is zero. If
$x_e\ne y_e$, this forces equality of the two cost values; if
$x_e=y_e$, that equality is immediate. Thus equilibrium resource costs,
and hence all $\lambda_i(c,\boldsymbol d)$, $i\in I$, are unique even when loads are not.
\ma{When costs are strictly increasing}, equal cost values also force equal loads.
\ma{Replacing $(c,\boldsymbol d)$ with $(\bar{c},\bar{\boldsymbol d})$ is called an \emph{improvement} if}
\begin{equation}\label{eq:improvement-order}
 \bar c_e(s)\le c_e(s)\text{ for all }e\in E,s\ge0\quad\text{and}\quad
 0<\bar d_i\le d_i\text{ for all }i\in I.
\end{equation}
\begin{definition}[The strong Braess paradox and immunity]
\label{def:strong-braess}\label{def:immunity}
A \emph{strong Braess paradox} occurs if an improvement satisfying
\eqref{eq:improvement-order} increases \ma{the equilibrium cost of some population, i.e.,
\begin{equation}\label{eq:paradox}
    \lambda_i(\bar{c},\bar{\boldsymbol d})>\lambda_i(c,\boldsymbol d)\text{ for some }i\in I.
\end{equation}
The (fixed) strategy} family $(\F_i)_{i\in I}$ is \emph{immune to the strong Braess paradox}, or simply \emph{immune},
if no admissible pair of instances \ma{$(c,\boldsymbol d)$ and $(\bar{c},\bar{\boldsymbol d})$ satisfies both \eqref{eq:improvement-order} and \eqref{eq:paradox}.} 
%Equivalently, \begin{equation}\label{eq:immunity}  \lambda_i(\bar{c},\bar{\boldsymbol d})\le\lambda_i(c,d)\quad(i\in I) \end{equation} for every choice of continuous nondecreasing nonnegative resource costs and positive demands satisfying \eqref{eq:improvement-order}.
\end{definition}
This is the fixed-sharing version of \citet[Definition~2.7]{fujishige}.
The comparison concerns the cost experienced by a unit of population $i$,
not its demand-weighted total or a sum over populations. Resource
identifications and strategy families remain fixed in both instances.
Throughout the paper, \emph{immune} and \emph{immunity} refer to this
definition unless a different paradox is explicitly specified. Thus
``strong'' qualifies the paradox being excluded.
A system is specified by its fixed strategy families and resource
identifications, whereas an instance additionally specifies costs and
demands. Nonimmunity means that at least one admissible pair of instances
produces a paradox; it does not mean that every improvement is harmful.

For comparison, a \emph{weak Braess paradox} occurs if some equilibrium
resource cost increases, that is,
$\bar c_e(\bar x_e)>c_e(x_e)$ for some $e\in E$. Clearly, a strong paradox implies a
weak \ma{one}. %: if every resource cost weakly decreased, the minimum cost over each fixed strategy family would also weakly decrease. 
Consequently,
immunity to the weak paradox implies immunity to the strong paradox,
as observed in \citet[Remark~2.8]{fujishige}; the converse fails,
as the explicit cost comparison in \cref{ex:network-two-paths} shows.

\subsection{Relevant menus}\label{sec:relevant-menus}
For each population $i\in I$, we call a nonempty subfamily of $\F_i$ a \emph{menu}. \ma{For a positive weight $\boldsymbol q\in\mathbb R^E_{>0}$, which assigns} each
resource \ma{$e\in E$} an additive weight $q_e>0$, the weight of strategy $S\in\mathcal F_i$ is the total weight of its resources ${\boldsymbol  q}^\top\boldsymbol  \chi^S=\sum_{e\in S}q_e$. We pay special attention to the {complete}
minimizing menu
\begin{equation}\label{eq:complete-menu}
 \HH_i(\boldsymbol q)=\operatorname*{arg\,min}_{S\in\F_i}{\boldsymbol  q}^\top\boldsymbol  \chi^S,
\end{equation}
where $\argmin$ denotes the set of all minimizers, including those carrying
zero flow.  \ma{We call a menu $\HH_i$ \emph{positively exposed} if there exists $\boldsymbol q\in\mathbb R^E_{>0}$ such that $\HH_i=\HH_i(\boldsymbol q)$. Equivalently, we say that $\boldsymbol q>\boldsymbol 0$ exposes $\mathcal H_i$. We call}
 a tuple of menus $(\mathcal{H}_i)_{i\in I}$ \emph{jointly positively exposed} if the same vector $\boldsymbol q>\boldsymbol 0$
exposes every \ma{$\mathcal{H}_i$, $i\in I$}. %population's menu. 
%The weights \ma{induce} %describe 
%an additive comparison of strategies; they need not come from a specified numerical game. 
At an equilibrium with positive resource costs, 
$(c_e(x_e):e\in E)$ exposes exactly the strategies \ma{with the} minimum cost.
A \emph{complete minimizing} menu \ma{consists of all strategies whose weights are the minimum}, %includes every strategy tied for minimum weight, 
rather than only the strategies used by a particular equilibrium
flow. \ma{Strategies that are the most cost-effective, even if currently unused, must be preserved in the structural test since they may be utilized upon improvement.} %Such unused tied strategies can become used after an improvement, so they must remain in the structural test.

\ma{An equivalent description can be given} %There is an equivalent description 
using resource counts alone. Take \ma{an equal number of}
%equally many 
old and new replacement strategies from each population. If \ma{every old strategy is minimum under $\boldsymbol q$} %the old strategies all minimize $\boldsymbol q$ 
and the replacements \ma{collectively} use no more of any
resource, then \ma{replacing the old strategies with new ones cannot decrease the total weight, because} %their total weight cannot decrease: 
each replacement has at least its population's minimum weight.
Positivity of $\boldsymbol q$ forces equal
resource counts, and completeness \ma{ensures that every replacement also belongs to} %forces every replacement to remain in
the menu. The following definition \ma{captures} %requires 
precisely this property,
without \ma{reference to} %prespecifying 
weights.

\begin{definition}[Trade-closed menus]\label{def:trade-single}\label{def:trade-multi}
A tuple of \ma{menus} $(\HH_i)_{i\in I}$ is \emph{jointly trade-closed} if, for any integers
$m_i\ge0$ and multisets $S_{i,j}\in\HH_i$, $T_{i,j}\in\F_i$, the implication
\begin{equation}\label{eq:trade-multi}
 \sum_{i\in I}\sum_{j=1}^{m_i}\boldsymbol \chi^{T_{i,j}}
 \le\sum_{i\in I}\sum_{j=1}^{m_i}\boldsymbol \chi^{S_{i,j}}
 \quad\Longrightarrow\quad
 \begin{cases}
 \text{the two resource-count vectors are equal,}\\
 T_{i,j}\in\HH_i\text{ for every }i,j
 \end{cases}
\end{equation}
holds coordinatewise. \ma{When $|I|=1$, we call $\mathcal{H}_1$} %For one population, call $\HH$ 
a \emph{trade-closed} menu.
\end{definition}
A multiset allows repeated strategies. \ma{In \eqref{eq:trade-multi},}  $m_i$ counts strategy
copies \ma{rather than} population demand, and the \ma{sum of incidence vectors} counts how often each
resource is used. The replacement preserves the number of copies in each
population separately.
The definition \ma{rules out both a strict saving of resources and an} %prohibits both strict resource saving and 
escape from the
menu at equal resource counts. \Cref{thm:positive-exposure,lem:trade-multi} \ma{show that trade-closed menus are exactly the positively exposed menus, and jointly trade-closed tuples are exactly the jointly positively exposed tuple.}
%prove the converse to the weight argument: these are exactly the positively exposed menus and jointly positively exposed tuples.

%\paragraph{The ambient family matters.} 
\ma{Complete exposure requires equal total weights for menu strategies and strictly greater total weights for outside strategies. Let $\HH$ be a menu. A positive resource-weight vector $\boldsymbol q$ that assigns the same total weight to every strategy in \(\mathcal H\) defines} 
%A positive weight that is constant on $\HH$ is an 
\emph{internal}
conservation law. \ma{For \(q\) to expose \(\mathcal H\) as a complete menu in \(\mathcal F\), every strategy in \(\mathcal F\setminus\mathcal H\) must have strictly greater total weight than any strategy in \(\mathcal H\).} %To expose $\HH$ as a complete menu in $\F$, all outside strategies must also have strictly larger weight. 
For example, \ma{assigning weight one to each resource gives both strategies
in $\mathcal H=\{ab,cd\}$ the same total weight two, but $\mathcal H$ is not
trade-closed in $\mathcal F=\{ab,cd,ac,bd\}$, since replacing the menu
strategies $ab$ and $cd$ by the outside strategies $ac$ and $bd$
preserves all resource counts.}
%$\HH=\{ab,cd\}$ has constant weight under $q=\one$ but is not trade-closed in $\F=\{ab,cd,ac,bd\}$, because $\chi^{ab}+\chi^{cd}=\chi^{ac}+\chi^{bd}$ replaces menu strategies by outsiders at equal counts. 
\Cref{cor:full-family} \ma{shows that $\mathcal H$ is trade-closed in itself if and
only if some strictly positive resource-weight vector assigns the same
total weight to every strategy in $\mathcal H$.} %separates the internal test from the complete-exposure requirement.
\ma{Although the immunity criterion tests only menus exposed by strictly
positive resource weights, it also applies to games whose resource
costs may be zero. \cref{app:regularity} establishes this extension through
an} 
%The use of strictly positive weights loses no generality for the immunity criterion, including zero costs; \cref{app:regularity} supplies the
order-preserving limiting argument.

\subsection{Matroid preliminaries}
%\paragraph{Matroid conventions.}
\label{sec:matroid-primer}
For a real matrix $A$, its \emph{column matroid} $M(A)$ has \ma{the set of the column labels
as its ground set; a set of labels is independent if and only if
the corresponding columns are linearly independent.} %the column labels as elements and linear independence as independence. 
A basis is a maximal
independent set, a circuit a minimal dependent set, a loop a zero column,
and a coloop an element in every basis. Deletion removes columns;
contraction takes the quotient by their span and removes their labels. 
\ma{Let $M$ be a matroid and $X$ a subset of its ground set.
The restriction $M|_X$ is obtained from $M$ by deleting all elements
outside $X$.}
A \emph{minor} is obtained by a sequence of deletions and contractions;
$M\setminus X$ and $M/X$ denote deletion and contraction of $X$.
The rank $r_M(X)$ is the maximum size of an independent subset of $X$;
we omit the subscript when the matroid is clear from the context.
A matrix whose row space is the full orthogonal complement of
$\operatorname{row}(A)$ represents the dual $M(A)^*$.
Cocircuits, the circuits of the dual, are the minimal nonempty supports of
row-space vectors. A pair \ma{of elements} is \emph{parallel} when it is a two-element
circuit, and \emph{in series} when it is a two-element cocircuit.
These conventions and the following standard facts can be found in
\citet{oxley}.
These definitions concern column dependence: the elements in the ground set are column labels, rather than strategies. A pair in series is
characterized by a minimal row-space support of size two, not by
proportional columns.
The uniform matroid $U_{r,n}$ has $n$ elements and a set
is independent if and only if it consists of at most $r$ elements. For a graph $G$,
$M(G)$ is its cycle matroid, in which edges are elements and forests are independent,
and therefore circuits are graph cycles. A parallel extension adds an element parallel
to an existing one; so the deletion of the new element recovers the original
matroid. A series extension is the dual operation, for which contraction of the new
element recovers the original matroid.

\begin{definition}[Connectivity and series-parallel structure]
\label{def:connected}\label{def:sp-matroid}
A nonempty matroid $M$ on ground set $F$ is \emph{connected} if no partition \ma{of $F$} into nonempty
sets $X$ and $Y$ such that $r(F)=r(X)+r(Y)$. \ma{When $|F|\ge2$, $M$ is connected if and only if} any two distinct
elements of $F$ lie in a common circuit of $M$. \ma{The components of $M$ are the inclusion-maximal subsets $C\subseteq F$
such that the restriction $M|_C$ is connected.} % Its components are the maximal connected restrictions. 
A connected matroid with at least two elements is
\emph{series-parallel} if it is obtained from $U_{1,2}$ by series and
parallel extensions.
\end{definition}
For column matroids, \ma{the columns indexed by each component span a
linear subspace, referred to as a \emph{component span}. These}  component spans 
form a direct sum, where every vector in
their sum has a unique decomposition into \ma{one vector from each component span}. % vectors from the component spans. 
In particular, a zero sum of column combinations splits into a zero
sum within each component \ma{span}. Matroid duality preserves connectivity and interchanges
series and parallel relations. A connected matroid with at least two elements has no
loops or coloops. A series-parallel matroid reduces to $U_{1,2}$ by \ma{iteratively} deleting
one member of a parallel pair or contracting one member of a series pair;
each reduction preserves connectivity. Any distinguished element may be
retained by removing \ma{(deleting or contracting)} the other member of a chosen pair.

Matroid $U_{2,4}$ has four elements, with every pair independent and every
triple dependent; $M(K_4)$ is the cycle matroid on the six edges of $K_4$. \Cref{fig:obstructions} illustrates both matroids. The following characterization can be found in \cite[Theorem~10.4.8 and Corollary~12.2.14]{oxley}. %, restricted to connected matroids with at least two elements.
\begin{theorem}[Standard excluded-minor characterization]
\label{thm:standard-sp}
For a connected matroid $M$ with at least two elements,
\begin{equation}\label{eq:sp-excluded-minors}
 M\text{ is series-parallel}
 \quad\Longleftrightarrow\quad
 M\text{ has neither a }U_{2,4}\text{ minor nor an }M(K_4)\text{ minor}.
\end{equation}
\end{theorem}

 We use the connected convention throughout. Excluding the above two minors alone
does not prove connectivity, which we establish separately for relevant menus.

\subsection{Characterization for single population}\label{sec:single-statement}
\ma{Since every strategy uses at least one resource, every nonempty menu
involves at least one resource.} For a menu $\HH$, we omit unused resources and define
\begin{equation}\label{eq:single-incidence}
 E_{\HH}\triangleq\bigcup_{S\in\HH}S,\qquad
 A_{\HH}\triangleq(\chi^S_e)_{S\in\HH,e\in E_{\HH}},\qquad
 B_{\HH}\triangleq[\,\one\mid A_{\HH}\,],\quad N_{\HH}\triangleq M(B_{\HH}).
\end{equation}
\ma{The matrix $A_\HH$ has its rows and columns indexed by strategies and resources, respectively. The first column of $B_{\mathcal H}$, the all-ones vector, is labelled
$g$ and does not correspond to any resource.} %The added column has label $g$ and is not a usable resource. 
It will be used to record
the sum of the coefficients in a combination of strategy rows (the rows of $A_\HH$). \ma{The column matroid $N_\HH$ is called the \emph{augmented incidence column matroid} associated with $\HH$.} Let $\Xi$ stand for
$\aff\{\boldsymbol \chi^S:S\in\HH\}$.Then we have
\begin{equation}\label{eq:affine-column-identity}
 \operatorname{row}(B_{\HH})=\spanop\{(1,\boldsymbol x^\top):\boldsymbol x\in \Xi\}\quad\text{and}\quad
 B_{\HH}\binom{-\rho}{\boldsymbol q}=\boldsymbol0\ \Longleftrightarrow\ {\boldsymbol  q}^\top \boldsymbol x=\rho\,\text{ for all } \boldsymbol x\in \Xi.
\end{equation}
%The right-hand condition means \emph{for every} $x\in U$; $\boldsymbol q$ and $\rho$ are arbitrary in this identity. 
\ma{Indeed, the %Here is the derivation. The 
rows of $B_{\HH}$ are $\bigl(1,\boldsymbol (\chi^S)^\top\bigr)$}, $S\in\HH$. If $\boldsymbol x\in\Xi$, i.e.,
$\boldsymbol x=\sum_{S\in\HH}\alpha_S\boldsymbol \chi^S$ with $\sum_{S\in\HH}\alpha_S=1$, then
\(
 (1,\boldsymbol x^\top)=\sum_{S\in\HH}\alpha_S\bigl(1,(\boldsymbol \chi^S)^\top\bigr)
       \in\operatorname{row}(B_{\HH})
\).
Conversely, each $\boldsymbol\chi^S$ belongs to $\Xi$, which proves equality for the
two spans. The normalization has the useful equivalent form
\[
 \operatorname{row}(B_{\HH})\cap
       \bigl(\{1\}\times\R^{E_{\HH}}\bigr)=\{(1,\boldsymbol x^\top):\boldsymbol x\in \Xi\}.
\]
For the second statement in \eqref{eq:affine-column-identity}, multiplication row by row gives
\[
 B_{\HH}(-\rho,\boldsymbol q^\top)\T=0
 \quad\Longleftrightarrow\quad {\boldsymbol  q}^\top\boldsymbol\chi^S=\rho\quad\text{for all }S\in\HH.
\]
These equalities extend to every affine combination $\sum_{S\in\HH}\alpha_S\boldsymbol\chi^S$ since
${\boldsymbol  q}^\top\sum_{S\in\HH}\alpha_S\boldsymbol\chi^S=\rho\sum_{S\in\HH}\alpha_S=\rho$.
The converse follows by taking $\boldsymbol x=\boldsymbol\chi^S$.
Thus a column dependence with coefficient $-\rho$ at position $g$ records the
constant term of an affine equation on resource loads. This is distinct
from asking whether the strategies are bases of a matroid
(see \cref{app:matroid-comparisons}).
\begin{theorem}[Single population]\label{thm:single}
For a %finite 
\ma{nonempty strategy} family
$\F\subseteq2^E\setminus\{\varnothing\}$, the following statements are
equivalent.
\begin{enumerate}[label=\textup{(\roman*)}]
\item $\F$ is immune to the strong Braess paradox.
\item For every trade-closed menu $\HH$ \ma{in $\F$}, the augmented incidence
column matroid $N_{\HH}$ is series-parallel.
\item  For every trade-closed menu $\HH$ \ma{in $\F$}, $N_{\HH}$ has \ma{neither $U_{2,4}$ nor 
$M(K_4)$ as a minor}.
\end{enumerate}
If the system is not immune, a strong Braess paradox can be
constructed on \ma{the original resource set $E$} %its original resources 
at demand one, with affine resource
costs having strictly positive slopes and intercepts, and a decrease in
the intercept of one resource only.
\end{theorem}
\ma{To check the immunity, testing the full strategy family $\mathcal F$ alone does not suffice;
the series-parallel condition must hold for every trade-closed menu
$\mathcal H\subseteq\mathcal F$. The} %The quantifier over all relevant menus cannot be replaced by a test of the full family alone: its
augmented column matroid \ma{of $\F$} can be
series-parallel while that of a positively exposed submenu is not
\ma{(see \cref{ex:submenus})}. %Every nonempty menu has a nonempty resource set under the standing assumption that strategies are nonempty.

\begin{example}[An immune system beyond matroid bases]\label{ex:abc}
Let $\F=\{ab,c\}$, where $ab$ denotes $\{a,b\}$. The full menu $\F$ is
exposed by $\boldsymbol q=(1,1,2)^\top$, and its augmented matrix is
\[
 B_{\F}=\begin{array}{c|rrrr}
 &g&a&b&c\\\hline
 ab&1&1&1&0\\
 c &1&0&0&1
 \end{array}.
\]
\ma{The columns labelled $a$ and $b$ are identical, and the column labelled
$g$ is the sum of the columns labelled $a$ and $c$.} %Here $a=b$ and $g=a+c$ as column vectors. 
Deleting one of the parallel
columns $a,b$ \ma{from $B_{\mathcal F}$ yields a matrix
whose column matroid is isomorphic to $U_{2,3}$. The matroid $N_{\mathcal F}$ is obtained from $U_{2,3}$ by a parallel
extension and is therefore series-parallel.} %leaves $U_{2,3}$, so the full matroid is series-parallel.
For a singleton menu $\{S\}$, omitting unused resources leaves a
single row of ones on $g$ and the resources of $S$. Its column matroid
is $U_{1,|S|+1}$, which is series-parallel. Together with the full menu $\F$,
these are all nonempty submenus. So \cref{thm:single} certifies immunity. \ma{Thus $\mathcal F$ is immune even though it is not a matroid base
family (note that its two strategies have different cardinalities).}
%The two strategies have different cardinalities and therefore cannot be the bases of a matroid.

%The same example anticipates the affine criterion used in the proof. Its affine equations are $x_a=x_b$ and $x_a+x_c=1$. Setting nonbasis coordinates to zero gives just $(1,1,0)$ and $(0,0,1)$ as distinct basic points. Both are nonnegative, although the whole affine line also contains signed points such as $(2,2,-1)$. The proof will test these basic points rather than all points of the affine hull.
\end{example}

\ma{Matroid duality relates the criterion in \Cref{thm:single} to the
series-parallel structure of networks. Let $G$ be a directed acyclic
network with terminals $s$ and $t$, with every vertex and edge lying on
a directed $s$-$t$ path. Let $\mathcal H$ be the family of all such
paths, and let $G^+$ be obtained by adding a return edge from $t$
to $s$ and then forgetting edge orientations. Then
$N_{\mathcal H}=M(G^+)^*$.
Under this duality, series and parallel extensions in the network
correspond to parallel and series extensions, respectively, in
$N_{\mathcal H}$. \Cref{sec:sp-prelim} proves this identity and explains
the correspondence.}
%\paragraph{Network versus column series-parallel structure.} For the full path family of an irredundant directed acyclic network, the precise identity is $N_{\HH}=M(G^+)^*$, where $G^+$ adds a return edge between the terminals. This duality reverses the elementary series and parallel operations. \Cref{sec:sp-prelim} proves the identity and develops the network interpretation.

\subsection{Characterization for multiple population}\label{sec:multi-statement}
\ma{Given a tuple of menus $\boldsymbol\HH=(\HH_i)_{i\in I}$, let $E_{\boldsymbol{\HH}}\triangleq\bigcup_{i\in I}\bigl(\bigcup_{S\in\HH_i}S\bigr)$ denote the set of resources used by the menus.} %and let $G=\{g_1,\ldots,g_k\}$ consists of demand labels. 
The \emph{marked incidence matrix} \ma{$B_{\boldsymbol{\HH}}$ has one row for each
labelled strategy $(i,S)$ with $i\in I$ and $S\in\mathcal H_i$}: % has one row per labelled strategy $(i,S)$:
\begin{equation}\label{eq:marked-matrix}
 (\ee_i\T,(\boldsymbol\chi^S|_{E_{\boldsymbol{\HH}}})\T),
\end{equation}
\ma{where $\ee_i$ has dimension $k=|I|$ for each $i\in I$. The first $k$ columns of $B_{\boldsymbol{\HH}}$ are labelled
$g_1,\ldots,g_k$, with $g_i$ corresponding to population $i$.
Let $G=\{g_1,\ldots,g_k\}$ denote this set of population labels, which are also called demand labels.
The remaining columns of $B_{\boldsymbol{\HH}}$ are labelled by the resources in
$E_{\mathcal H}$. We define three matroids}
\begin{equation}\label{eq:marked-matroids}
 N\triangleq M(B_{\boldsymbol{\HH}}),\qquad M\triangleq N^*,\qquad R\triangleq M\setminus G.
\end{equation}

Each demand column $g_i$ records population $i$'s own normalization
$\sum_{S\in\HH_i} f_{i,S}=d_i$. %Replacing these columns by one all-ones column would record only total demand and lose the separate population directions.
To interpret the dual, let $L$ be the span of all differences between the incidence vectors of two strategies in the same population menu. \ma{Choose a basis of the vector space $L^\perp$ and form a matrix $H$ by taking
the transposes of these basis vectors as its rows.} % and choose the rows of $H$ to be a basis of $L^\perp$. 
Then $\ker H=L$. \ma{Since $\boldsymbol\chi^S-\boldsymbol  \chi^T\in L=\ker H$ for all $S,T\in\mathcal H_i$,
we have $H\boldsymbol\chi^S=H\boldsymbol  \chi^T$. We denote this common image under $H$ by $\boldsymbol b_i$. Let $f$ be a feasible strategy flow supported on the menus
$(\mathcal H_i)_{i\in I}$, and let $\boldsymbol x$ be its induced aggregate
load vector, as defined in~\eqref{eq:resource-load}. Then
\[
H\boldsymbol x
=\sum_{i\in I}\sum_{S\in\mathcal H_i}f_{i,S}H\boldsymbol \chi^S
=\sum_{i\in I}d_i \boldsymbol b_i,
\]
where the last equality uses $H\boldsymbol \chi^S=\boldsymbol b_i$ for
$S\in\mathcal H_i$ and $\sum_{S\in\mathcal H_i}f_{i,S}=d_i$. Redistributing flow within each population's menu leaves $H\boldsymbol x$
unchanged.} % and every strategy vector of population $i$ has the same image $b_i$ under $H$. Multiplying the aggregate-load formula by $H$ therefore gives \[ Hx=\sum_i d_i b_i.\]
Thus $\boldsymbol b_i$ specifies how one unit of population $i$ enters the above affine
load equations\ma{, regardless of how that unit of flow is
distributed among the strategies in $\mathcal H_i$.} %after within-population strategy exchanges are removed.
The dual \ma{$M=N^*$} is represented by $[-B_{\mathrm{dem}}\mid H]$, where submatrix
$B_{\mathrm{dem}}=[\boldsymbol b_1\ \cdots\ \boldsymbol b_k]$ collects these demand directions
and is distinct from the marked incidence matrix
$B_{\boldsymbol{\HH}}$. \Cref{sec:multi-proof} proves this is indeed the representation of $M$. \ma{The matrix $H$ represents the resource matroid $R=M\setminus G$,
whose connected components partition the set of column labels of $H$.
The subspaces spanned by the columns indexed by these components
form a direct sum. The next definition associates each resource component with a marked
matroid obtained from $M$ by contracting all resources outside that
component and deleting the demand labels that become loops.}
%The resource columns of $H$ split into components with independent spans;the next definition retains each component together with the parts of the demand directions that act in it.

\ma{A \emph{resource block} is a connected component
$C\subseteq E_{\boldsymbol{\mathcal H}}$ of the resource matroid
$R=M\setminus G$. For each resource block $C$, we first contract all resources outside
$C$ in $M$ to obtain a matroid $\widehat M_C$ on $C\cup G$.
Then we delete all demand labels that are loops in $\widehat M_C$ to derive matroid $M_C$, which we call the \emph{marked block}
associated with $C$. Explicitly, 
\begin{equation}\label{eq:block-minor}
\widehat M_C\triangleq M/(E_{\boldsymbol{\mathcal H}}\setminus C)\quad\text{ and }\quad 
M_C\triangleq \widehat M_C|_{C\cup G_C},
\end{equation}
where $G_C\triangleq  \{g_i\in G : g_i \text{ is not a loop of }\widehat M_C\}$}

%\begin{definition}[Resource block]\label{def:resource-block} A \emph{resource block} is a connected component $C$ of the resource restriction $R=M\setminus G$. Its corresponding \emph{marked block} $M_C$ is obtained by contracting all resources outside $C$ and then discarding the demand labels that become loops: \begin{equation}\label{eq:block-minor}  \widehat M_C=(M/(E_{\boldsymbol{\HH}}\setminus C))|_{C\cup G},\quad  G_C=\{g_i:g_i\text{ is not a loop of }\widehat M_C\},\quad  M_C=\widehat M_C|(C\cup G_C). \end{equation} \end{definition}

In the dual representation $[-B_{\mathrm{dem}}\mid H]$, resource $e\in E$ \ma{labels} %has 
column $\boldsymbol h_e$, the $e$-th
column of $H$, and population $i\in I$ \ma{labels} %has 
demand column $-\boldsymbol b_i$. \ma{Each $\boldsymbol b_i$ has a unique decomposition $\boldsymbol b_i=\sum_{\bar C}\boldsymbol b_i^{(\bar C)}$, %$\boldsymbol b_i=\sum_{\bar C} \boldsymbol b_{i,\bar C}$,
where $\bar C$ ranges over the resource blocks of $R$ and $\boldsymbol b_i^{(\bar C)}
\in\operatorname{span}\{\boldsymbol h_e:e\in\bar C\}$. 
%$\boldsymbol b_{i,\bar C}\in\operatorname{span}\{\boldsymbol h_e:e\in \bar C\}$. 
After contracting all resources outside $C$, each demand label
$g_i$ is represented by the column $-\boldsymbol b_i^{(C)}$. Thus $g_i$
becomes a loop if and only if $\boldsymbol b_i^{(C)}=\boldsymbol0$.}
%Component spans form a direct sum, so each $b_i$ has a unique component in the span of block $C$. Contracting the other resources removes the other components. A demand label becomes a loop exactly when its component in $C$ is zero. 
Consequently, the retained demand labels are
parallel exactly when their nonzero component vectors \ma{$\boldsymbol b_i^{(C)}$} are proportional. \ma{The map $\boldsymbol b_i\mapsto\boldsymbol b_i^{(C)}$ is the projection onto
$\operatorname{span}\{\boldsymbol h_e:e\in C\}$ obtained by discarding all other
terms in the direct-sum decomposition. This projection need not
be orthogonal.}
%This is a projection along the other component spans; it need not be an orthogonal projection. 
The following main theorem \ma{imposes its conditions on the marked matroid
$M_C$ with ground set $C\cup G_C$, rather than on the resource
restriction $R|_C$ alone.} %tests the marked matroid $M_C$, which includes those retained labels, rather than the resource restriction on $C$ alone.

\begin{theorem}[\ma{Multiple} populations with fixed sharing]\label{thm:multi}
The fixed strategy system $(\F_i)_{i\in I}$ is immune to the strong Braess paradox if and only
if, for every jointly trade-closed menu tuple \ma{$(\HH_i)_{i\in I}$ with $\emptyset\neq\HH_i\subseteq\F_i$, $i\in I$} and every
corresponding resource block $C$, both of the following conditions hold:
\begin{enumerate}[label=\textup{(\roman*)}]
\item \ma{All} %nonloop 
demand elements of $M_C$ belong to \ma{a single} %one 
parallel class.
\item $M_C$ is series-parallel.
\end{enumerate}
% Every nonempty resource block has at least one retained demand label, and $M_C$ is automatically connected; connectivity is not another test. In \textup{(b)}, one may equivalently require the absence of $U_{2,4}$ and $M(K_4)$ minors. Once \textup{(a)} holds, one may also retain any one nonloop demand element and delete the others before performing the series-parallel test.
If the system is not immune, a strong Braess paradox can be
constructed on \ma{the original resource set $E$} with $d_i=1$ for every $i\in I$,
affine costs having strictly positive slopes and intercepts on every
resource, and a decrease in exactly one resource-cost intercept.
\end{theorem}

\ma{As shown in \Cref{sec:joint}, for every jointly trade-closed menu tuple
and every associated resource block $C$, the set $G_C$ is nonempty and $M_C$ is connected.
These properties hold without assuming either~\textup{(i)} or~\textup{(ii)}.
Consequently, condition~\textup{(ii)} is equivalent to the absence of
$U_{2,4}$ and $M(K_4)$ minors. Under condition~\textup{(i)}, $M_C$ is series-parallel if and only
if the matroid obtained by deleting all but one demand element
is series-parallel.}

The two conditions in \Cref{thm:multi} have different roles. The series-parallel structure in condition (ii)
controls response signs for one common demand direction. The parallel labels in condition (i)
require the populations acting in a block to have that same direction,
although with different magnitudes. This compatibility condition cannot
be omitted, even when the full marked matroid is series-parallel. Blocks
are defined by column dependencies, not simply by which populations
share a physical resource.
For $\F_1=\{ab,c\}$, $\F_2=\{a\}$, $\F_3=\{b\}$, the full marked
matroid is series-parallel but its demand labels occupy different parallel
classes in one resource block. In the quotient coordinates just described,
\begin{equation}\label{eq:marked-preview}
 H=\begin{pmatrix}1&0&1\\0&1&1\end{pmatrix},\qquad
 \boldsymbol b_1=\binom11,\quad\boldsymbol  b_2=\binom10,\quad \boldsymbol b_3=\binom01.
\end{equation}
\ma{The resource matroid has a single block $C$. The marked matroid
$M_C$ is series-parallel, so condition~\textup{(ii)} holds.
However, the three demand vectors are not all proportional,
so condition~\textup{(i)} fails. \Cref{ex:marked-first} constructs a strong
Braess paradox for this strategy system.}
%The resource columns form a connected rank-two matroid, but these three demand vectors are not proportional. Their sum $(2,2)\T$ passes the aggregate sign test, while $b_2=h_c-h_b$ has a negative coefficient on the basis $\{b,c\}$. This small matrix explains why the theorem must keep each demand label. \Cref{ex:marked-first} gives the full marked matrix and the actual equilibrium cost increase; the same resource matrix supplies the intuition before \cref{lem:ray}.

\begin{corollary}[Unit demands and affine costs suffice]\label{cor:affine-test}
For fixed finite strategy families, the following statements are equivalent:
\begin{enumerate}
\item[\textup{(i)}]
\ma{The strategy system is immune to the strong Braess paradox
under all reductions in resource costs and demands.}

\item[\textup{(ii)}]
\ma{For every positive demand vector and every choice of resource
cost functions, reducing resource costs while keeping demands
fixed does not increase any population's equilibrium cost.}

\item[\textup{(iii)}]
\ma{In every game on this strategy system with all demands equal
to one and affine resource costs having strictly positive
slopes and intercepts, lowering any one intercept while
keeping it strictly positive does not increase any
population's equilibrium cost.}
\end{enumerate}
\end{corollary}

%\begin{corollary}[Unit-demand affine tests suffice]\label{cor:affine-test}
%For fixed finite strategy families, the following are equivalent:
%\begin{enumerate}[label=\textup{(\roman*)}]
%\item immunity to all cost and demand reductions;
%\item immunity to all cost reductions with the demand vector held fixed,
%for every positive demand vector;
%\item with every demand equal to one, no population is harmed by decreasing
%one intercept of positive affine resource costs, provided all slopes and
%intercepts remain strictly positive.
%\end{enumerate}
%\end{corollary}
Indeed, the first condition implies the second, which in turn implies the third.
If the first fails, the witness assertion of \cref{thm:multi} violates the
third. For the single population case, the same conclusion follows from \cref{thm:single}.

%This equivalence concerns a fixed strategy system and quantifies over all costs in the specified classes. To disprove immunity, some unit-demand affine instance and one-intercept reduction always suffice. It does not follow that one chosen numerical test can certify immunity, or that an arbitrary given intervention is represented by that witness.

\section{Positive exposure and trade closure}\label{sec:menus}
\label{sec:positive-exposure}

The menus introduced in \cref{sec:model} contain all strategies of the
minimum weight for some $\boldsymbol q>\boldsymbol 0$. Their convex hull is the part of the
strategy polytope where ${\boldsymbol  q}^\top \boldsymbol x$ attains its minimum: a face exposed by
the positive weight vector $\boldsymbol q$. We prove that a specified menu has this
form exactly when no population-preserving replacement can save
resources or leave the menu at equal resource counts. This is the
trade-closure property. The equivalence turns a condition on arbitrarily
large multisets into finitely many linear constraints, leading to a
polynomial-time test for one menu relative to its full ambient family.

\subsection{Separation and the single-population \ma{case}%equivalence
}
We first establish the equivalence for single population. \ma{If no strictly positive resource weights make the menu exactly the set of minimum-weight strategies, we use the separation lemma below to construct a replacement that violates trade closure. This replacement preserves the number of strategy copies and either reduces some resource counts without increasing any, or preserves all resource counts while introducing an outside strategy.} %The separation lemma below converts failure of a positive exposing weight into a nonnegative relation, which will become a trade after clearing denominators.

\begin{lemma}[Strict separation modulo a subspace]\label{lem:strict-alternative}
Let $L\subseteq\R^E$ be a linear subspace and let $\boldsymbol  v_1,\ldots,\boldsymbol  v_r\in\R^E$,
where $r\ge1$. Exactly one of the following holds:
\begin{enumerate}[label=\textup{(\roman*)}]
\item There is $\boldsymbol  q\in L^\perp$ such that ${\boldsymbol  q}^\top \boldsymbol  v_j>0$ for every $j$.
\item There is $\boldsymbol  \eta\in\R_+^r\setminus\{\boldsymbol0\}$ such that
$\sum_{j=1}^r\boldsymbol  \eta_j\boldsymbol  v_j\in L$.
\end{enumerate}
\ma{If \(\boldsymbol  v_1,\ldots,\boldsymbol  v_r\) have rational coordinates and \(L\) admits a basis of rational vectors, then \(\boldsymbol  \eta\) in \textup{(ii)} can be chosen rational. The coordinates of \(\sum_{j=1}^r\boldsymbol \eta_j\boldsymbol v_j\) in any fixed rational basis of \(L\) are then rational as well.}
%For rational data, the coefficients in \textup{(ii)} and the corresponding representation in a rational basis of $L$ can be chosen rational.
\end{lemma}
%\paragraph{Why these are the two alternatives.}
A vector $\boldsymbol  q\in L^\perp$ assigns weight zero to every direction
in $L$. Projecting each $\boldsymbol  v_j$ onto $L^\perp$ therefore preserves its
weight under every admissible $\boldsymbol q$. If the projections have a convex
combination equal to zero, the corresponding combination of the
original vectors lies in $L$, giving (ii). Otherwise their convex hull
can be strictly separated from zero, giving (i). The alternatives
cannot coexist: the combination in (ii) would have both weight zero,
by orthogonality, and strictly positive weight, by (i).
Appendix~\ref{app:separation-proof} presents the elementary proof, including an
explicit separating vector and the rationality argument.

In the application of \Cref{lem:strict-alternative} below, $L$ \ma{is the linear span of all pairwise differences of the incidence vectors of strategies in the menu. Therefore, a resource weight vector orthogonal to \(L\) assigns the same total weight to every strategy in the menu. If alternative (i) holds, it gives strictly positive resource weights under which the menu is exactly the set of minimum-weight strategies. If (i) fails, alternative (ii) yields a relation with rational coefficients. Rearranging this relation and clearing denominators makes the strategy coefficients nonnegative integers. These integers specify two equally large multisets, one consisting of some strategies in the menu and the other consisting of available replacement strategies. The replacement either decreases at least one resource count without increasing any, or preserves all resource counts while introducing an outside strategy. Either scenario violates trade closure.} %consists of within-menu differences. Orthogonality to $L$ makes menu weights equal; the unit vectors require positive resource weights, and outside differences require strictly larger weights outside the menu. Rational coefficients allow a failed separation to become integer strategy multiplicities.

\begin{theorem}[Trade closure and positive exposure]
\label{thm:positive-exposure}
For a finite family $\F\subseteq2^E\setminus\{\varnothing\}$ and a nonempty menu
$\HH\subseteq\F$, the following are equivalent:
\begin{enumerate}[label=\textup{(\roman*)}]
\item $\HH$ is trade-closed in $\F$.
\item There are $\boldsymbol  q\in\R_{>0}^E$ and $\rho\in\R$ such that
\begin{equation}\label{eq:strict-exposure}
{\boldsymbol  q}^\top\boldsymbol\chi^S=\rho\quad\text{for all }S\in\HH\qquad\text{and}\qquad
{\boldsymbol  q}^\top\boldsymbol  \chi^T>\rho\quad\text{for all }T\in\F\setminus\HH.
\end{equation}
\item For some $\boldsymbol  q\in\R_{>0}^E$,
\begin{equation}\label{eq:exposed-menu}
\HH=\left\{S\in\F:{\boldsymbol  q}^\top\boldsymbol\chi^S=\min_{T\in\F}{\boldsymbol  q}^\top\boldsymbol  \chi^T\right\}
     =\operatorname*{arg\,min}_{S\in\F}{\boldsymbol  q}^\top\boldsymbol\chi^S.
\end{equation}
\end{enumerate}
%Here and throughout, $\argmin$ denotes the \emph{set of all minimizers}.
\end{theorem}
\begin{proof}
Statements (ii) and (iii) are clearly equivalent. %express the same condition: $\rho$ is the minimum weight and exactly the strategies in $\HH$ attain it. 
To prove
(ii)$\Rightarrow$(i), recalling \cref{def:trade-single}, we consider a replacement represented by a multiset $P$ of $m$ menu strategies and
a multiset $Q$ of $m$ available replacement strategies. Let
$\boldsymbol  n(P)$ and $\boldsymbol  n(Q)$ denote their resource-count vectors, respectively. We assume $\boldsymbol  n(Q)\le\boldsymbol  n(P)$.
Every available strategy has weight at least $\rho$, and every strategy in
$P$ has weight exactly $\rho$. Thus
\[
m\rho\le {\boldsymbol  q}^\top \boldsymbol  n(Q)\le {\boldsymbol  q}^\top \boldsymbol  n(P)=m\rho.
\]
The first inequality uses the minimum weight of each replacement;
the second uses the resource-count comparison; and the final equality
uses the weight of each old strategy. Therefore, equality holds throughout.
Because $\boldsymbol q>\boldsymbol 0$, the resource-count comparison has no strict coordinate,
giving $\boldsymbol  n(Q)=\boldsymbol  n(P)$. Every replacement must also have
weight $\rho$ and hence belong to the complete menu $\HH$, yielding (i).

To prove the reverse implication, we fix $S^0\in\HH$, and define
\[
L\triangleq\spanop\{\boldsymbol \chi^S-\boldsymbol \chi^{S^0}:S\in\HH\}.
\]
A weight vector is constant on $\HH$ exactly when it lies in
$L^\perp$. It is positive on every resource and gives every outsider
greater weight than $S^0$ exactly when it has positive inner product
with all vectors in the following list:
\[
\{\ee_e:e\in E\}\ \cup\
\{\boldsymbol  \chi^T-\boldsymbol \chi^{S^0}:T\in\F\setminus\HH\}.
\]
The list is nonempty because all strategies use resources. Suppose
that no such weight vector exists. \Cref{lem:strict-alternative} provides
rational coefficients
$\boldsymbol\alpha\in\R_+^E$ and $\beta_T\ge0$, not all zero, such that
\[
\boldsymbol\alpha+\sum_{T\notin\HH}\beta_T(\boldsymbol  \chi^T-\boldsymbol \chi^{S^0})\in L.
\]
The vector $\boldsymbol\alpha=\sum_e\alpha_e\ee_e$ collects the coefficients
of the resource unit vectors; the $\beta_T$'s are the coefficients of
outside-strategy differences. To \ma{construct a replacement that violates trade closure, we} %obtain a trade, 
write the representation
of this vector in $L$ as
$\sum_{S\in\HH\setminus\{S^0\}}\mu_S(\boldsymbol \chi^S-\boldsymbol \chi^{S^0})$.
Set $a_S=\mu_S$ for $S\ne S^0$, and set
$a_{S^0}=\sum_{T\notin\HH}\beta_T-
          \sum_{S\in\HH\setminus\{S^0\}}\mu_S$.
Collecting the $\boldsymbol \chi^{S^0}$ terms, we have
\begin{equation}\label{eq:rational-trade}
\sum_{S\in\HH}a_S\boldsymbol \chi^S
 =\sum_{T\notin\HH}\beta_T\boldsymbol  \chi^T+\boldsymbol\alpha,
\qquad
\sum_{S\in\HH}a_S=\sum_{T\notin\HH}\beta_T.
\end{equation}
Let us write $a_S=a_S^+-a_S^-$, where $a_S^+,a_S^-\ge0$.
The coefficient identity gives
$\sum_Sa_S^+=\sum_Sa_S^-+\sum_T\beta_T$.
Thus moving negative $a_S$ to the right preserves the number of strategy
copies on both sides. Clearing denominators, we see that the two sides define equally large multisets $P$ and $Q$, with $P$ drawn from
$\HH$, $Q$ drawn from $\F$, and $\boldsymbol  n(P)-\boldsymbol  n(Q)$ nonnegative. Their common
cardinality $|P|=|Q|$ is positive; otherwise all strategy coefficients vanish,
forcing \ma{\(\boldsymbol\alpha=\boldsymbol0\) and \(\beta_T=0\) for every \(T\in\mathcal F\setminus\mathcal H\).} %$\alpha=\beta=0$. 
If $\boldsymbol\alpha\ne\boldsymbol0$, \ma{replacing the menu multiset \(P\) by \(Q\) strictly decreases at least one resource count without increasing any other.} %the trade strictly saves a resource. 
If $\boldsymbol\alpha=\boldsymbol0$, \ma{replacing \(P\) by \(Q\) leaves every resource count unchanged, and some $\beta_T>0$, which implies that \(Q\) contains a strategy outside \(\mathcal H\).} %and it escapes the menu while preserving counts. 
Both contradict (i). Hence (ii) holds with
$\rho={\boldsymbol  q}^\top \boldsymbol \chi^{S^0}$.
\end{proof}

The following corollary is immediate from an application of \cref{thm:positive-exposure} with ambient family $\F=\HH$.
\begin{corollary}[The internal \ma{conservation law}%conservation-law test
]\label{cor:full-family}
A nonempty family $\HH$ is trade-closed in itself if and only if
there exist vector $\boldsymbol q>\boldsymbol 0$ and number $\rho$ with
${\boldsymbol  q}^\top\boldsymbol \chi^S=\rho$ for every $S\in\HH$.
\end{corollary}

\ma{Recall from the example in \Cref{sec:relevant-menus} (menu $\HH=\{ab,cd\}$ in family $\F=\{ab,cd,ac,bd\}$) that, for a proper menu, an internal conservation law alone does not ensure complete exposure; the strict inequalities for outside strategies in \eqref{eq:strict-exposure} are also required.}
%For a proper menu, constancy alone is insufficient: the example $\F=\{ab,cd,ac,bd\}$ and $\HH=\{ab,cd\}$ in \cref{sec:model} admits the constant weight $q=\one$ but fails the no-escape condition. The outside inequalities in \eqref{eq:strict-exposure} are therefore indispensable.
Geometrically, the affine hull of the menu lies in
$\{\boldsymbol x:{\boldsymbol  q}^\top \boldsymbol x=\rho\}$, and every outside strategy lies strictly on the
higher-weight side. Equivalently, $\conv(\HH)$, with strategies identified
with their incidence vectors, is a complete positively exposed face of
$\conv(\F)$.

\subsection{Recognizing a specified menu}
\ma{For a specified menu \(\mathcal H\subseteq\mathcal F\), we formulate the weight conditions in \Cref{thm:positive-exposure} as a feasibility problem of a linear equality system involving all strategies in \(\mathcal F\).} %The exposure characterization gives a finite test for a specified menu. We now write that test as a linear feasibility problem, making its dependence on the full ambient family explicit.

\begin{proposition}[Recognition of a specified menu]\label{prop:recognition}
Suppose that \ma{the family of strategies} $\F$ is given explicitly \ma{as a list of strategy incidence vectors, and let \(\varnothing\ne\mathcal H\subseteq\mathcal F\) be a menu.} %by its incidence table.
Fix $S^0\in\HH$. Then menu $\HH$ is trade-closed in $\F$ if and only if \ma{there exists a resource weight vector \(\boldsymbol q\in\mathbb R^E\) satisfying the following linear constraints:}
%the following linear system is feasible:
\begin{equation}\label{eq:single-recognition}
\begin{aligned}
{\boldsymbol  q}^\top(\boldsymbol \chi^S-\boldsymbol \chi^{S^0})&=0 &&\text{for all }S\in\HH\setminus\{S^0\},\\
q_e&\ge1 &&\text{for all }e\in E,\\
{\boldsymbol  q}^\top(\boldsymbol  \chi^T-\boldsymbol \chi^{S^0})&\ge1 &&\text{for all }T\in\F\setminus\HH.
\end{aligned}
\end{equation}
Therefore, this decision problem is solvable in time polynomial in
the length of explicit encoding $(\F,\HH)$.
\end{proposition}
\begin{proof}
Note that a feasible $\boldsymbol q$ to the linear system \eqref{eq:single-recognition} satisfies \eqref{eq:strict-exposure} with
$\rho\triangleq{\boldsymbol  q}^\top\boldsymbol \chi^{S^0}$. Conversely, an exposing weight has positive resource coordinates and positive weight gaps to every outside strategy.
There are finitely many such quantities. \ma{This collection is nonempty because it contains the resource weights \(q_e>0\), $e\in E$ and \(E\ne\varnothing\).} %, and at least the resource coordinates are present. 
\ma{Let \(\delta>0\) be the minimum of all resource weights and all weight gaps between outside strategies and the menu. Replace \(\boldsymbol q\) by \(\boldsymbol q/\delta\).} %Divide $\boldsymbol q$ by their smallest value. 
All lower
bounds become at least one while the within-menu equalities remain zero,
giving \eqref{eq:single-recognition}.
There are $|E|$ variables, $|\HH|-1$ equalities, and
$|E|+|\F\setminus\HH|$ inequalities. The coefficients belong to
$\{-1,0,1\}$ and right-hand sides to $\{0,1\}$. \ma{We can therefore determine in polynomial time whether the linear system \eqref{eq:single-recognition} is feasible;} %Polynomial-time feasibility of rational linear programs therefore applies; 
see
\citet[Chapters~10 and~13]{schrijverLP}.
\end{proof}

A rational vector $\boldsymbol q$ \ma{satisfying \eqref{eq:single-recognition} certifies that \(\mathcal H\) is trade-closed in \(\mathcal F\).}   An infeasibility
certificate for \eqref{eq:single-recognition}, as in \cref{lem:strict-alternative}, yields
\eqref{eq:rational-trade} and hence a \ma{strategy replacement that violates trade closure.} %violating trade. 
Standard rational
polyhedral bounds give certificates of polynomial \emph{binary encoding
length}; the integer multiplicities obtained by clearing denominators
need not be polynomially bounded in value. For a proper menu $\HH$, the input
must include the ambient family, and its size cannot be replaced by $|\HH|$
alone. \ma{Trade closure can be decided in polynomial time for a specified menu, but checking all menus individually may take exponential time, because the} %Likewise, this result checks one specified menu, not every menu simultaneously. The 
number of menus can be exponential.

%Positive exposure explains how dominated strategies are treated.
If $T$ strictly contains an available strategy $S$, then
${\boldsymbol  q}^\top\boldsymbol  \chi^T>{\boldsymbol  q}^\top\boldsymbol \chi^S$ for every $\boldsymbol q>\boldsymbol 0$. Such a $T$ cannot belong to
any positively exposed menu, and hence cannot belong to a trade-closed
menu. %The nonempty-strategy assumption does not exclude zero resource costs; those are covered by the order-preserving approximation in \cref{app:regularity}.

\subsection{Joint exposure and recognition}
For \ma{multiple} populations, \ma{the same strictly positive resource weight vector must expose each population’s menu.} %resource weights must expose all menus at once.
We extend the separation argument for single population by allowing exchanges within each
population while comparing aggregate resource counts. \ma{This establishes the equivalence between joint trade closure and joint positive exposure. For a specified menu tuple, we express the condition for joint exposure as a linear feasibility problem.}%; this yields both the joint characterization and its linear feasibility test.

\begin{lemma}[Joint trade closure]\label{lem:trade-multi}
A tuple $(\HH_i)_{i\in I}$ is jointly trade-closed if and only if there exists a
single vector $\boldsymbol  q\in\R_{>0}^E$ that satisfies
\begin{equation}\label{eq:joint-exposure}
 \HH_i=\operatorname*{arg\,min}_{S\in\F_i}{\boldsymbol  q}^\top\boldsymbol \chi^S
 \quad\text{for every }i\in I.
\end{equation}
\end{lemma}
\begin{proof}
To see that \eqref{eq:joint-exposure} implies trade closure property, let
$\rho_i$ be \ma{the minimum strategy weight for population \(i\) under \(\boldsymbol q\).} %population $i$'s minimum weight. 
Then the old strategies have
total weight $\sum_{i\in I} m_i\rho_i$, and the replacements have at least
that weight. \ma{Since the replacements use no more of any resource and \(\boldsymbol q>0\), their total weight is at most \(\sum_{i\in I} m_i\rho_i\).} %Taking the scalar product of the resource-count comparison with $\boldsymbol q>\boldsymbol 0$ gives the reverse inequality. 
Equality is therefore forced
both in every resource count and in every replacement strategy's
minimum-weight bound, proving joint trade closure.

To prove the converse, we choose $S_i^0\in\HH_i$, put $\boldsymbol a_i^0=\boldsymbol \chi^{S_i^0}$,
and define
\[
 L\triangleq\sum_i\spanop\{\boldsymbol \chi^S-\boldsymbol a_i^0:S\in\HH_i\}.
\]
We apply \cref{lem:strict-alternative} \ma{with this subspace \(L\) and the vector list consisting of the resource unit vectors together with \(\boldsymbol \chi^T-\boldsymbol a_i^0\) for all \(i\in I\) and \(T\in\mathcal F_i\setminus\mathcal H_i\).} %to this $L$, the resource unit vectors, and all outside differences. 
As in the proof of
\cref{thm:positive-exposure}, the failure of a separating $\boldsymbol  q\in L^\perp$ that is strictly positive on all
resource unit vectors and all outside differences $\boldsymbol  \chi^T-\boldsymbol a_i^0$ gives a
rational nonnegative combination
\[
 \boldsymbol \alpha+\sum_{i\in I}\sum_{T\in\F_i\setminus\HH_i}
          \beta_{i,T}(\boldsymbol  \chi^T-\boldsymbol a_i^0)\in L,
\]
with $\boldsymbol \alpha,\boldsymbol \beta$ not all zero. Represent this vector as
$\sum_{i\in I}\sum_{S\in\HH_i\setminus\{S_i^0\}}
\mu_{i,S}(\boldsymbol \chi^S-\boldsymbol a_i^0)$. For each population \ma{$i\in I$, we} %separately, 
set
$a_{i,S}\triangleq\mu_{i,S}$ for $S\ne S_i^0$ and 
$a_{i,S_i^0}\triangleq\sum_T\beta_{i,T}-\sum_{S\ne S_i^0}\mu_{i,S}$. \ma{The coefficients \(a_{i,S}\) are rational and satisfy the following identities}
%Collecting terms gives rational $a_{i,S}$ such that
\[
 \sum_{i\in I}\sum_{S\in\HH_i}a_{i,S}\boldsymbol \chi^S
  =\sum_{i\in I}\sum_{T\notin\HH_i}\beta_{i,T}\boldsymbol  \chi^T+\boldsymbol \alpha,
 \qquad
 \sum_{S\in\HH_i}a_{i,S}=\sum_{T\notin\HH_i}\beta_{i,T}
 \quad\text{for all }i\in I.
\]
 \ma{In the first identity, we move each term \(a_{i,S}\boldsymbol\chi^S\) with \(a_{i,S}<0\) to the right-hand side. For each population \(i\), the second identity ensures that the sums of the strategy coefficients on the two sides are equal after this rearrangement. Clearing a common denominator, we derive equal numbers of old and replacement strategy copies for each population.} %For each population separately, move negative $a_{i,S}$ terms to the right. Its coefficient-sum equality guarantees equally many old and replacement copies. Clearing a common denominator gives integer multiplicities. 
 If $\boldsymbol\alpha\ne\boldsymbol0$, the replacements strictly save a
resource; if $\boldsymbol\alpha=\boldsymbol0$, \ma{the replacement preserves every aggregate resource count. At least one coefficient \(\beta_{i,T}\) is positive, which implies that the replacement for population \(i\) contains a strategy \(T\notin\mathcal H_i\).} %some $\beta_{i,T}>0$ replaces a menu strategy by an outsider at equal counts. 
Either possibility contradicts
\cref{def:trade-multi}. Thus a common exposing vector $\boldsymbol q>0$ exists.
\end{proof}

\begin{corollary}[Recognition of a specified joint tuple]
\label{cor:joint-recognition}
\ma{Suppose that each \(\mathcal F_i\), $i\in I$, is given explicitly by the incidence vectors of its strategies. For each \(i\in I\), let \(\varnothing\ne\mathcal H_i\subseteq\mathcal F_i\) and fix any \(S_i^0\in\mathcal H_i\). Then \((\mathcal H_i)_{i\in I}\) is jointly trade-closed if and only if there exists a resource weight vector \(\boldsymbol q\in\mathbb R^E\) satisfying the following linear constraints}
%For explicitly listed families $\F_i$, a specified tuple $(\HH_i)$ is jointly trade-closed if and only if the following system is feasible, where $S_i^0\in\HH_i$ is fixed for each $i$:
\begin{equation}\label{eq:recognition}
\begin{aligned}
 q_e&\ge1 &&\text{for all }e\in E,\\
 {\boldsymbol  q}^\top(\boldsymbol \chi^S-\boldsymbol\chi^{S_i^0})&=0 &&\text{for all }S\in\HH_i\setminus\{S_i^0\},\\
 {\boldsymbol  q}^\top(\boldsymbol  \chi^T-\boldsymbol\chi^{S_i^0})&\ge1 &&\text{for all }T\in\F_i\setminus\HH_i.
\end{aligned}
\end{equation}
\ma{Feasibility of \eqref{eq:recognition} can be decided in time polynomial in the total binary encoding length of \((\mathcal F_i,\mathcal H_i)_{i\in I}\).}
%The feasibility test is polynomial in the total explicit input size.
\end{corollary}
\begin{proof}\ma{If \eqref{eq:recognition} is feasible, \cref{lem:trade-multi} provides joint trade closure. Conversely, let \(\boldsymbol q>0\) expose all menus and let \(\delta>0\) be the minimum of all resource weights and all weight gaps for outside strategies across populations. Then \(\boldsymbol q/\delta\) satisfies \eqref{eq:recognition}, as in \cref{prop:recognition}. The system \eqref{eq:recognition} has \(|E|\) variables and at most \(|E|+\sum_{i\in I}|\mathcal F_i|\) constraints, with coefficients in \(\{-1,0,1\}\) and right-hand sides in \(\{0,1\}\). Its feasibility can therefore be decided in polynomial time.}
%Use \cref{lem:trade-multi} and scale by the minimum of all positive coordinates and outside gaps, exactly as in \cref{prop:recognition}. The system has $|E|$ variables and at most $|E|+\sum_{i\in I}|\F_i|$ constraints with coefficients in $\{-1,0,1\}$.
\end{proof}

\subsection{\ma{Matroid connectivity}%Why connectivity is automatic
}
The following fact justifies the connected-matroid convention in
\cref{thm:single}. It uses only the internal no-saving condition, not the
outside inequalities needed for complete exposure.
\begin{proposition}[Automatic connectivity from no resource saving]
\label{prop:automatic-connected}
\ma{Let \(\mathcal H\) be a nonempty menu. Suppose that, for every integer \(m\ge1\) and every choice of strategies \(S_1,\ldots,S_m,T_1,\ldots,T_m\in\mathcal H\), with repetitions allowed, the coordinatewise inequality \(\sum_{j=1}^m\boldsymbol\chi^{T_j}\le\sum_{j=1}^m\boldsymbol\chi^{S_j}\) implies \(\sum_{j=1}^m\boldsymbol\chi^{T_j}=\sum_{j=1}^m\boldsymbol\chi^{S_j}\)}
%Let $\HH$ be a nonempty menu. Suppose that, for every two equally large multisets of strategies \emph{both drawn from $\HH$}, \[ \sum_j\chi^{T_j}\le\sum_j\chi^{S_j} \quad\Longrightarrow\quad \sum_j\chi^{T_j}=\sum_j\chi^{S_j}.\]
Then there exist $\boldsymbol  q\in\R_{>0}^{E_{\HH}}$ and $\rho>0$ such that
$A_{\HH}\boldsymbol q=\rho\one$, and the matroid $N_{\HH}$ is connected.
In particular, every trade-closed menu has a connected augmented
incidence column matroid. %Neither replacements from outside the menu nor the no-escape clause in \cref{def:trade-single} are needed for this conclusion.
\end{proposition}

\begin{proof}
The hypothesis is exactly trade closure of $\HH$ in itself. %, because there are no outside strategies in this internal test. 
By
\cref{cor:full-family}, $A_{\HH}\boldsymbol q=\rho\one$ for some $\boldsymbol q>\boldsymbol 0$.
\ma{Since every menu strategy \(S\) is nonempty and all resource weights are strictly positive, \(\rho=\sum_{e\in S}q_e>0\).}%Every menu strategy is nonempty, so its strictly positive weight gives $\rho>0$.

The conservation equation is precisely
$B_{\HH}(-\rho,\boldsymbol q)\T=\boldsymbol 0$. \ma{In this linear dependence, the all-ones column labelled \(g\) has coefficient \(-\rho<0\), while each resource column labelled \(e\) has coefficient \(q_e>0\).} %: the negative coefficient belongs to the demand column $g$, and all resource coefficients are positive.
Suppose a component $C$ did not contain $g$. Since column spans of
components form a direct sum, restricting this dependence to $C$ gives
$\sum_{e\in C}q_e(A_{\HH})_e=0$. Each resource column is nonzero and
nonnegative and every $q_e$ is positive, so that sum cannot vanish. \ma{Hence all elements of \(N_{\mathcal H}\) lie in the same connected component as \(g\), which implies that \(N_{\mathcal H}\) is connected.}
%Thus every element belongs to the component of $g$.
\end{proof}

\ma{It is worth noting that the hypothesis of \Cref{prop:automatic-connected} is trade closure of \(\mathcal H\) in itself and imposes no condition on strategies outside \(\mathcal H\).}

\ma{Adding the all-ones column alone does not guarantee connectivity. For \(\mathcal H=\{a,b,ab\}\), the matrix \(B_{\mathcal H}\) is nonsingular. So \(N_{\mathcal H}\) has three coloops and is disconnected.}
%Adding a column of ones alone does not guarantee connectivity. For $\HH=\{a,b,ab\}$ the augmented matrix is nonsingular, so its three columns are coloops. 
The positive conservation law, rather than augmentation by
itself, supplies the needed connection to $g$.

\section{Immunity for single population}\label{sec:single-proof}

By \cref{thm:positive-exposure}, the trade-closed menus are exactly the
complete minimizer menus of positive resource weights. In this section, we prove that
these menus determine immunity for single population. First, \cref{lem:bridge} \ma{shows that the augmented column matroid is series-parallel if and only if all basic solutions of the menu’s affine load equations have nonnegative coordinates.} %identifies series-parallel column structure with nonnegativity of the menu's affine basic points. 
Second, \cref{lem:cb,prop:single-analytic} turn \ma{the coordinates of these basic solutions} %those basic coordinates 
into coefficients of equilibrium cost changes. A negative
coefficient produces a paradox, while nonnegative coefficients prevent any
population-cost increase. Finally, we extend \ma{the conclusion that equilibrium costs do not increase to the full improvement path, allowing changes in the minimizing menu. An approximation argument then extends this conclusion to general continuous nondecreasing resource cost functions (see Appendix \ref{app:regularity} for more details).} %\Cref{app:regularity}this local calculation through changes of menu and to all continuous nondecreasing costs allowed by the model.

\subsection{\ma{A bridge between affine geometry and matroid structure}%The affine--matroid bridge
}
This subsection \ma{establishes our} %contains the 
main combinatorial lemma. The reader may keep the
following picture in mind. \ma{For each \(e\in E\), the corresponding \emph{coordinate hyperplane} is \(\{\boldsymbol x\in\mathbb R^E:x_e=0\}\), the set of vectors whose \(e\)-th coordinate is zero.} %A coordinate hyperplane is the set $x_e=0$.
The arrangement of these hyperplanes in an affine space $U$ divides $U$
into \emph{chambers}, the connected regions left after removing the
nonempty hyperplanes. Within a chamber, every nonconstant coordinate has
a fixed nonzero sign; its closure allows the corresponding coordinates
to be zero. A chamber is bounded if its closure is bounded.
The equations $H\boldsymbol x=\boldsymbol b$ describe the affine hull of a
complete menu. Setting selected coordinates to zero \ma{restricts attention to solutions of the menu’s affine load equations supported on the remaining resources.} %asks which resources can be removed from the support of a solution of the menu's affine equations.
Such a solution may have negative coordinates and therefore need not be
a feasible load. The vertices of the arrangement are the affine basic
points defined below. \ma{The identity \(\boldsymbol q^\top \boldsymbol x=\rho\) on \(U\), with \(\boldsymbol q>\boldsymbol0\) and \(\rho>0\), ensures that the nonnegative part of \(U\) is bounded.} %The positive equation $q^\top x=\rho$ makes the nonnegative part bounded. 
The lemma identifies series-parallel augmented
column structure precisely with the existence of only one bounded
chamber, namely the chamber whose coordinates are all positive.

Let $H$ be an $m\times n$ real matrix of full row rank $m$, and \ma{let \(\boldsymbol h_e\) denote the column of \(H\) indexed by resource \(e\).} %denote its resource columns by $h_e$. Write 
Recall that $E=\{1,\ldots,n\}$ consists of 
these column indices. The role of $H$ is to provide equations for the
menu's affine hull. It is different from the strategy-incidence matrix
$A_{\HH}$, whose rows list the strategies. Consider
\begin{equation}\label{eq:positive-affine}
 U\triangleq\{\boldsymbol x\in\R^n:H\boldsymbol x=\boldsymbol b\}\text{ with }
 U\cap\R_{>0}^n\ne\varnothing,
 \quad {\boldsymbol  q}^\top \boldsymbol x=\rho\;\forall\boldsymbol x\in U,\text{ where }\boldsymbol q>\boldsymbol 0,\ \rho>0.
\end{equation}
For a set $J$ of column indices of $H$, let $H_J$ denote the submatrix consisting of those
columns and $\boldsymbol z_J$ the restriction of a vector $\boldsymbol z$ to coordinates in
$J$, in the same order. A column basis has $m$ elements; its square
matrix $H_J$ is nonsingular. For every such basis, we define the affine
basic point $\boldsymbol z^J$ by
\begin{equation}\label{eq:basic}
 \boldsymbol z^J_J\triangleq H_J^{-1}\boldsymbol b\qquad\text{and}\qquad\boldsymbol  z^J_{E\setminus J}\triangleq\boldsymbol 0.
\end{equation}
\ma{Given a column basis \(J\) of \(H\), set \(z_e=0\) for \(e\notin J\) and solve \(H_J\boldsymbol z_J=\boldsymbol b\). The resulting vector \(\boldsymbol z\) is the basic point associated with \(J\). A basic point may have negative coordinates, and}
%A basic point solves $H\boldsymbol z=\boldsymbol b$ after setting every nonbasis coordinate to zero. It can be signed; 
it need not be a feasible equilibrium load.
The distinct basic points are exactly the points of $U$ with
\emph{inclusion-minimal support}, i.e., no other point of $U$ has support
properly contained in theirs.  \ma{Ineed, let \(S=\operatorname{supp}\boldsymbol z\). If the columns of \(H\) indexed by \(S\) are linearly dependent,} %If the columns on $S=\supp z$ are dependent, 
there is a nonzero $\boldsymbol v\in\ker H$ supported
on $S$. Choose $e\in E$ such that $v_e\ne0$. Then
$\boldsymbol z-(z_e/v_e)\boldsymbol v\in U$. \ma{Its \(e\)-th coordinate is zero, and all coordinates outside \(S\) remain zero. Therefore, its support is a proper subset of \(S\).} %has coordinate $e$ zero and opens no coordinate outside $S$, reducing support. 
Conversely, if $\boldsymbol z'\in U$ has smaller
support contained in $S$, then $\boldsymbol z-\boldsymbol z'$ is a nonzero kernel vector
supported on $S$, yielding dependence. Thus minimal support is equivalent
to independence of the columns in $S$. Extending $S$ to a basis $J$
gives precisely the solution in \eqref{eq:basic}; a basic solution
cannot have smaller support because its supporting columns are independent.

\ma{To illustrate the definition, we consider again the family
$\mathcal F=\{ab,c\}$ in \Cref{ex:abc}. At unit demand, its feasible
load vectors form the segment joining $(1,1,0)$ and $(0,0,1)$,
with coordinates ordered as $a,b,c$. The affine hull of this segment
is the line $(t,t,1-t)$, $t\in\mathbb R$, described by
\[
\begin{pmatrix}
1 & -1 & 0\\
1 &  0 & 1
\end{pmatrix}
\begin{pmatrix}
x_a\\ x_b\\ x_c
\end{pmatrix}
=
\begin{pmatrix}
0\\ 1
\end{pmatrix}.
\]
The columns of this equation matrix are labelled $a,b,c$, and every
pair forms a column basis. Choosing the basis $\{a,b\}$ and setting
$x_c=0$ gives $(1,1,0)$. Choosing $\{a,c\}$ or $\{b,c\}$ and setting
$x_b=0$ or $x_a=0$, respectively, gives $(0,0,1)$.
Thus there are only two distinct basic points, both nonnegative,
although the affine line also contains points with negative
coordinates, such as $(2,2,-1)$.}

We homogenize $U$ by introducing a scalar coordinate $t$. The resulting linear space is
\[
 \widehat U\triangleq\{(t,\boldsymbol x\T)\T:H\boldsymbol x=t\boldsymbol b\}
          =\spanop\{(1,\boldsymbol x\T)\T:\boldsymbol x\in U\}=:V.
\]
To verify the second equality, each $(1,\boldsymbol x)$ satisfies the homogeneous
\ma{system \(H\boldsymbol x-t\boldsymbol b=0\).} %equations. 
Conversely, if $t\ne0$, then $(t,\boldsymbol x)=t(1,\boldsymbol x/t)$ with $\boldsymbol x/t\in U$.
If $t=0$, \ma{then \(\boldsymbol x\in\ker H\). Taking \(\boldsymbol x^0\in U\), we have \(\boldsymbol x^0+\boldsymbol x\in U\). Both \((1,\boldsymbol x^0+\boldsymbol x)\) and \((1,\boldsymbol x^0)\) are therefore among the spanning vectors. Their difference is \((0,\boldsymbol x)\), which therefore} %choose $\boldsymbol x^0\in U$; a vector $x\in\ker H$ satisfies $(0,\boldsymbol x)=(1,\boldsymbol x^0+\boldsymbol x)-(1,\boldsymbol x^0)$ and also 
belongs to the displayed span.
Let $N(U)$ denote the column matroid represented by any matrix with row space
$V$, where the first column, corresponding to $t$, is labeled by the demand element $g$. The
orthogonal equations of $\widehat U$ are $H\boldsymbol x-\boldsymbol bt=\boldsymbol0$.
The matrix $[-\boldsymbol b\mid H]$ has full row rank $m$, and its kernel is $V$. Therefore,
its row space is exactly $V^\perp$. By the dual representation
convention, it represents the dual matroid. This is the formal version of the ``append an
all-ones demand column'' construction in \cref{sec:single-statement}.

\begin{lemma}[\ma{Bridge between affine geometry and matroids}]\label{lem:bridge}
Under \eqref{eq:positive-affine}, \ma{both matroids \(N(U)\) and \(M([-\boldsymbol b\mid H])\) are connected, and} the following are equivalent:
\begin{enumerate}[label=\textup{(\roman*)}]
\item Every affine basic point $\boldsymbol z^J$ is nonnegative.
\item \ma{\(\{\boldsymbol x\in U:x_e=0\}\), \(e\in E\), form a hyperplane arrangement in \(U\) with} %The coordinate hyperplane arrangement $x_e=0$ in $U$ has
exactly
one bounded chamber.
\item $N(U)$ is a connected series-parallel matroid.
\item $M([\,-\boldsymbol b\mid H\,])$ is a connected series-parallel matroid.
\end{enumerate}
%Both matroids are connected independently of \textup{(i)}.
\end{lemma}

%\paragraph{The implication cycle.} The proof separates the geometric count from the positivity argument: \[  \text{\textup{(i)}}\ \Longrightarrow\  text{\textup{(ii)}}  \ \Longleftrightarrow\ \text{\textup{(iii)}}  \ \Longrightarrow\ \text{\textup{(i)}},  \qquad \text{\textup{(iii)}}\ \Longleftrightarrow\ \text{\textup{(iv)}}. \] The chamber count identifies the matroid class. Positive elimination supplies the last arrow, showing why that class forces \emph{all} basic coordinates to have the correct sign. This last step uses both positivity assumptions in \eqref{eq:positive-affine}; counting bounded chambers alone does not establish it.
\begin{proof}
We first prove connectivity, which is required before applying
the connected-matroid characterizations below. By \eqref{eq:positive-affine}, we may take a positive point
$\boldsymbol x^+\in U$. \ma{Since the identity \(\boldsymbol q^\top\boldsymbol x=\rho\) holds for every \(\boldsymbol x\in U\),} %The equation ${\boldsymbol  q}^\top \boldsymbol x=\rho$ on all of $U$ says that 
$(-\rho,\boldsymbol q)$ is a dependence
among the columns representing $N(U)$, while $(1,\boldsymbol x^+)$ belongs to its
row space and has every coordinate strictly positive.
If a connected component did not contain $g$, the restriction of that
dependence to the component would be a positive dependence among its
resource columns. The coefficient vector of this restricted dependence belongs to the
kernel of the representing matrix. So it is orthogonal to every row-space
vector. Its pairing with $(1,\boldsymbol x^+)$, however, is
$\sum_{e\in C}q_ex_e^+>0$, a contradiction. \ma{Hence all elements of \(N(U)\) lie in the connected component containing \(g\), saying that \(N(U)\) is connected. Note that the dual matroid \(N(U)^*\) is represented by \([-\boldsymbol b\mid H]\). Since matroid duality preserves connectivity, \(M([-\boldsymbol b\mid H])\) is also connected.} %Thus all columns belong to the component of $g$.

\ma{Moreover, (iii) and (iv) are equivalent, because series-parallel matroids are
closed under duality because series and parallel extensions are interchanged.}

%Duality preserves connectivity. We now 

\ma{Subsequently, we prove that (i) implies (ii).}
%\emph{1. Basic points and bounded chambers.}
Set $Q=U\cap\R_+^n$. The \ma{identity \(q^\top\boldsymbol x=\rho\)} %positive equation 
in
\eqref{eq:positive-affine} implies $0\le x_e\le\rho/q_e$ for every $\boldsymbol x\in Q$. So $Q$ is bounded. \ma{The point \(\boldsymbol x^+\) lies in \(\operatorname{ri}Q=U\cap\mathbb R_{>0}^n\). This set is the chamber of the coordinate arrangement in which every coordinate is strictly positive.} %The assumed positive point lies in its relative interior, which is exactly the chamber of the coordinate arrangement having all signs positive. 
The vertices of the arrangement are its zero-dimensional
intersections. At a point~$\boldsymbol z$, fixing every coordinate outside
$S=\supp \boldsymbol z$ to zero leaves the linear system $H_S\boldsymbol x_S=\boldsymbol b$. Its solution
is unique precisely when the columns of $H_S$ are linearly independent.
By the support argument above, \ma{a point \(\boldsymbol z\in U\) is an affine basic point if and only if the columns of \(H\) indexed by \(S=\operatorname{supp}\boldsymbol z\) are linearly independent.} %this is exactly the basic-point condition.
Thus the arrangement vertices are precisely the distinct basic points $\boldsymbol z^J$.

Assuming (i), every vertex of every bounded chamber is then nonnegative. The
closure of a bounded chamber is a bounded polyhedron and hence the convex hull
of its vertices. Its closure must therefore be contained in $Q$. But the
relative interior of $Q$ is one chamber of the arrangement, and two distinct
chambers have disjoint relative interiors. Hence no second bounded chamber can
exist, yielding (ii). %This proves (i)$\Rightarrow$(ii).

\ma{We next show that the number of bounded chambers equals the beta invariant of \(N(U)\).} %\emph{2. From bounded chambers to the beta invariant.}
The restricted coordinate arrangement in $U$ is \emph{essential},
meaning that no nonzero direction is invisible to all its coordinate
\ma{functions.} %: a vector with every coordinate zero is the zero vector.
After homogenization, each coordinate equation \ma{\(x_e=0\), \(e\in E\), defines a linear hyperplane through the origin in \(V\). We also include the hyperplane defined by \(t=0\).} %and the equation $t=0$ becomes a linear hyperplane through the origin in $V$. 
We choose a basis
of $V$\ma{, and form a matrix \(B\) whose rows are the chosen basis vectors. Thus \(B\) has row space \(V\), and every vector \(\boldsymbol y=(t,\boldsymbol x)\in V\) can be written as \(\boldsymbol y=B^\top \boldsymbol a\), where \(\boldsymbol a\) is its coordinate vector in this basis. The coordinate functionals \(\boldsymbol y\mapsto t\) and \(\boldsymbol y\mapsto x_e\), \(e\in E\), therefore have coefficient vectors given by the corresponding columns of \(B\). These columns are the normals of the corresponding hyperplanes in the \(\boldsymbol a\)-coordinates, and their} %The coefficient vectors, or normals, of these coordinate functionals are exactly the columns of a matrix whose row space is $V$. Their 
dependence matroid is $N(U)$. Let $d=\dim U$. \ma{The dependence matroid of these normals is therefore \(N=N(U)\), with ground set \(F=E\cup\{g\}\) and rank \(r\triangleq d+1\). Let \(\psi_N(t)\) denote its characteristic polynomial.} %and let $\psi_N(t)$ be the characteristic polynomial of the rank-$(d+1)$ normal matroid. 
\ma{Let \(\mathcal A\) denote the arrangement in \(U\) consisting of the nonempty coordinate hyperplanes \(\{x\in U:x_e=0\}\), \(e\in E\).} Coning and deconing give $\psi_N(t)=(t-1)\psi_{\mathcal A}(t)$; % for the affine arrangement $\mathcal A$ in $U$; 
see the discussion following
\citet[Corollary~4.8]{stanley}. Zaslavsky's bounded-region
formula \citep{zaslavsky}, stated in \citet[Theorem~2.5]{stanley}, gives
\begin{equation}\label{eq:bounded-beta}
\#\{\text{bounded chambers}\}
 =(-1)^d\psi_{\mathcal A}(1)
 =(-1)^d\psi_N'(1)=\beta(N),
\end{equation}
where $\beta(N)$ is the beta invariant of a rank-$r$ matroid $N$ on
$F$, defined by
$\beta(N)\triangleq(-1)^r\sum_{X\subseteq F}(-1)^{|X|}r_N(X)$.
Its characteristic polynomial has the subset expansion
\[
 \psi_N(t)=\sum_{X\subseteq F}(-1)^{|X|}t^{r-r_N(X)}.
\]
Differentiating at one gives
$\psi_N'(1)=-\sum_{X\subseteq F}(-1)^{|X|}r_N(X)$, because
$\sum_{X\subseteq F}(-1)^{|X|}=0$ for nonempty ground set. Since $r=d+1$, we have 
$(-1)^d\psi_N'(1)=\beta(N)$, proving the last identity in
\eqref{eq:bounded-beta}. The affine characteristic polynomial
$\psi_{\mathcal A}$ is the arrangement polynomial in the cited
coning/deconing and region-count formulas; the identity above specifies
the matroid polynomial used to evaluate it. \ma{If \(x_e=\kappa\) throughout \(U\), then \(\kappa>0\), so the zero set \(\{\boldsymbol x\in U:x_e=0\}\) is empty. After homogenization, \(x_e=\kappa t\) on \(V\). Hence the column indexed by \(e\) is \(\kappa\) times the column indexed by \(g\); moreover, \(e\) and \(g\) are parallel elements of \(N(U)\).}
%A coordinate that is constant on $U$ is a positive constant, so its zero set in $U$ is empty and its homogenized column is parallel to $g$.
Discard these empty hyperplanes and retain just one copy of each repeated
proper hyperplane. This simplifies parallel classes of the central normal
matroid, which preserves its characteristic polynomial and the chamber count. Brylawski's characterization
\citep[Theorem~7.6]{brylawski}, also stated in
\citet[Theorem~2.1]{oxleybeta}, gives $\beta(N)=1$ precisely for connected series-parallel
matroids in this setting. Hence, we see that (ii) and (iii) are equivalent.  
If $\dim U=0$, then
$U$ consists of a single positive point and every resource $e$ is parallel to $g$, and
the conclusion is immediate.

%\emph{3. Duality.} The dual representation is $[\,-b\mid H\,]$. Series-parallel matroids are closed under duality because series and parallel extensions are interchanged. Thus (iii)$\Leftrightarrow$(iv).

%\emph{4. Complete the implication cycle.}
To prove that (iii) implies (i), we retain $g$ during a series-parallel reduction. \ma{Each reduction produces an affine space \(U'\) on fewer resource coordinates. It contains a strictly positive point and satisfies \((\boldsymbol q')^\top\boldsymbol  y=\rho'\) for every \(\boldsymbol y\in U'\), for some \(\boldsymbol q'>0\) and \(\rho'>0\).}
%Each step gives a smaller affine space that still has a positive point and a positive constant equation. 
\Cref{tab:positive-elimination} records
the four possible steps and how minimal supports lift. In the final
one-resource space the only point is $\rho/q_1>0$; lifting through
the reductions proves (i). Appendix~\ref{app:elimination-proof} verifies each
correspondence, including contraction as an affine slice and the case
where a resource is in series with $g$. This completes the proof of the lemma.
\end{proof}

\begin{table}[htbp]
\centering\small
\begin{tabular}{@{}>{\raggedright\arraybackslash}p{.23\textwidth}>{\raggedright\arraybackslash}p{.30\textwidth}>{\raggedright\arraybackslash}p{.39\textwidth}@{}}
\toprule
Pair in $N(U)$ & Reduced affine coordinates & Why basic-point signs survive\\
\midrule
Resources $e,f$ parallel
 & Delete $e$, with $x_e=\kappa x_f$.
 & Both coordinates vanish together; restoration multiplies by $\kappa>0$.\\[5pt]
Resource $e$ parallel to $g$
 & Delete the fixed coordinate $x_e=\kappa>0$.
 & Every support contains $e$; the remaining positive equation has a positive right-hand side.\\[5pt]
Resources $e,f$ in series
 & Contract $e$ and use $y=\alpha x_e+\beta x_f$.
 & A minimal support uses at most one of $e,f$; its lift is $y/\alpha$ or $y/\beta$.\\[5pt]
Resource $e$ in series with $g$
 & Contract $e$ by taking the slice $x_e=0$.
 & The remaining basic points lie on that slice; the exceptional point has sole positive coordinate $x_e=\rho/q_e$.\\
\bottomrule
\end{tabular}
\caption{Positive elimination for the final implication of
\cref{lem:bridge}. Here $\kappa,\alpha,\beta$ are strictly positive.
The reduced positive points and equations, and the precise support
correspondences, are constructed in Appendix \ref{app:elimination-proof}.}
\label{tab:positive-elimination}
\end{table}

\begin{figure}[htbp]
\centering
\begin{tikzpicture}[scale=3]
\fill[safe!16](0,0)--(1,0)--(0,1)--cycle;
\fill[bad!15](1,0)--(1,1)--(0,1)--cycle;
\draw[gray!70,-{Stealth}](-.12,0)--(1.26,0)node[right]{$u$};
\draw[gray!70,-{Stealth}](0,-.12)--(0,1.2)node[above]{$v$};
\draw[resource](1,-.1)--(1,1.16)(-.1,1)--(1.16,1);
\draw[resource](-.1,1.1)--(1.1,-.1);
\node[safe]at(.3,.3){$Q$};
\node[bad]at(.73,.73){$R$};
\fill[bad](1,1)circle(.65pt);
\node[bad,above right]at(1,1){$z$};
\node[anchor=west,align=left]at(1.52,.77){$\boldsymbol x=(1-v,u,v,1-u,1-u-v)$};
\node[anchor=west,align=left]at(1.52,.43){$\boldsymbol z=(0,1,1,0,-1)$};
\end{tikzpicture}
\caption{The Wheatstone menu has two bounded chambers. The positive
chamber $Q$ is the triangle of feasible unit-demand loads. The second
chamber $R$ has a vertex $\boldsymbol z$ with a negative resource coordinate.
Series-parallel menus have exactly one bounded chamber.}
\label{fig:chambers}
\end{figure}

\subsection{Basic solutions and sensitivity signs}\label{sec:basic}

The next step connects affine basic points to equilibrium cost changes. \ma{For the local sensitivity calculation, suppose that each resource-cost function \(c_e\) is differentiable at the equilibrium load \(x_e\), with \(c'_e(x_e)>0\). Define
\[
D\triangleq\operatorname{diag}\bigl(c'_e(x_e):e\in E\bigr)
\qquad\text{and}\qquad W\triangleq D^{-1}.
\]
%At a differentiable instance, the diagonal entries of $D$ are resource-cost slopes, and we write $W=D^{-1}$. 
The vector $k(W,\boldsymbol b)$ defined in \Cref{lem:cb}} below
contains the coefficients multiplying direct resource-cost changes in
\eqref{eq:single-sensitivity}; it is not itself assumed to be a feasible
load. The averaging identity expresses $k(W,\boldsymbol b)$ as a convex combination
of \ma{the affine basic points \(\boldsymbol z^J\) of \(U=\{\boldsymbol x:H\boldsymbol x=\boldsymbol b\}\), as defined in \eqref{eq:basic}.} %basic points. 
Conversely, \ma{for any column basis \(J\) of \(H\), suitable choices of \(W\) make \(k(W,\boldsymbol b)\) approach the associated basic point \(\boldsymbol z^J\). Hence all coordinates of \(k(W,\boldsymbol b)\) are nonnegative for every diagonal matrix \(D\) with strictly positive diagonal entries if and only if all affine basic points of \(U\) are nonnegative. This reduces the former sign condition to the latter examination of finitely many basic points.} %concentrating the weights on one basis approaches its basic point. Hence checking response signs for every positive slope matrix reduces to checking finitely many basic points.

%The identity is valid for any full-row-rank $H$ and any right-hand side $b$; no combinatorial assumption is needed.

\ma{Throughout this subsection, \(W>0\) means that \(W\) is diagonal with strictly positive diagonal entries.}

\begin{lemma}[Positive averaging of basic solutions]\label{lem:cb}
\ma{Let \(H\in\mathbb R^{m\times n}\) have full row rank, let \(\boldsymbol b\in\mathbb R^m\), and let \(W=\operatorname{diag}(w_e:e\in E)\) with \(w_e>0\). Set} %For a diagonal matrix $W=\operatorname{diag}(w_e)$ with $w_e>0$, set
\[
 R\triangleq HWH\T,\qquad k(W,\boldsymbol b)\triangleq WH\T R^{-1}\boldsymbol b.
\]
Then \ma{\(R\) is positive definite, and}
\begin{equation}\label{eq:cb}
 k(W,\boldsymbol b)=\sum_{J}\omega_J(W)\boldsymbol z^J \quad\text{with}
 \quad
 \omega_J(W)\triangleq\frac{\det(H_J)^2\prod_{e\in J}w_e}{\det(HWH\T)}\text{ for all }J,
\end{equation}
where $J$ ranges over \ma{the index sets of} column bases \ma{of $H$} and $\det$ denotes determinant.
%Here $W>0$ means positive diagonal entries, not that its off-diagonal entries are positive. 
\ma{The coefficients in \eqref{eq:cb} satisfy
\[
\omega_J(W)>0\quad\text{for every column basis }J,
\quad\text{and}\quad
\sum_J\omega_J(W)=1.
\]}
%The inverse exists: for $y\ne0$, $y\T HWH\T y=\sum_e w_e(h_e\T y)^2>0$ because $H$ has full row rank. The weights are positive and sum to one. Consequently,
\ma{Moreover, the following equivalence holds:}
\begin{equation}\label{eq:k-basic-equivalence}
 k(W,\boldsymbol b)\ge0\text{ for every }W>0
 \quad\Longleftrightarrow\quad \boldsymbol z^J\ge0\text{ for every column basis index set }J.
\end{equation}
\end{lemma}

%The averaging formula in \eqref{eq:cb} is a standard generalized-inverse identity. To match \citet[Theorem~5]{derezinskiwarmuth}, take $X=W^{1/2}H\T$, transpose the size-$\rank H$ volume-sampling identity, and multiply by $W^{1/2}$ and $\boldsymbol b$. The sampled basis probabilities are proportional to $\det(H_J)^2\prod_{j\in J}w_j$, and the resulting sampled vectors are $\boldsymbol z^J$. The proof in Appendix~\ref{app:averaging-proof} derives the averaging identity directly by Cauchy-Binet formula and explains the concentration step used to construct a paradox.

\noindent\emph{Proof sketch.} \ma{Since \(H\) has full row rank, $\boldsymbol y\T HWH\T \boldsymbol y=\sum_{e\in E} w_e(h_e\T y)^2>0$ for every $\boldsymbol y\ne\boldsymbol0$, saying that \(R\) is positive definite. The Cauchy–Binet formula expands \(\det R\) as a sum over column bases, giving the positive weights \(\omega_J(W)\) that sum to one. Differentiating this expansion after perturbing a column of \(H\) yields \eqref{eq:cb}. For a fixed basis \(J\), set \(w_e=1\) for \(e\in J\) and \(w_e=\varepsilon\) otherwise. As \(\varepsilon\to 0^+\), we obtain \(\omega_J(W)\to1\) and \(k(W,\boldsymbol b)\to\boldsymbol z^J\). The averaging identity and this limiting argument establish the two implications in \eqref{eq:k-basic-equivalence}. In particular, a negative basic coordinate yields a negative resource-cost response coefficient for some positive diagonal \(W\).}

\ma{The averaging identity can also be obtained from the size-\(\operatorname{rank}H\) volume-sampling identity of \citet[Theorem~5]{derezinskiwarmuth}. Set \(X=W^{1/2}H^\top\), transpose their identity, and multiply on the left by \(W^{1/2}\) and on the right by \(\boldsymbol b\). The sampling probabilities are then \(\omega_J(W)\), and the transformed sampled vectors are \(\boldsymbol z^J\). Appendix~\ref{app:averaging-proof} provides the direct derivation using the Cauchy-Binet formula and determinant differentiation, together with the concentration argument.}
%Cauchy--Binet expands $\det(HWH\T)$ as a sum over column bases, giving positive weights of total one. Differentiating the same identity after perturbing one column identifies each coordinate of $k$. For a fixed basis $J$, taking $w_e=1$ on $J$ and $w_e=\varepsilon$ off $J$ makes its weight tend to one and gives $k(W,\boldsymbol b)\to \boldsymbol z^J$. Thus nonnegative basic points give nonnegative averages, while a negative basic coordinate yields a negative response for some strictly positive weights. \Cref{app:averaging-proof} spells out the determinant derivatives and the concentration argument.

For an exposed menu $\HH$ from \cref{thm:positive-exposure}, we put
$G=\conv\{\boldsymbol \chi^S:S\in\HH\}$, discard unused coordinates, and write
$\aff G=\{\boldsymbol x:H\boldsymbol x=\boldsymbol b\}$. To verify \eqref{eq:positive-affine}, we take the average of all
menu incidence vectors. Every retained resource occurs in at least one
strategy. Therefore, this average is strictly positive. The exposing vector
$\boldsymbol q>\boldsymbol 0$ gives all menu strategies the same weight $\rho$, and hence
${\boldsymbol  q}^\top \boldsymbol x=\rho$ holds throughout their affine hull. Since every strategy is
nonempty, $\rho>0$. Finally, \eqref{eq:affine-column-identity} gives
$N(\aff G)=N_{\HH}$. Hence the affine lemma applies to the actual
augmented incidence matroid.

\begin{lemma}[Regularity reduction]\label{lem:regularity-reduction}
Fix finite strategy families. To prove immunity, it suffices to prove the
endpoint comparison for ordered pairs of strictly positive piecewise-affine
costs with strictly positive slopes, on an interval containing all feasible
loads. Under linear interpolation of such a pair and of two positive demand
vectors, the population costs at equilibrium are continuous and piecewise
rational with finitely many pieces. Hence nonpositive derivatives on the
open pieces imply the required endpoint comparison.
\end{lemma}
\begin{proof}
For the finite-piece assertion, \ma{let \(t\in[0,1]\) parameterize the linear interpolation of the cost functions and demand vectors.} \cref{lem:finite-branches} \ma{considers each possible choice of the complete minimizing menu for each population and the affine cost segment containing each resource load. For each such choice, it constructs a candidate load branch whose coordinates are rational functions of the interpolation parameter \(t\). Since there are finitely many strategies and affine cost segments, there are only finitely many such choices. After merging identical candidate branches, any two remaining branches intersect at only finitely many values of \(t\).} %constructs rational candidate load branches from every complete-menu and cost-segment state. There are finitely many states; distinct branches intersect only finitely often. 
Continuity, proved in \cref{lem:cost-continuity}, joins the
pieces, and integration on each piece gives the endpoint inequality. \ma{If a resource load remains at a fixed breakpoint of its cost function over an interval of the interpolation parameter, its derivative on that interval is zero. Either adjacent slope can therefore be used in the sensitivity calculation; see Appendix~\ref{apx:piecewise}.}
%Persistent cost kinks cause no difficulty because the corresponding load derivative is zero, as verified in \cref{app:regularity}.

For the reduction to these costs, interpolate an arbitrary ordered cost
pair on a common increasingly fine grid of the bounded feasible-load
interval. \ma{For each grid, add the same affine term \(\varepsilon(1+s)\) to both interpolants, where \(s\) denotes resource load and \(\varepsilon>0\) tends to zero as the grid is refined. This ensures strictly positive cost values and affine-piece slopes while preserving the order of the cost pair.} %and add the same term $\varepsilon(1+s)$ to both interpolants. They remain ordered, have positive values and slopes, and converge uniformly to the original pair. 
By \Cref{lem:cost-continuity}, \ma{the equilibrium population costs for the approximating cost pairs converge to those for the original pair. Taking limits therefore yields the required cost comparison for the original costs.} %passes the endpoint inequality to the limit. The last subsection of \cref{app:regularity} 
Appendix~\ref{apx:order} gives the full construction, including the
extension to $[0,\infty)$. None of these regularity facts uses the sign
criterion or the matroid characterization.
\end{proof}

\ma{For each positively exposed menu \(\mathcal H\), let \(U_{\mathcal H}\) stand for the affine hull of its strategy-incidence vectors, with unused resource coordinates omitted.}
\begin{proposition}[Single-population analytic criterion]\label{prop:single-analytic}
The \ma{strategy family} %system 
$\F$ is immune if and only if\ma{, for every positively exposed menu \(\mathcal H\subseteq\mathcal F\), all affine basic points of \(U_{\mathcal H}\) are coordinatewise nonnegative.} %every affine basic point of every positively exposed complete menu is nonnegative.
If some basic point has a negative coordinate, one can construct on the
original resource set a strong Braess paradox with unit demand and positive affine
costs by lowering a single resource-cost intercept, as stated in
\cref{thm:single}.
\end{proposition}

\ma{The nonnegativity requirement in \Cref{prop:single-analytic} applies only to affine basic points; other points of the corresponding affine hulls may have negative coordinates. For example, the affine line \(x_a+x_b=1\) contains points with negative coordinates, such as \((2,-1)\), although its two affine basic points, \((1,0)\) and \((0,1)\), are nonnegative.}

For sufficiency, \ma{we connect the initial and final instances by linear interpolation of the resource-cost functions and demand.} %a cost or demand reduction is followed continuously from its initial to its final instance. 
Nonnegative basic points \ma{ensures that the equilibrium cost has a nonpositive derivative on each smooth piece of this interpolation path.} 
%make the equilibrium-cost derivative nonpositive on each smooth piece.
For necessity, we choose costs for which one negative basic coordinate
produces a negative response coefficient. Lowering the corresponding
intercept then increases the equilibrium population cost. %These are statements about all admissible instances of the fixed strategy system; a signed basic point is an algebraic certificate, not a negative physical load.

\begin{proof}[Proof of \Cref{prop:single-analytic}]
For sufficiency, \cref{lem:regularity-reduction} allows us to take an
ordered pair of positive piecewise-affine costs with positive slopes.
We interpolate between the two instances by
\[
    c^{(t)}\triangleq(1-t) c+t\bar{{c}} \quad\text{and}\quad \boldsymbol d^{(t)}\triangleq(1-t)\boldsymbol d+t\bar{\boldsymbol{d}},
 \quad\text{where } 0\le t\le1.
\]
\ma{Fix an open subinterval of the finite partition of \([0,1]\) supplied by Lemma 4.3. If a resource load remains at a fixed cost breakpoint throughout this subinterval, either adjacent slope may be used in the derivative calculation, since the load derivative is zero. Fix any \(t\) in this open subinterval. Let \(\boldsymbol x\) be the equilibrium load vector for the instance \(( c^{(t)},\boldsymbol d^{(t)})\), set \(\boldsymbol p=c^{(t)}(\boldsymbol x)\triangleq( c^{(t)}_e(x_e):e\in E)\) to the resource-cost vector, and let \(\mathcal H\) be the corresponding complete minimizing menu. For this \(\mathcal H\), we omit unused resource coordinates and set  \(G\triangleq\operatorname{conv}\{\boldsymbol\chi^S:S\in\mathcal H\}\). We then choose a full-row-rank matrix \(H\) and a vector \(\boldsymbol b\) such that \(\operatorname{aff}G=\{\boldsymbol z:H\boldsymbol z=\boldsymbol b\}\). Define} %Work on an open piece of the finite subdivision in that lemma, choosing either adjacent slope when a load remains at a cost kink. At the point in question write $p=  c^{(t)}(x)$, let $\HH$ be its complete minimizing menu, and use $H,b$ as above. Put
\[
 D\triangleq\operatorname{diag}\bigl(\partial_x   c^{(t)}_e(x_e):e\in E\bigr)>0,
 \quad W\triangleq D^{-1},\quad\text{and}\quad
 \boldsymbol h\triangleq\partial_t   c^{(t)}(\boldsymbol x)=\bar c(\boldsymbol x)-c(\boldsymbol x)\le\boldsymbol 0,
\]
%Primes denote derivatives with respect to $t$. The partial derivative $h$ holds the load fixed; 
\ma{where \(\boldsymbol h\) is obtained by differentiating the resource costs with respect to \(t\) while holding the load vector fixed;} it is the direct cost improvement. The total
cost derivative also includes the effect of changing loads, namely
$\boldsymbol p'=D\boldsymbol x'+\boldsymbol h$. \ma{For nearby parameter values \(\tau\), write \(\boldsymbol p(\tau)\) and \(\lambda(\tau)\) for the equilibrium resource-cost vector and minimum strategy cost, respectively. For a strategy in the current minimizing menu $\HH$, with incidence vector \(\boldsymbol a\), the gap \(\boldsymbol a^\top\boldsymbol p(\tau)-\lambda(\tau)\) is nonnegative and vanishes at \(\tau=t\). It therefore has a local minimum there, and differentiability gives $\boldsymbol a^\top\boldsymbol p'(t)=\lambda'(t)$.}
%For any strategy in $\HH$, let $\boldsymbol a$ denote its incidence vector. Its gap $\boldsymbol a^\top p(t)-\lambda(t)$ from the minimum strategy cost is nonnegative at every nearby parameter value and zero at the current one. At a differentiability point this is a local minimum of the gap, so its derivative is zero: $\boldsymbol a^\top \boldsymbol p'=\lambda'$. 
Subtracting the equalities for two strategies in $\HH$
shows that $\boldsymbol p'=\boldsymbol p'(t)$ is orthogonal to the differences of their incidence vectors. These differences
span the direction space of $\aff G$, which is $\ker H$; hence
$\boldsymbol p'\in(\ker H)^\perp$.
Every strategy outside $\HH$ has a strictly positive cost gap.
Continuity preserves these finitely many gaps nearby. Thus, nearby
positive-flow strategies all belong to the current menu. Since each
of their incidence vectors satisfies $H\boldsymbol a=\boldsymbol b$, their aggregate load
satisfies \ma{$H\boldsymbol x(\tau)=\boldsymbol b\,\boldsymbol d^{(\tau)}$
for all \(\tau\) sufficiently close to \(t\).} %$H\boldsymbol x=\boldsymbol b\,\boldsymbol d^{(t)}$ locally. 
Differentiation \ma{at $\tau=t$} gives $H\boldsymbol x'=\boldsymbol b\,\boldsymbol d'$\ma{, where $\boldsymbol{d}'=\bigl(\boldsymbol d^{(t)}\bigr)'=-\boldsymbol d+\bar{\boldsymbol  d}$}.
We spell out the solution of these equations. Since $\boldsymbol p'\in(\ker H)^\perp$ and
$(\ker H)^\perp$ is the row space of $H$, we may write $\boldsymbol p'=H\T \boldsymbol y$ for some
vector $\boldsymbol y$. Using $R=HWH\T$ and $\boldsymbol p'=D\boldsymbol x'+\boldsymbol h$, we obtain
\[
 \boldsymbol x'=WH\T\boldsymbol y-W\boldsymbol h,\qquad
 R\boldsymbol y-HW\boldsymbol h=\boldsymbol b\,\boldsymbol d',\qquad
\boldsymbol  y=R^{-1}(\boldsymbol b\,\boldsymbol d'+HW\boldsymbol h).
\]
For any strategy in $\HH$ with incidence vector $\boldsymbol a$, since $H\boldsymbol a=\boldsymbol b$, we have
$\lambda'(t)=\boldsymbol a^\top \boldsymbol p'=\boldsymbol b\T\boldsymbol y$. Substitution and symmetry of $R^{-1}$ provide
\begin{equation}\label{eq:single-sensitivity}
 \lambda'(t)=k(W,\boldsymbol b)\T\boldsymbol  h+
\boldsymbol  b\T(HWH\T)^{-1}\boldsymbol b\,\boldsymbol d'.
\end{equation}
\ma{Since all affine basic points of \(\operatorname{aff}G\) are nonnegative by hypothesis,} %The assumed basic nonnegativity and 
\cref{lem:cb} gives $k(W,\boldsymbol b)\ge\boldsymbol 0$.
It follows that $k(W,\boldsymbol b)\T\boldsymbol  h\le0$. Recall from \Cref{lem:cb} that $R$ is positive definite, which gives
$\boldsymbol b\T R^{-1}\boldsymbol b\ge0$. Moreover, $\boldsymbol d'\le\boldsymbol 0$ because \ma{$\bar{\boldsymbol d}\le\boldsymbol d$}.
Both terms in \eqref{eq:single-sensitivity} are therefore nonpositive.
Since all strategies are nonempty, \ma{each current minimizing menu involves at least one resource. The preceding derivative argument therefore applies to every such menu. The endpoint comparison follows by integrating the derivative inequalities over the smooth subintervals and using continuity. The approximation argument in \Cref{lem:regularity-reduction} extends this comparison to arbitrary continuous nondecreasing resource-cost functions.} %every current menu has at least one used resource and is covered by this calculation. \Cref{lem:regularity-reduction} joins the smooth pieces to obtain the endpoint comparison and then passes that comparison to arbitrary continuous nondecreasing costs.

To prove necessity, \ma{we assume that, for some positively exposed menu \(\mathcal H\subseteq\mathcal F\), the affine hull of its strategy-incidence vectors contains an affine} %suppose a menu has a 
basic point with a negative coordinate. 
By \cref{lem:cb}, we can choose $W>0$ and $e^*\in E$ such that 
$k(W,\boldsymbol b)_{e^*}<0$. Take $\boldsymbol x^0\in\ri G$, which means that $x^0_e>0$ on all used resources $e$. \ma{Let \(\boldsymbol q\in\mathbb R^E_{>0}\) be the vector positively exposing $\HH$, which satisfies $\mathcal H
=\operatorname*{arg\,min}_{S\in\mathcal F}
\boldsymbol q^\top\boldsymbol\chi^S$.}
%and let $\boldsymbol q>\boldsymbol 0$ expose exactly $G$. 
Note that scaling $W$ by a large constant leaves
$k$ unchanged and give positive %: both $W$ and $HWH\T$ scale by that constant, which cancels in $WH\T(HWH\T)^{-1}b$. It makes $D=W^{-1}$ sufficiently small that
$\boldsymbol \gamma\triangleq \boldsymbol q-D\boldsymbol x^0>0$. Set
\begin{equation}\label{eq:affine-witness-costs}
 c_e(s)\triangleq D_{ee}s+\gamma_e\;\;\text{ for each used resource }e \text{ and }s\ge0.
\end{equation}
\ma{For each resource \(e\) that does not appear in any strategy of \(\mathcal H\), we take arbitrary \(\alpha_e>0\) and define $c_e(s)\triangleq\alpha_e s+q_e$
for all $s\ge0$. 
Since \(x_e^0=0\), its initial cost satisfies \(c_e(x_j^0)=q_e\).}
%For an unused resource $e$, assign any positive slope and intercept $q_j$; its load is zero, so its initial cost is also $q_j$.
At unit demand, $\boldsymbol x^0\in G$ can be realized by using strategies in $\HH$, all
of which minimize \ma{the strategy cost under} the cost vector $\boldsymbol q$. Therefore, $\boldsymbol x^0$ is an equilibrium.
We now lower only the intercept of resource $e^*$'s cost by $\delta>0$,
keeping every slope and every other intercept fixed.\ma{For sufficiently small \(\delta>0\), we define the improved resource-cost functions as
\[
\bar c_e(s)\triangleq
\begin{cases}
c_e(s)-\delta,&e=e^*,\\
c_e(s),&e\ne e^*.
\end{cases}
\] Recalling} $R=HWH\T$, we set
\begin{align}
 \boldsymbol v\triangleq W\ee_{e^*}-WH\T R^{-1}HW\ee_{e^*}, \qquad\boldsymbol x^\delta=\boldsymbol x^0+\delta \boldsymbol v, \qquad
  \boldsymbol p^\delta\triangleq \boldsymbol q-\delta H\T R^{-1}HW\ee_{e^*}.\label{eq:witness-xp}
\end{align}
\ma{It follows that} %The two identities needed for this construction follow directly:
\[
 H\boldsymbol v=HW\ee_{e^*}-RR^{-1}HW\ee_{e^*}=\boldsymbol0\qquad\text{and}\qquad
 D\boldsymbol v=\ee_{e^*}-H\T R^{-1}HW\ee_{e^*}.
\]
The first identity $ H\boldsymbol v=\boldsymbol{0}$ means that $\boldsymbol v$ is a direction of $\aff G$. Since $\boldsymbol x^0$ is
in the relative interior, a sufficiently small $\delta$ leaves
$\boldsymbol x^\delta\in\ri G$. The second identity implies that \ma{the resource-cost vector at \(\boldsymbol x^\delta\) after the intercept reduction is} %the improved costs at this load are
$\bar c(\boldsymbol x^\delta)=\boldsymbol q+\delta D\boldsymbol v-\delta\ee_{e^*}=\boldsymbol p^\delta$. %, as asserted.
The change $\boldsymbol p^\delta-\boldsymbol q$ lies in the row space of $H$. Recall that every menu in $\HH$ with incidence vector $\boldsymbol a$ satisfies $H\boldsymbol a=\boldsymbol b$. It follows that every strategy in $\HH$
has the same cost under the resource-cost vector $\boldsymbol{p}^\delta$. \ma{By continuity, each strategy outside \(\mathcal H\) remains strictly more costly than the menu strategies for sufficiently small \(\delta>0\). Since there are only finitely many strategies outside \(\mathcal H\), a single sufficiently small \(\delta>0\) can be chosen so that every strategy outside \(\mathcal H\) has strictly greater cost than every strategy in \(\mathcal H\) under \(\boldsymbol p^\delta\). We additionally require \(\delta<\gamma_{e^*}\), ensuring that the reduced intercept \(\gamma_{e^*}-\delta\), and hence all resource-cost intercepts, are strictly positive.} %The finitely many strict outside gaps and the positive intercepts remain positive for sufficiently small $\delta$.
Because $\boldsymbol x^\delta\in G$ is realizable by menu strategies, it follows from %these facts verify 
the Wardrop conditions that $\boldsymbol x^\delta$ is the equilibrium load of the improved instance $(\bar c,\boldsymbol{d})$.
For any strategy in $\HH$ with incidence vector~$\boldsymbol{a}$, it follows from $H\boldsymbol a=\boldsymbol b$ that its cost
change is $\boldsymbol a^\top(\boldsymbol p^\delta-\boldsymbol q)=-\delta \boldsymbol b\T R^{-1}HW\ee_{e^*}$. \ma{Using the symmetry of \(W\) and \(R^{-1}\) and the definition of \(k(W,\boldsymbol b)\), we obtain $-\delta \boldsymbol b\T R^{-1}HW\ee_{e^*}=-\delta k(W,\boldsymbol b)_{e^*}$.
%The last scalar equals $-\delta k_e(W,b)$ by transposition. 
Thus
\begin{equation}\label{eq:single-cost-increase}
 \lambda(\bar c,1)-\lambda(c,1)=-\delta k(W,b)_{e^*}>0.
\end{equation}
This} is the required one-resource affine witness for the strong Braess paradox.
\end{proof}

\begin{proof}[Proof of \cref{thm:single}]
\ma{By \cref{thm:positive-exposure} and \cref{prop:single-analytic}, the strategy system \(\mathcal F\) is immune if and only if, for every trade-closed menu \(\mathcal H\subseteq\mathcal F\), all affine basic points of the affine hull of the incidence vectors of its strategies are coordinatewise nonnegative. By \Cref{lem:bridge}, the nonnegativity condition is equivalent to requiring that \(N_{\mathcal H}\) be series-parallel for every trade-closed menu \(\mathcal H\subseteq\mathcal F\).}
%By \cref{thm:positive-exposure}, the menus in the theorem are exactly the positively exposed menus. By \cref{prop:single-analytic}, immunity is equivalent to nonnegativity of every basic point on every such menu. The affine--matroid bridge then makes this equivalent to series-parallel structure of each $N_{\HH}$. 
Connectivity holds independently by
\cref{prop:automatic-connected}. It follows from \cref{thm:standard-sp} that 
this equivalent to excluding the two minors in (iii). Finally, the
necessity part of \cref{prop:single-analytic} supplies the stated
unit-demand, one-intercept witness for the paradox whenever the conditions fail.
\end{proof}

%The criterion tests affine basic points, not every point of an affine hull. A negative coordinate at an arbitrary affine point does not by itself certify a paradox. Even the safe line $x_a+x_b=1$ contains negative points, but its basic points are $(1,0)$ and $(0,1)$. 
%Every weighted response is a convex combination of basic points, and each basic point can be approached by concentrating the weights on its basis. These two facts, rather than a characterization of all possible limits, justify the finite sign test. 

\section{Immunity for \ma{multiple} populations}\label{sec:multi-proof}
Our proof of immunity characterization for multiple populations has two steps. First, we express immunity as a sign test on
basic solutions, keeping a separate demand direction for each population.
Second, we show that, within one connected resource block, all directions
passing this sign test must lie on the same ray. This is the reason the
nonloop demand labels must be parallel.

\subsection{Joint exposure and the dual representation}\label{sec:joint}
By \cref{lem:trade-multi}, a jointly trade-closed tuple is exactly a
tuple of complete menus exposed by one common positive resource-weight
vector. Fix such a tuple. Restrict the coordinates to the union of the
resources used by its strategies. This union is nonempty because every
strategy is nonempty. We identify load vectors that differ only by
within-population strategy exchanges. %The resulting quotient is described by an equation matrix $H$, as in the single-population proof, but now the exchange directions of all populations must be included. 
For every $i\in I$, we set
\[G_i\triangleq\conv\{\boldsymbol \chi^S:S\in\HH_i\}\quad\text{and}\quad
 L_i\triangleq\spanop(G_i-G_i),\]
 where $G_i-G_i=\{\boldsymbol u-\boldsymbol v:\boldsymbol u,\boldsymbol v\in G_i\}$ consists of the load changes obtained
by replacing one unit of population $i$'s strategy distribution by another. \ma{Choosing a basis of \((\sum_{i\in I} L_i)^\perp\), let \(H\) be the matrix whose rows are the transposes of these basis vectors. Then \(H\) has full row rank and}
\begin{equation}\label{eq:joint-affine-data}
\ker H=L\triangleq\sum_{i\in I} L_i.
\end{equation}
%Their linear spans add to $L$. Choose the rows of $H$ to be a basis of $L^\perp$. Then $H$ has full row rank and $\ker H=L$: applying $H$ removes precisely the joint exchange directions.

Taking an arbitrary $\boldsymbol a_i\in G_i$, we define $\boldsymbol b_i\triangleq H\boldsymbol a_i$. It follows that $H\boldsymbol a'_i=\boldsymbol b_i$ for all $\boldsymbol a'_i\in G_i$, because $\boldsymbol a_i-\boldsymbol a_i'\in L_i\subseteq\ker H$.
In particular, every feasible load $\boldsymbol x$ generated by these menus satisfies the linear system of equations:
\[
 H\boldsymbol x=\sum_{i\in I} d_i \boldsymbol b_i.
\]
Thus $\boldsymbol b_i$ records how population $i$'s demand enters the equations that
are unchanged by strategy exchanges. Define matrix
$B_{\mathrm{dem}}\triangleq[\boldsymbol b_1\ \cdots\ \boldsymbol b_k]$ and vector $\boldsymbol b\triangleq\sum_{i\in I} \boldsymbol b_i$, which is the right-hand side of the above system of equations when all demands equal one. \ma{Recall from \eqref{eq:marked-matroids} that \(N=M(B_{\mathcal H})\), where \(B_{\mathcal H}\) is the marked incidence matrix of the fixed menu tuple \(\mathcal H=(\mathcal H_i)_{i\in I}\). We now show that its dual has the representation} %These definitions give the dual representation

\begin{equation}\label{eq:dual-H}
 M=N^*=M([\,-B_{\mathrm{dem}}\mid H\,]).
\end{equation}
To verify the dual representation, a row \ma{$(\boldsymbol\eta^\top,\boldsymbol z^\top)$} is orthogonal
to every marked strategy row \ma{of $B_\HH$} precisely when
$\eta_i+\boldsymbol z^\top\boldsymbol \chi^S=0$ for every $S\in\HH_i$.
Subtracting two equations for the same population gives $\boldsymbol z\perp L_i$
for every $i$, implying $\boldsymbol z\in L^\perp$. Hence, there exists a vector $\boldsymbol{y}$ such that $\boldsymbol z=H\T \boldsymbol y$. \ma{Substituting \(\boldsymbol z=H^\top\boldsymbol y\) into the orthogonality equations $\eta_i+\boldsymbol z^\top\boldsymbol \chi^S=0$, $S\in\HH_i$ and using \(H\boldsymbol\chi^S=\boldsymbol b_i\) for every \(S\in\mathcal H_i\) gives \(\eta_i=-\boldsymbol b_i^\top\boldsymbol y\) for each population \(i\).}
%One equation for each population now gives $\eta_i=-\boldsymbol a_i\T H\T \boldsymbol y=-\boldsymbol b_i\T \boldsymbol y$.
Therefore, the orthogonal rows \ma{$(\boldsymbol\eta^\top,\boldsymbol z^\top)$} are exactly
$\boldsymbol y^\top[-B_{\mathrm{dem}}\mid H]$, proving that this matrix represents the dual.

Since the exposing vector is orthogonal to every exchange direction, there exists a vector $\boldsymbol{\ell}$ such that
${\boldsymbol  q}^\top={\boldsymbol{\ell}}^\top H$. It follows that
\begin{equation}\label{eq:halfspace}
 {\boldsymbol{\ell}}^\top \boldsymbol{h}_e>0\quad\text{for all }e\in E_{\boldsymbol{\HH}}.
\end{equation}
In particular, no resource column $\boldsymbol{h}_e$ is a zero vector. So the resource
matroid $M(H)$ has no loops. Let $\boldsymbol a_i^+$ be the barycenter of all
strategy-incidence vectors in $\HH_i$. Clearly, it belongs to $\ri G_i$. Every used
resource occurs in at least one of these strategies. Thus, its coordinate
in the aggregate barycenter is strictly positive. Therefore
\begin{equation}\label{eq:aggregate-positive}
 \boldsymbol u^+=\sum_{i\in I} \boldsymbol a_i^+>\boldsymbol 0\qquad\text{and}\qquad H\boldsymbol u^+=\boldsymbol b.
\end{equation}
For a resource component $C$, let $V_C\triangleq\spanop\{\boldsymbol{h}_e:e\in C\}$. \ma{The column space of \(H\) decomposes as the direct sum of the subspaces \(V_C\), where \(C\) ranges over the connected components of the resource matroid \(M(H)\).}
%The column space is the direct sum of the $V_C$. 
Thus
$\boldsymbol b_i=\sum_C \boldsymbol b_i^{(C)}$ and $\boldsymbol b=\sum_C \boldsymbol b^{(C)}$ have unique component
decompositions. These component projections are along the sum of the
other component spans; they need not be Euclidean orthogonal projections.
Contracting resources outside $C$ takes exactly this quotient. \ma{For each resource component \(C\), we choose a basis of \(V_C\). Let \(H_C\) be the matrix whose columns are the coordinate vectors of \(\boldsymbol h_e\), \(e\in C\), in this basis. Since these original vectors span \(V_C\), the matrix \(H_C\) has full row rank. Since} %Choose a basis of each $V_C$ to write its columns as a full-row-rank matrix $H_C$ in component coordinates.  This decomposition is important conceptually: 
different components have no linear exchange
interaction, comparative statics can be checked block by block.  In
compatible row coordinates, \eqref{eq:block-minor} is represented by
\begin{equation}\label{eq:block-H}
 [\,-\boldsymbol b_{1,C}\ \cdots\ -\boldsymbol b_{k,C}\mid H_C\,],
\end{equation}
with zero demand columns deleted exactly as in \eqref{eq:block-minor}. Moreover
$\boldsymbol b_i^{(C)}=H_C(\boldsymbol a_i^+|_C)$ and $\boldsymbol b^{(C)}=H_C(\boldsymbol u^+|_C)$.
The restriction of ${\boldsymbol{\ell}}$ to a block evaluates every resource
column positively. Since $\boldsymbol u^+|_C>0$, it follows that
${\boldsymbol{\ell}}^\top \boldsymbol b^{(C)}=\sum_{e\in C}u_e^+{\boldsymbol{\ell}}^\top \boldsymbol{h}_e>0$, implying $\boldsymbol b^{(C)}\ne\boldsymbol 0$.
Similarly, $\boldsymbol b_i^{(C)}$ is a nonnegative combination of resource columns;
if it is nonzero, at least one coefficient is positive and
${\boldsymbol{\ell}}^\top \boldsymbol b_i^{(C)}>0$. As $\boldsymbol b^{(C)}=\sum_{i\in I} \boldsymbol b_i^{(C)}$, some demand label is
nonloop in every block. Moreover, proportional nonzero $\boldsymbol b_i^{(C)}$
vectors must have positive proportionality factors. %: the common positive functional rules out opposite orientations.

Each marked block is connected. Its resource restriction is one
connected component by construction. A nonzero demand column is in the
span of the resource columns. A minimal set of resources spanning that
column, together with the demand label, is a circuit, so the demand label
joins the same component. \ma{Thus every marked block is connected without any series-parallel assumption. \Cref{thm:standard-sp} therefore yields the equivalent excluded-minor formulation of} %This proves connectivity before any series-parallel test and justifies the excluded-minor version of
condition~(ii) in \cref{thm:multi}.

\subsection{The joint affine \ma{geometric} criterion}
The next proposition extends \cref{prop:single-analytic} while
retaining each population's demand separately. Reducing one population's demand can change another population's equilibrium cost. We measure these effects by cross
coefficients in \eqref{eq:multi-sensitivity} below. \ma{Furthermore, we will show that, if the affine basic points defined below are coordinatewise nonnegative for every jointly exposed menu tuple, then reducing resource costs or population demands cannot increase any population’s equilibrium cost.}
%we will show that the same basic-solution sign test controls both these coefficients and the response to resource-cost reductions.

Let matrix $H$ be constructed as in \Cref{sec:joint}.
For each column basis $J$ of $H$ and population $i\in I$, we define basic point $ \boldsymbol z^{J,i}$ as follows:
\begin{equation}\label{eq:joint-basic}
 \boldsymbol z^{J,i}_J\triangleq H_J^{-1}\boldsymbol b_i,\qquad \boldsymbol z^{J,i}_{E\setminus J}\triangleq\boldsymbol0.
\end{equation}
\ma{The vectors \(\boldsymbol z^{J,i}\) are the affine basic points of the affine space $\{\boldsymbol z:H\boldsymbol z=\boldsymbol b_i\}$.
Its direction space is \(\ker H=L=\sum_{j\in I}L_j\), which includes strategy-exchange directions from all populations, whereas \(\operatorname{aff}G_i\) has direction space \(L_i\), involving only population \(i\).}
%These are basic points of $a_i+L$, not merely of $\aff G_i=a_i+L_i$. The larger space contains the exchange directions supplied by \emph{all} populations. This is exactly where resource sharing enters the sensitivity test: a perturbation of population $i$ may be compensated by strategy exchanges of population $j$, and therefore the relevant affine space is the joint one.

\begin{proposition}[Joint basic-solution criterion]\label{prop:multi-analytic}
Immunity to the strong Braess paradox is equivalent to the following condition: for every jointly positively exposed tuple of menus \((\mathcal H_i)_{i\in I}\), %for every jointly exposed complete menu tuple:
\begin{equation}\label{eq:all-basic-positive}
 \boldsymbol z^{J,i}\ge\boldsymbol 0
 \quad\text{for every column basis }J\text{ of }H\text{ and every }i\in I.
\end{equation}
If the condition is violated, \ma{a strong Braess paradox can be constructed with unit demand for every population and affine resource-cost functions having strictly positive slopes and intercepts, by decreasing only the intercept of one resource-cost function.} %there is a strong Braess paradox with all demands equal to one and strictly positive affine costs, obtained by lowering one resource intercept, as asserted in \cref{thm:multi}.
\end{proposition}
\begin{proof}
Let $W>0$ be diagonal, $R_W\triangleq HWH\T$, and $\boldsymbol k_i\triangleq WH\T R_W^{-1}\boldsymbol b_i$.
By \cref{lem:cb}, condition \eqref{eq:all-basic-positive} is equivalent
to $\boldsymbol k_i\ge\boldsymbol 0$ for every $W>0$ and every $i\in I$.

For sufficiency, we apply \cref{lem:regularity-reduction} and interpolate the
piecewise-affine costs and demands as in the proof for single population. 
%On an open piece of the finite subdivision, 
\ma{Fix \(t\) in an open subinterval of the finite partition supplied by \Cref{lem:regularity-reduction}. Let \(\boldsymbol x\) be the aggregate equilibrium load vector for the interpolated instance at \(t\).} Let $\boldsymbol p\triangleq c^{(t)}(\boldsymbol x)$ be the
resource-cost vector, and let primes denote derivatives with respect to
$t$. Write $D=W^{-1}$ for the positive diagonal matrix of resource-cost
slopes. The partial derivative $\boldsymbol h\triangleq\partial_t   c^{(t)}(\boldsymbol x)\le\boldsymbol 0$ is the
change in costs with the load held fixed, and the total derivative is
$\boldsymbol p'=D\boldsymbol x'+\boldsymbol h$.

\ma{For each population \(i\in I\), let \(\lambda_i\) denote its equilibrium cost for the interpolated instance at \(t\), so that $\lambda_i
=\min_{S\in\mathcal F_i}
(\boldsymbol\chi^S)^\top\boldsymbol p$. For each currently minimizing strategy of population \(i\), with incidence vector \(\boldsymbol a\), its gap from the minimum strategy cost is \(\boldsymbol a^\top\boldsymbol p-\lambda_i=0\).} %For each current minimizing strategy $\boldsymbol a$ of population $i$, its gap $\boldsymbol a^\top p-\lambda_i$ is nonnegative and zero. 
At a differentiability point, we have % its derivative is therefore zero, so 
$\boldsymbol a^\top \boldsymbol p'=\lambda_i'$.
Subtracting these equalities within each population, we see that $\boldsymbol p'\perp L$.
\ma{For each population \(j\), let \(\mathcal H_j\) be its complete minimizing menu at \(t\), and fix the corresponding matrix \(H\) and vectors \(\boldsymbol b_j\) as constructed in \Cref{sec:joint}. The cost differences between strategies outside these menus and their respective minimum strategy costs are strictly positive at \(t\). Since these differences are continuous and finite in number, they remain strictly positive for all parameter values \(\tau\) sufficiently close to \(t\). The Wardrop equilibrium condition therefore implies that every strategy carrying positive flow at such a parameter value belongs to the corresponding fixed menu. By \(H\boldsymbol\chi^S=\boldsymbol b_j\) for every \(S\in\mathcal H_j\), the aggregate equilibrium load at \(\tau\) satisfies $H\boldsymbol x(\tau)
=\sum_{j\in I}\boldsymbol b_j\,d_j(\tau)$.
Differentiating at \(\tau=t\), we obtain} %The strict gaps to outside strategies persist nearby, so nearby loads are generated by the current menus and satisfy $H\boldsymbol x=\sum_j \boldsymbol b_jd_j$. Differentiating gives
$H\boldsymbol x'=\sum_{j\in I} \boldsymbol b_jd_j'$\ma{, where $d_j'=-d_j+\bar d_j$ for all $j\in I$.} Finally, $\boldsymbol p'\perp L$ enables us to write $\boldsymbol p'=H\T \boldsymbol y$ for some $\boldsymbol y$, which, together with $\boldsymbol p'=D\boldsymbol x'+\boldsymbol h$, gives

\[
\boldsymbol  x'=WH\T\boldsymbol  y-W\boldsymbol h,\qquad
 R_W\boldsymbol y=HW\boldsymbol h+\sum_j\boldsymbol b_jd_j',\qquad \lambda_i'=\boldsymbol b_i\T \boldsymbol y.
\]
It follows from $\boldsymbol y=R_W^{-1}(HW\boldsymbol h+\sum_j\boldsymbol b_jd_j')$ that
\begin{equation}\label{eq:multi-sensitivity}
 \lambda_i'
  =\boldsymbol k_i\T \boldsymbol h+\sum_{j\in I} \boldsymbol b_i\T R_W^{-1}\boldsymbol b_j\,d_j'
  =\boldsymbol k_i\T\boldsymbol  h+\sum_{j\in I} \boldsymbol k_i\T D\boldsymbol k_j\,d_j',
\end{equation}
where the second equality uses the fact that $WDW=W$ and 
\[
\boldsymbol  k_i\T D\boldsymbol k_j
 =\boldsymbol b_i\T R_W^{-1}HWDWH\T R_W^{-1}\boldsymbol b_j
 =\boldsymbol b_i\T R_W^{-1}\boldsymbol b_j.
\]
If all $\boldsymbol k_i$, $i\in I$, are nonnegative, then
$\boldsymbol k_i\T D\boldsymbol k_j=\sum_{e\in E}D_{ee}{k_{i}}_e{k_{j}}_e\ge0$.
The term $\boldsymbol k_i\T \boldsymbol h$ is nonpositive because $\boldsymbol k_i\ge\boldsymbol 0$ and
$\boldsymbol h\le\boldsymbol 0$. Each demand derivative $d_j'$ is nonpositive, while its
coefficient $\boldsymbol k_i\T D\boldsymbol k_j$ is nonnegative. Hence
$\lambda_i'\le0$ for every population $i\in I$. By
\cref{lem:regularity-reduction}, \ma{each population $i$’s equilibrium cost is continuous on \([0,1]\). Its derivative $\lambda_i'$ is nonpositive on every open subinterval of the finite partition. Therefore, the equilibrium cost is nonincreasing on \([0,1]\). Hence each population’s equilibrium cost in the final instance is no greater than in the initial instance. The approximation argument in \cref{lem:regularity-reduction} extends this conclusion to arbitrary continuous nondecreasing resource-cost functions.} %these local inequalities join across menu changes to give the endpoint comparison, and the comparison extends to arbitrary continuous nondecreasing costs.

Conversely, \ma{suppose that condition \eqref{eq:all-basic-positive} fails for a jointly positively exposed tuple of menus \((\mathcal H_j)_{j\in I}\). Let  \(\boldsymbol q\in\mathbb R^E_{>0}\) be a common exposing vector, which satisfies $\mathcal H_j
=
\operatorname*{arg\,min}_{S\in\mathcal F_j}
\boldsymbol q^\top\boldsymbol\chi^S$ for every $ j\in I$.} Because $\boldsymbol z^{J,i}_{e^*}<0$ for some $e^*\in E$, we may take $W$ such that ${k_{i}}_{e^*}<0$.
Set all populations' demands to one. We choose
$\boldsymbol x^0\in\ri(G_1+\cdots+G_k)$ as the sum of relative interior points, which ensures that \ma{$x^0_e>0$ for all
resources $e$ used by strategies in $\cup_{i\in I}\HH_i$.} By scaling $W$, we can construct the positive
affine costs \eqref{eq:affine-witness-costs} so that their vector at
$\boldsymbol x^0$ is the exposing vector $\boldsymbol q>\boldsymbol 0$. Lowering the intercept of $c_{e^*}$ by $\delta$, we derive an improved resource costs $\bar c$.  Moreover, we define $\boldsymbol v$, $\boldsymbol x^\delta$ and $\boldsymbol p^\delta=\bar c(\boldsymbol x^\delta)$ as in
\eqref{eq:witness-xp} with matrix $H$ being constructed as in \Cref{sec:joint}.
Because $\boldsymbol v\in L$, for sufficiently small $\delta>0$ the new load $\boldsymbol x^\delta$ remains
in the relative interior of $G_1+\cdots+G_k$. \ma{Let \(\boldsymbol p^\delta=\bar{\boldsymbol c}(\boldsymbol x^\delta)\). It follows from \(H\boldsymbol\chi^S=\boldsymbol b_j\), \(S\in\mathcal H_j\), that all strategies in each menu \(\mathcal H_j\) have the same cost under \(\boldsymbol p^\delta\). Since \(\boldsymbol q\) jointly exposes the menus and \(\boldsymbol p^\delta\to\boldsymbol q\) as \(\delta\to0^+\), continuity and finiteness of the strategy families imply that, for sufficiently small \(\delta>0\), every strategy in \(\mathcal F_j\setminus\mathcal H_j\) has strictly greater cost than every strategy in \(\mathcal H_j\), for all \(j\in I\). Our choice of  \(\delta>0\) further ensures that it is smaller than the intercept of $c_{e^*}$, which guarantees that all resource-cost intercepts of $\bar c$ remain strictly positive. Since \(\boldsymbol x^0\) and \(\boldsymbol x^\delta\) both belong to \(\operatorname{ri}(G_1+\cdots+G_k)\), each can be realized by a feasible flow with unit demand for every population, with each population using only strategies from its corresponding menu. Every strategy carrying positive flow has minimum cost among all strategies available to its population. Therefore, the feasible flows realizing \(\boldsymbol x^0\) and \(\boldsymbol x^\delta\) are Wardrop equilibria of the initial and improved instances, respectively.  It is easy to check that $\lambda_i(\bar c,\boldsymbol 1)-\lambda_i(c,\boldsymbol 1)=-\delta{k_{i}}_{e^*}>0$, exhibiting a strong Braess paradox of the properties as asserted.} %All menu ties persist, all outside gaps remain positive, and every intercept stays positive. The relative-interior condition ensures that both aggregate loads have feasible decompositions into the populations' menu flows. The preserved ties and outside gaps make these decompositions Wardrop flows. Thus the construction gives actual equilibria, rather than only formal solutions of the affine equations. Population $i$'s equilibrium cost increases by \[  \lambda_i^\delta-\lambda_i^0=-\delta k_{i,e}>0.\]
\end{proof}

\subsection{A ray lemma for an indecomposable resource block}
We now isolate the structural fact that turns the \ma{joint basic-solution criterion in \Cref{prop:multi-analytic}} %analytic criterion 
into the
parallel-label condition. \ma{We state the following construction for a general full-row-rank matrix; the resulting lemma will be applied to the matrix constructed in \Cref{sec:joint}. Let \(H\) be a full-row-rank matrix whose columns \(\boldsymbol h_e\) are indexed by \(e\in E\). In the definition below, \(J\subseteq E\) ranges over all sets of column indices for which \(H_J\) is square and nonsingular.  We define cone
\begin{equation}\label{eq:K}
K(H)\triangleq
\bigcap_{\substack{J\subseteq E\\ H_J\text{ square and nonsingular}}}
\operatorname{cone}\{\boldsymbol h_e:e\in J\}
=
\left\{
\boldsymbol v:
H_J^{-1}\boldsymbol v\ge0
\text{ for every such }J
\right\}.
\end{equation} %For a full-row-rank matrix $H$, whose columns are $\boldsymbol h_e$, $e\in E$, we define the cone
%\begin{equation}\label{eq:K}
% K(H)=\bigcap_{J\text{ a column basis}} \cone\{\boldsymbol{h}_e:e\in J\}=\{v:H_J^{-1}v\ge0\text{ for every basis }J\}.
%\end{equation}
A} vector belongs to $K(H)$ precisely when it has nonnegative coordinates in
\emph{every} column basis.  This property is much stronger than belonging to the
ordinary cone generated by all columns, which only asks for a nonnegative
representation. For a fixed basis \ma{index set} $J$, \ma{every vector \(\boldsymbol v\) in the column space of \(H\) has a unique expression as a linear combination of the columns \(\boldsymbol h_e\), \(e\in J\). Its coefficient vector is \(H_J^{-1}\boldsymbol v\). This proves} %the representation is unique: $v=H_J\alpha$ with $\alpha=H_J^{-1}v$. Hence membership in its cone is exactly $H_J^{-1}v\ge0$, proving 
the equality in \eqref{eq:K}.  The distinction is the source of the rigidity below.  By
\cref{lem:bridge}, if
$\boldsymbol b=H\boldsymbol u^+$ for some $\boldsymbol u^+>\boldsymbol 0$ and \ma{there exists a vector \(\boldsymbol\ell\) such that \(\boldsymbol\ell^\top\boldsymbol h_e>0\) for every \(e\in E\), then} %the columns lie in an open halfspace,
\begin{equation}\label{eq:K-SP}
 \boldsymbol b\in K(H)
 \quad\Longleftrightarrow\quad
 M([\,-\boldsymbol b\mid H\,])\text{ is connected and series-parallel}.
\end{equation}

%\paragraph{Why should only one ray survive?}
A basis is a choice of independent resource columns with which to express
a demand direction $\boldsymbol v$. \ma{The condition \(\boldsymbol v\in\operatorname{cone}\{\boldsymbol h_e:e\in E\}\) means that there exists at least one vector \(\boldsymbol u\ge0\) satisfying \(H\boldsymbol u=\boldsymbol v\). In contrast, membership in \(K(H)\) requires \(H_J^{-1}\boldsymbol v\ge0\) for \emph{every} basis matrix \(H_J\).} %Ordinary feasibility asks for one nonnegative representation. Membership in $K(H)$ requires a nonnegative representation on \emph{every} basis. 
Each basis therefore imposes its own cone of allowed
directions, and $K(H)$ is their intersection. \ma{For \(\boldsymbol v=\boldsymbol b_i\), \cref{lem:cb} shows that each basic point \(\boldsymbol z^{J,i}\) can be approached by the resource-cost response vector \(\boldsymbol k_i\) through suitable choices of the positive diagonal weights. These basic points are algebraic solutions of \(H\boldsymbol z=\boldsymbol b_i\) and need not be realizable as equilibrium loads.} %By \cref{lem:cb}, these basis coefficients are limiting sensitivity responses; they are not assumed to be actual equilibrium loads.

\ma{Consider again the three-population example in~\eqref{eq:marked-preview}.
The resource columns are $\boldsymbol h_a=(1,0)^\top$, $\boldsymbol h_b=(0,1)^\top$,
and $\boldsymbol h_c=(1,1)^\top$, and any two of them form a basis.
For $\boldsymbol  v=(u,v_2)^\top$, the representations using these three
bases are $\boldsymbol v=u\boldsymbol h_a+v_2\boldsymbol h_b$, $\boldsymbol v=(u-v_2)\boldsymbol h_a+v_2\boldsymbol h_c$, and
$\boldsymbol v=(v_2-u)\boldsymbol h_b+u\boldsymbol h_c$. All coefficients in all three
representations are nonnegative if and only if $u=v_2\ge0$.
Thus $K(H)$ consists exactly of the nonnegative multiples
of $(1,1)^\top$. The population vectors are $\boldsymbol b_1=(1,1)^\top$,
$\boldsymbol b_2=(1,0)^\top$, and $\boldsymbol b_3=(0,1)^\top$.
Their sum $\boldsymbol b=\boldsymbol b_1+\boldsymbol b_2+\boldsymbol b_3=(2,2)^\top$ belongs to $K(H)$
and has the strictly positive representation $\boldsymbol b=\boldsymbol h_a+\boldsymbol h_b+\boldsymbol h_c$.
For population~2, however, using the basis $(\boldsymbol h_b\;\boldsymbol h_c)$
gives $\boldsymbol b_2=-\boldsymbol h_b+\boldsymbol h_c$, corresponding to the basic solution
$(0,-1,1)^\top$ of $H\boldsymbol x=\boldsymbol b_2$.
Its negative coordinate violates the condition in \Cref{prop:multi-analytic}. Hence,
 the strategy system is not immune. From this example, we see that the nonnegativity requirement must be checked for each
population vector, even when it holds for their sum.}

%The smallest nontrivial picture already explains the restriction. Use the
%same resource matrix as in \eqref{eq:marked-preview}, with columns
%\[
% h_a=(1,0)\T,\qquad h_b=(0,1)\T,\qquad h_c=(1,1)\T.
%\]
%All three pairs are bases. For $v=(u,v_2)\T$, their coefficient vectors are
%\[
%\begin{array}{c|c|c}
%\text{basis}&\text{coefficients}&\text{nonnegativity condition}\\\hline
%(a,b)&(u,v_2)&u\ge0,\ v_2\ge0\\
%(a,c)&(u-v_2,v_2)&u\ge v_2\ge0\\
%(b,c)&(v_2-u,u)&v_2\ge u\ge0
%\end{array}
%\]
%The last two conditions force $u=v_2\ge0$. Thus $K(H)$ is exactly the
%ray through $(1,1)\T$, although the ordinary resource cone is the whole
%first quadrant. The vector $(1,1)\T$ has the strictly positive
%representation $(h_a+h_b+h_c)/2$. Moving off its ray, for instance to
%$(2,1)\T$, gives a negative coefficient in the basis $(b,c)$.
%Even $(2,1)\T$ and $(1,2)\T$ have a safe sum $(3,3)\T$ while failing the
%individual tests. This is why a safe aggregate demand does not suffice.

The following lemma generalizes this picture, \ma{providing conditions under which \(K(H)\) is a single ray when the column matroid \(M(H)\) is connected.} 
We \ma{assume that \(\boldsymbol b\in K(H)\) and that \(H\boldsymbol u^+=\boldsymbol b\) for some strictly positive vector \(\boldsymbol u^+\).} %$b\in K(H)$, so the aggregate direction passes every basis sign test, and a representation $b=Hu^+$ with every resource coefficient $u_e^+>0$. 
The latter is stronger than membership in a single basis
cone. Positivity concerns these resource coefficients, not the
coordinates chosen for $\boldsymbol b$ in the equation space. Every basis cone
contains zero \ma{vector}, but connectivity alone does not guarantee any nonzero
direction in their intersection. For instance, the connected rank-two
matrix \ma{$H$} with columns $(1,0)\T,(1,1)\T,(1,2)\T,(1,3)\T$ has
$K(H)=\{\boldsymbol0\}$: the first pair requires $0\le v_2\le v_1$, while
the last pair requires $2v_1\le v_2\le3v_1$, with $v_1\ge0$. Under the hypothesis of the lemma, \ma{\(M([-\boldsymbol b\mid H])\) is connected and series-parallel, as stated in \eqref{eq:K-SP}. The proof below uses series or parallel pairs in this matroid to carry out induction on the number of resource columns.}
% the affine-matroid bridge supplies a series-parallel reduction.
The proof shows that each reduction preserves only multiples of $\boldsymbol b$ in a
connected block. Across distinct components, by contrast, a basis is a union
of independent component bases; it follows that each component's multiple can be chosen
separately. This explains both the single ray and the direct sum in the
statement below.

\begin{lemma}[One ray on each resource block]\label{lem:ray}
Let $H$ be a full-row-rank real matrix. Suppose \ma{there exists a vector \(\boldsymbol\ell\) such that} ${\boldsymbol\ell}^\top \boldsymbol{h}_e>0$ for every column $\boldsymbol h_e$ of $H$, $e\in E$, and that
vector ${\boldsymbol b}\in K(H)$ has a representation ${\boldsymbol b}=H\boldsymbol u^+$ with $\boldsymbol u^+>\boldsymbol 0$.
Let $\mathscr C$ consist of the connected components of $M(H)$. For each $C\in\mathscr C$, let ${\boldsymbol b}^{(C)}$ be
the component of ${\boldsymbol b}$ in their direct-sum column space. Then
\begin{equation}\label{eq:ray}
 K(H)=\bigoplus_{C\in\mathscr C}\R_{\ge0}\boldsymbol b^{(C)}\triangleq\left\{\sum_{C\in\mathscr C}\alpha_C\boldsymbol b^{(C)}:\alpha_C\ge0\right\}.
\end{equation}
%The direct-sum notation here means $\{\sum_{C\in\mathscr C}\alpha_Cb^{(C)}:\alpha_C\ge0\}$; the component spans are independent, so each coefficient can be chosen separately.
\ma{In particular}, when $M(H)$ is connected, $K(H)$ is a single ray.
\end{lemma}
%\paragraph{Two mechanisms behind the rigidity.} 
\ma{The proof of \Cref{lem:ray} proceeds by induction on the number of columns of \(H\). If \(M(H)\) is disconnected, the conditions defining \(K(H)\) separate across its connected components. Each component inherits the hypotheses of the lemma and has fewer columns; therefore induction applies to each one. When \(M(H)\) is connected and \(H\) has more than one column, \eqref{eq:K-SP} implies that the augmented matroid \(M([-\boldsymbol b\mid H])\) is connected and series-parallel, and therefore it contains a series or parallel pair. 
\begin{itemize}
    \item If the pair is parallel and both elements correspond to columns of \(H\), we delete one column, which leaves \(K(H)\) unchanged.
    \item If the parallel pair contains the demand element, we delete the column of \(H\) corresponding to the other element. This deletion may split \(M(H)\) into several components. By induction, we derive a nonnegative multiplier for each component of \(\boldsymbol b\); in turn, basis exchanges involving the deleted column force these multipliers to agree.
    \item If the pair is in series and both elements correspond to columns of \(H\), the nonnegativity conditions confine \(K(H)\) to a hyperplane. The quotient map for contracting one of the two elements is injective on that hyperplane; so the proportionality obtained by induction in the quotient also holds in the original space. 
    \item If the series pair contains the demand element, the other element would be a coloop of \(M(H)\), which is impossible because \(M(H)\) is connected and has more than one element.
\end{itemize}}

%After we reduce to a connected resource matroid, \eqref{eq:K-SP} makes the augmented matroid $M([-b\mid H])$ series-parallel. The easy reduction deletes one of two parallel resource columns. The two substantive reductions need more explanation. A series pair of resources forces every vector in $K(H)$ into the hyperplane $\ker w$; the quotient used in Case~2 is injective there, so its image determines the original direction. In Case~3, deleting a resource parallel to the demand element may split the remaining resources into components. Bases containing the deleted resource then compare those components' scale factors and force them to agree. Case~4 is ruled out by resource connectivity. These observations organize the induction below.
\begin{proof}[Proof of \Cref{lem:ray}]
We prove by induction on the number $n$ of resource columns. If $n=1$, the
resource cone is the ray generated by its only column, and
$\boldsymbol b= u_1^+\boldsymbol h_1$ is a positive multiple of that column. The assertion
follows. So we assume $n\ge 2$. If $M(H)$ is disconnected, every basis is the union of
bases of its components. Therefore $K(H)$ is the direct product of their
cones. Each component contains fewer than $n$ columns. \ma{For each connected component \(C\) of \(M(H)\), the vector \(\boldsymbol b^{(C)}\) belongs to the cone generated by the columns indexed by any basis of \(C\), and} %Its projected $\boldsymbol b^{(C)}$ lies in every component basis cone, 
has the strictly positive
resource coefficients inherited from $\boldsymbol u^+$. \ma{The restriction of \(\boldsymbol\ell^\top\) to the span of the columns indexed by \(C\) takes a strictly positive value on each of those columns.} %and is evaluated positively by the restriction of $\boldsymbol\ell$. 
Thus the hypotheses hold componentwise
and induction applies. We can therefore assume that $M(H)$ is connected and $n>1$.
In particular, \ma{\(M(H)\) has no coloops.} %no resource column is a coloop.

By \eqref{eq:K-SP}, $\widetilde M\triangleq M([\,-\boldsymbol b\mid H\,])$ is connected
and series-parallel. We write its distinguished demand element as $g$.
Because a connected series-parallel matroid with at least two elements has a
series or parallel pair, we can select such a pair $(e,f)$ in $\widetilde M$ with $e\in E$, and consider
four possibilities according to \ma{whether the pair is parallel or series} and whether the pair consists of two resources or
one resource together with the distinguished demand element $g$.

\emph{Case 1:  The pair is parallel and $\{e,f\}\subseteq E$.}
We may write $\boldsymbol h_e=\kappa \boldsymbol h_f$, \ma{where $\kappa=\boldsymbol h_e/\boldsymbol f_e=\boldsymbol{\ell}^\top\boldsymbol h_e/\boldsymbol{\ell}^\top\boldsymbol f_e>0$ is implied by} the
open-halfspace condition. \ma{Let \(H'\) be the matrix obtained from \(H\) by deleting the column indexed by \(e\). Since \(M(H)\) is connected and \(e\) is parallel to \(f\), the column matroid \(M(H')=M(H)\setminus e\) is connected. Since \(\boldsymbol b=H\boldsymbol u^+\) and \(\boldsymbol h_e=\kappa\boldsymbol h_f\), we can express \(\boldsymbol b\) using the columns of \(H'\) as
\(
\boldsymbol b
=
(u_f^++\kappa u_e^+)\boldsymbol h_f
+
\sum_{j\in E\setminus\{e,f\}}u_j^+\boldsymbol h_j\).
All coefficients in this representation are strictly positive because \(\boldsymbol u^+>\boldsymbol0\) and \(\kappa>0\).} %Deleting $e$ gives a connected resource matrix $H'$. Replacing $u_f^+$ by $u_f^++\kappa u_e^+$ preserves a strictly positive representation of $b$. 
Every basis containing $e$
can exchange it for $f$, with only a positive rescaling of its
coefficient. Thus $K(H')=K(H)$, and the induction hypothesis implies the assertion.

\emph{Case 2: The pair is parallel and  $f=g$.} Since \(\boldsymbol b=H\boldsymbol u^+\) and \(\boldsymbol u^+>\boldsymbol0\), we may 
write $\boldsymbol h_e=\kappa \boldsymbol b$ with $\kappa>0$. \ma{Let \(H'\) be the matrix obtained from \(H\) by deleting the column indexed by \(e\).  Since \(e\) is not a coloop of \(M(H)\), the deletion preserves the rank of \(H\). It follows that \(H'\) has full row rank.} %Since $e$ is not a coloop of $H$, the deletion $H'=H\setminus e$ still has full row rank.
Furthermore,
\[
 (1-\kappa u_e^+)\boldsymbol b=\sum_{j\ne e}u_j^+\boldsymbol h_j.
\]
The right-hand side of the identity has a strictly positive evaluation under the transformation
$\boldsymbol \ell^\top$, as does $\boldsymbol b$. It follows that $1-\kappa u_e^+>0$.
Dividing both sides of the identity by $1-\kappa u_e^+$ gives a strictly positive representation of $\boldsymbol b$
using all columns of $H'$: \ma{\(\boldsymbol b=\sum_{j\in E\setminus\{e\}}\tilde u_j^+\boldsymbol h_j\), where $\tilde u_j^+>0$ for all $j\in E\setminus\{e\}$.} Every basis of $H'$ is also a basis of
$H$, so $K(H)\subseteq K(H')$ and $\boldsymbol b\in K(H')$.

%\emph{Couple the component scales by a basis exchange.}
\ma{The column matroid \(M(H')\) may have several connected components. Let \(\mathscr D\) denote the set of these components. For each $D\in\mathscr D$,  we define vector \(\boldsymbol b^{(D)}\triangleq\sum_{j\in D}\tilde u_j^+\boldsymbol h_j\), which is nonzero because $\tilde u_j^+>0$ and $\boldsymbol{\ell}^\top\boldsymbol h_j>0$ for all $j\in D$. Clearly,
\(\boldsymbol b=\sum_{D\in\mathscr D}\boldsymbol b^{(D)}\).
%The deletion may split into components $D$. Induction gives, for 
For any
$\boldsymbol v\in K(H)$,} it follows from the induction that 
\[
 \boldsymbol v=\sum_{D\in \mathscr D}\alpha_D \boldsymbol b^{(D)},
 \quad\text{where } \alpha_D\ge0\text{ for all }D\in\mathscr D.
\]
\ma{For each connected component \(D\in\mathscr D\), we choose a basis \(J_D\) of the restriction of \(M(H')\) to \(D\). Then \(J\triangleq\bigcup_{D\in\mathscr D} J_D\) is a basis of \(M(H')\). For each \(D\in\mathscr D\), let \(z_j\), \({j\in J_D}\) be the coefficients in the expression of \(\boldsymbol b^{(D)}=\sum_{j\in J_D}z_j\boldsymbol h_j\) by the columns indexed by \(J_D\). Since \(\boldsymbol b=\sum_{D\in \mathscr D}\boldsymbol b^{(D)}\), its expansion in the columns indexed by \(J\) is
\(\boldsymbol b=\sum_{j\in J} z_j\,\boldsymbol h_j\). Since \(\boldsymbol b\in K(H')\) and \(J\) is a basis of \(M(H')\), all coefficients $z_j$, $j\in J$, are nonnegative. For each $D\in\mathscr D$, since} %Choose a basis in every $D$, whose union is a basis $J$ of $H'$. The coefficients of $\boldsymbol b$ on $J$ are the concatenation of those of the $b^{(D)}$. Let $z^D\ge0$ be the basis coefficients of $b^{(D)}$. Each 
$\boldsymbol b^{(D)}$ is a
nonzero vector, there exists $j\in J_D$ such that $z_j>0$. Replacing $j$ in $J$ by $e$ produces a basis of $M(H)$, because the
$j$-coefficient of $\boldsymbol h_e=\kappa \boldsymbol b$ on $J$ is $\kappa z_j\ne0$, \ma{which implies that $\boldsymbol h_j$ can be expressed as a linear combination of $\boldsymbol h_e$ and other columns in $H_J$. Let \(J'\triangleq(J\setminus\{j\})\cup\{e\}\) be the new basis of \(M(H)\). Since \(\boldsymbol b=\sum_{F\in\mathscr D}\boldsymbol b^{(F)}\) and \(\boldsymbol h_e=\kappa\boldsymbol b\), we have
\[
\begin{aligned}
\boldsymbol v
&=\sum_{F\in\mathscr D}\alpha_F\boldsymbol b^{(F)}\\
&=\alpha_D\boldsymbol b
  +\sum_{F\in\mathscr D\setminus\{ D\}}(\alpha_F-\alpha_D)\boldsymbol b^{(F)}\\
&=\frac{\alpha_D}{\kappa}\boldsymbol h_e
  +\sum_{F\in\mathscr D\setminus\{ D\}}\sum_{p\in J_F}
      (\alpha_F-\alpha_D)z_p\,\boldsymbol h_p.
\end{aligned}
\]
This is the expansion of \(\boldsymbol v\) in the columns indexed by \(J'\), with zero coefficients on the columns indexed by \(J_D\setminus\{j\}\).} %We express $\boldsymbol v$ using this basis. the coefficient of $e$ in $v$ is $\alpha_D/\kappa$, and the coefficients in every other component $F$ are \[  (\alpha_F-\alpha_D)z^F.\] The removed $j$-coordinate fixes the coefficient of $e$ to $\alpha_D/\kappa$. Subtracting that contribution removes $\alpha_Db$ from \emph{every} component, yielding the displayed coefficients on $F$. 
The membership in $\boldsymbol v\in K(H)$ requires the coefficients in the expression to be
nonnegative. For each $F\in\mathscr D\setminus\{D\}$, because there exists $p\in J_F$ such that $z^p>0$, it follows that
$\alpha_F\ge\alpha_D$. We can choose the removed coordinate in any
component $D$, so interchanging $D$ and $F$ gives the reverse inequality.
All scale factors are consequently equal, and $\boldsymbol v\in\R_{\ge0}\boldsymbol b$ as desired.

\emph{Case 3: The pair is series and $\{e,f\}\subseteq E$.}
Since \(\{e,f\}\) is a cocircuit of \(\widetilde M\), there exists a column vector \(\boldsymbol w\) such that
\[
 \boldsymbol w^\top\boldsymbol h_e=a\ne0,\quad \boldsymbol w^\top\boldsymbol h_f=c\ne0,\quad \boldsymbol w^\top \boldsymbol h_j=0\text{ for all }j\in E\setminus\{ e,f\}\quad\text{and}\quad
\boldsymbol w^\top\boldsymbol b=0.
\]
From $\boldsymbol b=H\boldsymbol u^+$ we obtain $au_e^++cu_f^+=0$. 
Rescaling $w$ if necessary, we may assume $a>0$ and $c<0$. For any $\boldsymbol v\in K(H)$,
the expression of $v$ via a basis avoiding $e$ gives $\boldsymbol w^\top \boldsymbol v\le0$, while the expression via a basis avoiding $f$ gives
$\boldsymbol w^\top \boldsymbol v\ge0$. Such bases exist because neither $e$ nor $f$ is a coloop.
%Indeed, a basis avoiding $e$ expresses $v$ with nonnegative coefficients and only $h_f$ can contribute to $wv$, with negative value $c$; the basis avoiding $f$ gives the opposite inequality. 
Consequently
\begin{equation}\label{eq:wvzero}
 \boldsymbol w^\top \boldsymbol v=0\qquad\text{for all }\boldsymbol v\in K(H).
\end{equation}

Let \ma{$\pi:\operatorname{span}\{\boldsymbol h_j:j\in E\}\to\operatorname{span}\{\boldsymbol h_j:j\in E\}/\operatorname{span}\{\boldsymbol h_e\}$} be the quotient map that identifies vectors differing
by a multiple of $\boldsymbol h_e$, \ma{i.e., \(\pi(\boldsymbol x)=\pi(\boldsymbol y)\) if and only if
\(\boldsymbol x-\boldsymbol y\in\operatorname{span}\{\boldsymbol h_e\}\). In the quotient space, we} define matrix $H^\star\triangleq(\pi (\boldsymbol h_j))_{j\in E\setminus\{ e\}}$ and
vector $\boldsymbol b^\star\triangleq\pi(\boldsymbol b)$. \ma{The matrices \(H^\star\) and \([-\boldsymbol b^\star\mid H^\star]\) represent \(M(H)/e\) and \(\widetilde M/e\), respectively. Since \(M(H^\star)=M(H)/e\), every basis \(J^\star\) of \(M(H^\star)\) extends to a basis \(J^\star\cup\{e\}\) of \(M(H)\).} %In matroid terms this contracts $e$. Every basis $J'$ of $H'$ lifts to the basis $J'\cup\{e\}$ of $H$.
For any $\boldsymbol v\in K(H)$, \ma{the definition of \(K(H)\) gives an expansion $\boldsymbol v=\alpha_e\boldsymbol h_e+\sum_{j\in J^\star}\alpha_j\boldsymbol h_j$, where $\alpha_e\ge0$ and $\alpha_j\ge0$ for every $j\in J^\star\cup\{e\}$.
Since  \(\boldsymbol h_e\in \operatorname{span}\{\boldsymbol h_e\}\), we have \(\pi(\boldsymbol h_e)\) equal to the zero element of the quotient space. Applying \(\pi\) to both sides of the identity gives
\[
\pi(\boldsymbol v)=\sum_{j\in J^\star}\alpha_j\pi(\boldsymbol h_j).
\]
Thus \(\pi(\boldsymbol v)\) has nonnegative coefficients in its expansion in the columns of \(H^\star\) indexed by \(J^\star\).} %its coefficients on that lifted basis are nonnegative. Applying $\pi$ removes the $e$ term and leaves a nonnegative representation of $\pi v$ on $J'$. 
Since this holds for
every $J^\star$, we obtain 
\[\{\pi(\boldsymbol v):\boldsymbol v\in K(H)\}\subseteq K(H')\] and in particular
$b^\star\in K(H^\star)$. \ma{Applying \(\pi\) to \(\boldsymbol b=\sum_{j\in E}u_j^+\boldsymbol h_j\), we have $\pi(\boldsymbol b)
=\sum_{j\in E\setminus\{e\}}u_j^+\pi(\boldsymbol h_j)$.
In the chosen coordinates of the quotient space, this gives \(\boldsymbol b^\star=H^\star\boldsymbol u^+_{E\setminus\{e\}}\). Since \(u_j^+>0\) for every \(j\in E\setminus\{e\}\), the representation uses every column of \(H'\) with a strictly positive coefficient. Define
\[
\boldsymbol\ell'
\triangleq\boldsymbol\ell-\frac{\boldsymbol\ell^\top\boldsymbol h_e}{a}\,\boldsymbol w.
\]
Using \(\boldsymbol w^\top\boldsymbol h_e=a>0\), \(\boldsymbol w^\top\boldsymbol h_f=c<0\), and \(\boldsymbol w^\top\boldsymbol h_j=0\) for \(j\in E\setminus\{e,f\}\), we obtain
\[
\begin{aligned}
(\boldsymbol\ell')^\top\boldsymbol h_e&=0,\\
(\boldsymbol\ell')^\top\boldsymbol h_f
&=\boldsymbol\ell^\top\boldsymbol h_f
-\frac{\boldsymbol\ell^\top\boldsymbol h_e}{a}\,c
>\boldsymbol\ell^\top\boldsymbol h_f>0,\\
(\boldsymbol\ell')^\top\boldsymbol h_j
&=\boldsymbol\ell^\top\boldsymbol h_j>0
\qquad \text{for all }j\in E\setminus\{e,f\}.
\end{aligned}
\]
Since \((\boldsymbol\ell')^\top\boldsymbol h_e=0\), the rule \(\pi(\boldsymbol x)\mapsto(\boldsymbol\ell')^\top\boldsymbol x\) defines a linear functional on the quotient space. This functional is strictly positive on \(\pi(\boldsymbol h_j)\) for every \(j\ne e\).}
%The original positive coefficients on the remaining columns represent $b'$ strictly positively. The functional \[  \ell'=\ell-\frac{\ell h_e}{a}w \] vanishes on $h_e$ and is strictly positive on all other columns: its value on $h_f$ increases and the remaining values are unchanged. It descends to the required separating functional on the quotient.

\ma{The support of \(\boldsymbol w^\top H\) is \(\{e,f\}\). Therefore, \(\{e,f\}\) contains a cocircuit of \(M(H)\). Since \(M(H)\) has no coloops, it has no singleton cocircuit. Therefore \(\{e,f\}\) itself is a cocircuit of \(M(H)\), and \(e\) and \(f\) are parallel in the connected dual matroid \(M(H)^*\). Deleting \(e\) preserves connectedness by the same argument as in Case 1. Taking the dual shows that \(M(H')=M(H)/e\) is connected.}
%The pair $e,f$ is also a series pair in $M(H)$, since a proper singleton cocircuit would be a coloop. Contracting a member of a series pair preserves connectedness here. 
\ma{Since \(\boldsymbol h_e\ne\boldsymbol 0\), the relations \((\boldsymbol\ell')^\top\boldsymbol h_e=0\) and \((\boldsymbol\ell')^\top\boldsymbol h_f>0\) imply that \(\boldsymbol h_e\) and \(\boldsymbol h_f\) are linearly independent. If these were the only two columns of \(H\), both elements would belong to the unique basis of \(M(H)\) and hence would be coloops. Thus \(H\) has at least three columns.}
%The two-element exception cannot occur: two connected resource columns have rank one, and the additional nonzero demand column would make all three parallel, without a two-element cocircuit.

%\emph{Recover the direction after quotienting.}
By induction, $K(H')=\R_{\ge0}\boldsymbol b'$. Thus, for any $\boldsymbol v\in K(H)$,
we have $\pi(\boldsymbol  v)\in K(H')$. Hence, there exists $\alpha\ge0$ such that  $\pi(\boldsymbol  v)=\alpha\pi(\boldsymbol  b)$. \ma{It follows that $\boldsymbol v=\alpha\boldsymbol b+t\boldsymbol h_e$ for some $t\in \mathbb R$.} %Ordinarily this only determines a vector up to a multiple of $\boldsymbol h_e$. 
Note from
\eqref{eq:wvzero} that both $\boldsymbol v$ and $\boldsymbol b$ belong to $\ker\boldsymbol w^\top$, \ma{implying  $\boldsymbol v-\alpha \boldsymbol b\in \ker\boldsymbol w^\top$. Therefore
\(0=\boldsymbol w^\top(\boldsymbol v-\alpha\boldsymbol b)
=t\boldsymbol w^\top\boldsymbol h_e=ta\).
Because \(a\ne0\), we have \(t=0\), and hence \(\boldsymbol v=\alpha\boldsymbol b\).}
%The map $\pi$ is therefore injective on that hyperplane: if $y\in\ker w$ and $\pi y=0$, then $y=\tau h_e$ for some scalar $\tau$, and $0=\boldsymbol w^\top\boldsymbol y=\tau \boldsymbol w^\top\boldsymbol h_e$ forces $\tau=0$. Applying this to $\boldsymbol v-\alpha \boldsymbol b$ gives $v=\alpha b$.

\emph{Case 4: The pair is \ma{series} and  $f=g$.} \ma{Since \(\{e,g\}\) is a cocircuit of \(\widetilde M\), there exists a column vector \(\boldsymbol w\) such that \(\boldsymbol w^\top[-\boldsymbol b\mid H]\) has support \(\{e,g\}\). Deleting the demand column indexed by \(g\) from $[-\boldsymbol b\mid H]$ leaves \(H\). The corresponding row-space vector \(\boldsymbol w^\top H\) has support \(\{e\}\).}
%Its two-element cocircuit has a row functional nonzero only on $h_e$ after deleting $g$. 
This implies that $e$ is a coloop of $H$, contrary
to connectedness and $n>1$. Thus this case cannot occur.

In each possible case we have proved
$K(H)\subseteq\R_{\ge0}\boldsymbol b$. The reverse inclusion follows from
$\boldsymbol b\in K(H)$ and the fact that $K(H)$ is a cone. This completes the proof. 
\end{proof}

\ma{For any fixed resource block \(C\), recall from \Cref{sec:joint} that \(\boldsymbol {b}^{(C)}=\sum_{i\in I}\boldsymbol b_i^{(C)}\), where \(\boldsymbol b_i^{(C)}\) is the \(V_C\)-component of \(\boldsymbol b_i\) in the direct-sum decomposition. Under \eqref{eq:halfspace} and \eqref{eq:aggregate-positive}, the positivity hypotheses of \Cref{lem:ray} hold on the resource block \(C\). Then the lemma gives
\[
K(H_C)=\{\alpha\boldsymbol {b}^{(C)}:\alpha\ge0\}.
\]
If \(\boldsymbol b_i^{(C)}\in K(H_C)\) for every \(i\in I\), each \(\boldsymbol b_i^{(C)}\) is therefore a nonnegative multiple of \(\boldsymbol {b}^{(C)}\). Membership of the sum \(\boldsymbol {b}^{(C)}\) in \(K(H_C)\) alone does not imply that each summand belongs to \(K(H_C)\).}

\ma{In the representation \eqref{eq:block-H}, the demand element \(g_i\) corresponds to the column \(-\boldsymbol b_i^{(C)}\). If \(\boldsymbol b_i^{(C)}=\boldsymbol 0\), then \(g_i\) is a loop and is deleted when forming \(M_C\). All retained demand columns are positive multiples of \(-\boldsymbol {b}^{(C)}\); so the retained demand elements lie in a single parallel class of \(M_C\), as required by condition (i) of  \cref{thm:multi}.}

%We can now read the additional compatibility condition directly. Once the aggregate direction passes the sign test on a connected block, its entire cone of directions passing every basis test is a single ray. If each population direction also passes, all nonzero $b_i^{(C)}$ must lie on that ray. Testing their sum alone would not establish this. In the marked representation \eqref{eq:block-H}, this is exactly parallelism of the nonloop demand labels. The positive functional in \eqref{eq:halfspace} ensures positive proportionality factors; a label with $b_i^{(C)}=0$ is a loop and is discarded from that block.

\subsection{\ma{The matroid characterization}%Completion of the combinatorial proof
}
It remains to prove  \cref{thm:multi}.
Starting from immunity, the joint analytic criterion places each $\boldsymbol b_i$, $i\in I$,
in $K(H)$, and the ray lemma forces parallel demand labels on every
resource block. The bridge between affine geometry and matroids establishes the series-parallel
condition. Conversely, conditions (i) and (ii) of the theorem \ma{imply the nonnegativity condition \eqref{eq:all-basic-positive} for each population, and immunity follows from the joint basic-solution criterion.} %recover nonnegative basic solutions for every population, which imply immunity by the analytic criterion. All statements below apply to each jointly trade-closed menu tuple, not just to the full strategy families.

\begin{proof}[Proof of \cref{thm:multi}]
Suppose first that the system is immune. By
\cref{lem:trade-multi,prop:multi-analytic}, for every jointly
trade-closed tuple, each $\boldsymbol b_i$ belongs to $K(H)$. Thus
$\boldsymbol b=\sum_{i\in I} \boldsymbol b_i\in K(H)$. Let \(C\) be a connected component of the resource matroid \(M(H)\). Inequalities in
\eqref{eq:halfspace} and \eqref{eq:aggregate-positive} verify \ma{the positivity} %the other
hypotheses of \cref{lem:ray}. Since \(\boldsymbol b\in K(H)\), the lemma shows that the component in \(V_C\) of each \(\boldsymbol b_i\in K(H)\) is a nonnegative multiple of \(\boldsymbol b^{(C)}\). Let \(\alpha_{i,C}\) denote the corresponding scalar. Then 
\begin{equation}\label{eq:alpha}
 b_i^{(C)}=\alpha_{i,C}\boldsymbol b^{(C)} \text{ with }\alpha_{i,C}\ge0\text{ for all }i\in I\qquad\text{and}\qquad \sum_{i\in I}\alpha_{i,C}=1,
\end{equation}
\ma{where the last equality follows by summing over \(i\) and using \(\sum_{i\in I}\boldsymbol b_i^{(C)}=\boldsymbol b^{(C)}\ne\boldsymbol 0\).} This is exactly the parallelism of the nonloop demand elements of
$M_C$. \ma{The positivity hypotheses of \eqref{eq:K-SP} hold on \(C\) by \eqref{eq:halfspace} and \eqref{eq:aggregate-positive}. Since \(\boldsymbol b^{(C)}\in K(H_C)\), \eqref{eq:K-SP} implies that \(M([-\boldsymbol b^{(C)}\mid H_C])\) is connected and series-parallel.} %Also $\boldsymbol b^{(C)}\in K(H_C)$, with a strictly positive representation and a positive separating functional. By \cref{lem:bridge}, $M([\,-\boldsymbol b^{(C)}\mid H_C\,])$ is connected and series-parallel. 
Adjoining
parallel copies of its demand element, and rescaling columns, gives
the representation \eqref{eq:block-H} of $M_C$. Hence (i) and (ii)
hold.

Conversely, suppose (i) and (ii) hold for every tuple and block. \ma{By \eqref{eq:halfspace}, the functional \(\boldsymbol\ell^\top\) takes a strictly negative value on every nonzero demand column \(-\boldsymbol b_i^{(C)}\). These columns are proportional by condition (i), and the common sign of their functional values ensures that the proportionality factors are positive. Then $\sum_{i\in I}\boldsymbol b^{(C)}_i=\boldsymbol b^{(C)}\ne\boldsymbol0$ and condition (i) imply} %The nonzero demand columns in \eqref{eq:block-H} have the same orientation under the positive functional. Their sum is $\boldsymbol b^{(C)}\ne0$, so (a) gives 
\eqref{eq:alpha}. \ma{By condition (ii), \(M_C\) is series-parallel. Deleting all but one of its parallel demand elements preserves this property. The retained demand column is a positive multiple of \(-\boldsymbol b^{(C)}\), so rescaling it to \(-\boldsymbol b^{(C)}\) gives a representation of the same matroid. Hence} %Removing repeated demand columns from (b) shows that 
$M([\,-\boldsymbol b^{(C)}\mid H_C\,])$ is series-parallel. \ma{The resulting matroid is connected and series-parallel, and (5.3) and (5.4) verify the positivity hypotheses. Thus from \eqref{eq:K-SP}, we derive \(\boldsymbol b^{(C)}\in K(H_C)\). It follows from} %The affine--matroid bridge gives $\boldsymbol b^{(C)}\in K(H_C)$, whence $b_i^{(C)}\in K(H_C)$ for every $i$. Bases decompose across resource components, so $b_i\in K(H)$ for every $i$. Apply
\cref{prop:multi-analytic} that the system is immune.

%Connectedness of each $M_C$ was established after \eqref{eq:block-H}, so the excluded-minor formulation follows from the standard theorem. Adding parallel copies, or deleting repeated nonloop demand columns, preserves the series-parallel property. 
Finally, if some jointly
trade-closed tuple violates either condition (i) or (ii), \ma{then for some population \(i\) and some basis \(J\) of \(M(H)\), the affine basic point \(\boldsymbol z^{J,i}\) defined in \eqref{eq:joint-basic} has a negative coordinate. By} %the equivalences just proved imply a negative basic coordinate for some population.
\Cref{prop:multi-analytic}, we can construct the stated strong Braess
paradox with unit demands, positive affine costs, and a single intercept
reduction.
\end{proof}

For $k=1$ the parallelism condition (i) of \Cref{thm:multi} is trivial, \ma{and each marked block \(M_C\) is represented by \([-\boldsymbol b^{(C)}\mid H_C]\). By \eqref{eq:K-SP}, condition (ii) is equivalent to \(\boldsymbol b^{(C)}\in K(H_C)\) for every resource block \(C\). Since every basis of \(M(H)\) is formed by taking a basis from each resource block, these block conditions are equivalent to \(\boldsymbol b\in K(H)\). This is precisely the nonnegativity condition in \Cref{prop:single-analytic} for the affine basic points of \(\{\boldsymbol x:H\boldsymbol x=\boldsymbol b\}\). Thus, for single population,  \Cref{thm:single,thm:multi} give equivalent characterizations in terms of the augmented incidence matroid and the marked blocks of its dual, respectively.} % The preceding proof then reduces to the single-population criterion: nonnegativity of the basic solutions on all blocks is equivalent to that for the full affine space. Thus the two theorems use compatible representations, even though the first is most simply stated with incidence columns and the second with their marked dual.  
This also clarifies the role of the demand
labels (columns), which contain no extra information in the single-population case and
become essential only when \ma{a marked block contains demand elements for two or more populations whose demands can vary independently.} %several independently variable demands share a resource block.

\section{ \ma{Graphical and cographical} %Networks and associated column 
series-parallel structure}
\label{sec:sp-prelim}

In the context of path systems, the adjective ``series-parallel'' can be applied both to the underlying physical
network or to the column matroid of its augmented path-incidence matrix.
\ma{The network’s series-parallel structure can be characterized both by series and parallel compositions of edges and by how simple source-sink paths traverse edges in the underlying undirected graph. In contrast, the paths in the strategy family index the rows of the augmented path-incidence matrix, and the column matroid records the linear dependencies among its columns. } %The network uses paths and edge operations, whereas the matroid records dependencies among incidence columns. 
For a complete
acyclic path system, \ma{we add a return arc from the sink to the source, and then ignore all edge orientations to obtain an undirected multigraph. The cycle matroid of this multigraph is dual to the column matroid of the augmented path-incidence matrix. The duality} %adding a return edge between the terminals relates them by matroid duality. This relation 
recovers the network case of
\cref{thm:single}. \ma{Series and parallel extensions of the network correspond to parallel and series extensions of the column matroid, respectively.}  %and explains why a network series extension becomes a column-matroid parallel extension, and conversely.
We first prove the incidence identity and then give the elementary
operations in \cref{tab:sp-dictionary}.

\subsection{Two-terminal networks and terminal closure}
A two-terminal series-parallel network is obtained from a single $s$-$t$
edge by iteratively applying series and parallel compositions. A series composition \ma{identifies the sink of the first network with the source of the second, whereas a} %joins the
%sink of one component to the source of another; 
parallel composition
identifies their source and their sink. Equivalently, edges may be
successively subdivided or replaced by parallel pairs. In the directed
case, every composition uses the consistent source-to-sink orientation. \ma{The graph is only part of the data of a two-terminal network. One must also specify which vertex is the source \(s\) and which is the sink \(t\). An undirected two-terminal network in which every vertex and edge lies on a \(s\)-\(t\) path is series-parallel if and only if, for each edge, all \(s\)-\(t\) paths using that edge traverse it in the same direction \cite[Proposition 2]{milchtaich}.}
%The terminals are part of the data: an unrooted graph condition alone need not describe a two-terminal network.

\begin{proposition}[The full-path incidence identity]\label{prop:path-dictionary}
Let $G=(V,E)$ be a directed acyclic network with distinct terminals $s$ and $t$ such that every vertex and edge lie on at least a directed $s$-$t$ path. Let $\HH$ be the
family of \emph{all} directed $s$-$t$ paths in $G$. %Add a return edge $g$ from $t$ to $s$. 
\ma{Let \(G^+\) be the directed graph obtained from \(G\) by adding the return arc \(g\) from \(t\) to \(s\). Let \(\bar G^+\) denote its underlying undirected multigraph. Then} %, under the natural edge labels,
\begin{equation}\label{eq:path-dictionary}
N_{\HH}\triangleq M([\one\mid A_{\HH}])=M(\bar G^+)^*
\end{equation}
Both matroids \ma{$N_\HH$ and $M(G^+)^*$} % in this identity 
are connected. In particular,
\begin{equation}\label{eq:path-sp-equivalence}
G\text{ is two-terminal series-parallel}
\quad\Longleftrightarrow\quad
N_{\HH}\text{ is series-parallel}.
\end{equation}
The identity \eqref{eq:path-dictionary} also applies to an undirected two-terminal series-parallel
network after its edges are consistently oriented along a series-parallel construction.
\end{proposition}
\begin{proof}
\ma{The vertex–arc incidence matrix of \(G\) has rows indexed by \(V\) and columns indexed by \(E\). For an arc \(e=(u,v)\) directed from \(u\) to \(v\), its column has entry \(-1\) in row \(u\), entry \(1\) in row \(v\), and zero entries elsewhere. Fix a vertex \(v_0\in V\), and let \(D\) be the reduced incidence matrix obtained by deleting the row indexed by \(v_0\). Let \(\boldsymbol b\) be indexed by the same vertices as the rows of \(D\). For each \(v\in V\setminus\{v_0\}\), set \(b_v=-1\) if \(v=s\), \(b_v=1\) if \(v=t\), and \(b_v=0\) otherwise. Let \(P\in\mathcal H\) be a directed \(s\)–\(t\) path, and let \(\boldsymbol\chi^P\) be its arc-incidence vector. In the full vertex–arc incidence matrix, summing the columns indexed by the arcs of \(P\) gives a vector with entry \(-1\) at \(s\), entry \(1\) at \(t\), and zeros elsewhere. Deleting the coordinate indexed by \(v_0\) gives \(D\boldsymbol\chi^P=\boldsymbol b\).}
%Let $D$ be a reduced node-arc incidence matrix of $G$: each arc column has $-1$ at its tail and $1$ at its head, and one vertex row is removed. The underlying graph is connected, so the remaining rows are independent.Use this incoming-minus-outgoing convention and let $b$ be the reduced vector with $-1$ at $s$ and $1$ at $t$. Every path vector satisfies $D\chi^P=b$.
Conversely, every nonnegative unit $s$-$t$ flow in this acyclic graph
is a convex combination of its paths' incidence vectors, where %: repeatedly extract a positive $s$-$t$ path and its minimum arc flow. 
acyclicity excludes any residual
circulation. \ma{Let \(\boldsymbol x^+\) be the average of the arc-incidence vectors \(\boldsymbol\chi^P\) over all \(P\in\mathcal H\). Since every edge \(e\in E\) lies on at least one path in \(\mathcal H\), we have \(x_e^+>0\) for every \(e\in E\). Since \(\boldsymbol x^+>\boldsymbol0\), we have \(\boldsymbol x^++\boldsymbol v>\boldsymbol0\) for every \(\boldsymbol v\in\ker D\) of sufficiently small norm. For each such \(\boldsymbol v\), we also have \(D(\boldsymbol x^++\boldsymbol v)=\boldsymbol b\). Hence \(\boldsymbol x^++\boldsymbol v\) represents a feasible unit \(s\)-\(t\) flow and is a convex combination of the arc-incidence vectors \(\boldsymbol\chi^P\) for \(P\in\mathcal H\). For every \(\boldsymbol v\in\ker D\) of sufficiently small norm, both \(\boldsymbol x^+\) and \(\boldsymbol x^++\boldsymbol v\) lie in the affine hull of the path-incidence vectors. Their differences \(\boldsymbol v\) belong to its direction space and span \(\ker D\). Hence the affine hull contains the full solution set \(\boldsymbol x^++\ker D=\{\boldsymbol x:D\boldsymbol x=\boldsymbol b\}\).} %The average of all path vectors is positive on every edge. Call this average $x^+$. Every sufficiently small perturbation $x^++v$, with $v\in\ker D$, remains positive and is still a unit flow, hence a convex combination of paths. 
These perturbations span
all directions in $\ker D$, so the path affine hull contains every
direction of $\{x:Dx=b\}$. The reverse inclusion follows from \ma{the fact that \(D\boldsymbol\chi^P=\boldsymbol b\) for every \(P\in\mathcal H\).} %the path equations. 
Therefore
\[
\aff\{\boldsymbol \chi^P:P\in\HH\}=\{\boldsymbol x:D\boldsymbol x=\boldsymbol b\}.
\]
Homogenizing as in \Cref{eq:affine-column-identity} gives
\[
\operatorname{row}(B_{\HH})
 =\{( z,\boldsymbol x):D\boldsymbol x=\boldsymbol bz\}=\ker[-\boldsymbol b\mid D].
\]
By the orthogonal-row-space description of duality in
\cref{sec:matroid-primer}, $[-\boldsymbol b\mid D]$ represents $N_{\HH}^*$.
It is also a reduced incidence matrix of the terminal-closed graph
$G^+$, where the return arc $g$ labels the column $-\boldsymbol b$. Forgetting arc orientations does
not change its cycle matroid, because reversing an orientation only
multiplies one column by $-1$. This proves \eqref{eq:path-dictionary}.
Every original arc in $G$ lies on a directed $s$-$t$ path whose union with $g$ forms a directed cycle in $G^+$.
Thus every arc belongs to the same matroid component as $g$, proving
connectivity. Furthermore, duality preserves that connectivity.

It remains to \ma{show that \(G\) is two-terminal series-parallel with source \(s\) and sink \(t\) if and only if \(M(\bar G^+)\) is series-parallel.} %connect this matroid identity to the prescribed terminals $s,t$. 
The return arc $g$ must be retained in this argument,
since an unrooted graph construction need not specify the required terminals. \ma{Suppose \(G\) is two-terminal series-parallel with source \(s\) and sink \(t\). We add the return arc \(g\) to the initial arc $(s,t)$ and leave \(g\) fixed throughout the construction of \(G\). The initial undirected multigraph consists of two parallel edges and has cycle matroid \(U_{1,2}\). Each series or parallel operation on an edge induces an extension of the same type in the cycle matroid. Hence \(M(\bar G^+)\) is series-parallel.} %If $G$ is two-terminal series-parallel, add $g$ to its construction: the starting edge and $g$ form $U_{1,2}$, and each network extension becomes a graphic series or parallel extension. 
Conversely,
if $M(\bar G^+)$ is series-parallel, then $\bar G^+$ is a block and has no
subdivision of $K_4$. By \citet[Theorem~5.4.10]{oxley}, $\bar G^+$ admits a graphical
series-parallel construction; see also \citet{duffin}.

\ma{We prove by induction on the construction of \(\bar G^+\) that deleting any specified edge \(\tilde g\) from \(\bar G^+\) leaves a two-terminal series-parallel network whose terminals are the end-vertices of \(\tilde g\). The initial graph consists of two parallel edges. Deleting either edge leaves a single edge joining its end-vertices. 
Let \(Q\) be the undirected graph before the last operation in the construction of \(\bar G^+\), where an edge \(e\) is either subdivided into a pair of two edges in series or replaced by a pair of two parallel edges. If \(\tilde g\) lies outside that pair, then \(Q\setminus \tilde g\) is two-terminal series-parallel with the end-vertices of \(g\) as terminals by induction. Performing the same operation on \(e\) produces \(\bar G^+\setminus \tilde g\) that is two-terminal series-parallel with those terminals. If \(\tilde g\) belongs to the pair, let $f$ denote the other edge in the pair. By induction, \(Q\setminus e\) is two-terminal series-parallel with the end-vertices of \(e\) as terminals. If the last operation is a parallel replacement, then $f$ joins the end-vertices of \(e\), and $\bar G^+\setminus g$  is the parallel composition of \(Q\setminus e\) and $f$. If the last operation is a subdivision, then $\bar G^+\setminus \tilde g$ is a series composition of $Q\setminus e$ and $f$, whose terminals are the non-common end-vertices of $e$ and $f$, i.e., $\tilde g$'s end-vertices.}

\ma{We now have $\bar G^+\setminus g$ a two-terminal series parallel network with terminals $s$ and $t$, which are end-vertices of $g$. The series-parallel structure allows us orient this undirected network along paths from source \(s\) to sink \(t\), where, for each edge, all \(s\)-\(t\) paths using that edge traverse it in the same direction \cite[Proposition 2]{milchtaich}.  Since every arc of \(G\) lies on a directed \(s\)-\(t\) path, its direction agrees with the orientation of $\bar G^+\setminus g$. Thus \(G\) is two-terminal series-parallel with terminals $s$ and $t$. Together with \eqref{eq:path-dictionary} and the preservation of the series-parallel property under duality, this completes the proof of \eqref{eq:path-sp-equivalence}.}
%Deleting any prescribed edge $g$ from such a nontrivial block leaves a two-terminal series-parallel network with terminals the ends of $g$. This follows by induction on the construction. If the last operation leaves $g$ untouched, perform that operation after deleting $g$ from the smaller graph. If $g$ is a newly added parallel copy of an edge $e$, the remaining network is the parallel composition of $e$ and the old graph with $e$ removed. If $g$ is one segment of a subdivision of $e$, the remaining network is the series composition of the other segment and the old graph with $e$ removed. The initial two-edge cycle supplies the base case. Thus the prescribed terminals, not only the unrooted graph, are accounted for. In the acyclic directed network, every arc's membership in an $s$--$t$ path forces each constituent to have the consistent orientation. Finally, duality preserves the series-parallel class and interchanges its extensions, proving \eqref{eq:path-sp-equivalence}.
\end{proof}

The all-ones column in $[\one\mid A_{\HH}]$ corresponds to the \emph{return edge through the
dual identity}; players cannot use it as an additional strategy resource.
The proof of \Cref{prop:path-dictionary} uses the complete path family to identify all feasible flows
and uses acyclicity to exclude residual circulations. \ma{In an undirected network, two \(s\)-\(t\) paths may traverse the same edge \(e\) in opposite directions. Both paths have entry \(1\) in column \(e\) of the path-incidence matrix \(A_{\mathcal H}\). Note that} %For an arbitrary undirected graph, different paths can traverse an edge in opposite directions, while the unsigned path table records only whether the edge is used. Thus 
\eqref{eq:path-dictionary} need not hold when \(A_{\mathcal H}\) is formed from all  \(s\)-\(t\) paths in an arbitrary undirected network. %does not assert the same identity for arbitrary undirected simple-path tables.

\begin{corollary}[The reversal of the elementary operations]
\label{cor:network-reversal}
Under the hypotheses of \cref{prop:path-dictionary}, subdividing an \ma{arc into two consecutive arcs \(e\) and \(f\)} %edge into $e,f$ 
produces a parallel extension of the augmented column matroid \(N_{\mathcal H}\).
Replacing an \ma{arc by a pair of parallel arcs $e$ and $f$ with the same head and tail as that arc} %edge by a directed parallel pair $e,f$ 
produces a series
extension of \(N_{\mathcal H}\). More generally, a two-edge circuit of $M(\bar G^+)$
is a series pair in $N_{\HH}$, and a two-edge cocircuit of $M(\bar G^+)$ is a
parallel pair in $N_{\HH}$.
\end{corollary}
\begin{proof}
The last assertion follows immediately from duality. Two direct incidence
checks explain it without a graph representation of the column matroid. \ma{After subdivision, a directed \(s\)-\(t\) path uses the new arc \(e\) if and only if it uses the new arc \(f\). At least one path uses both because the original arc lies on a directed \(s\)-\(t\) path. Thus the columns indexed by \(e\) and \(f\) are identical and nonzero. It follows that \(e\) and \(f\) are parallel elements of the resulting augmented column matroid.}
%After subdivision, every path uses either both new edges or neither. Their columns are identical and nonzero, hence parallel.

For a parallel replacement, \ma{we consider a directed \(s\)-\(t\) path that uses the original arc. Replacing that arc by \(e\) or \(f\) gives two paths that agree on all other arcs. In the augmented incidence matrix of the new network, we subtract the row for the path using \(f\) from the row for the path using \(e\). The difference has entry \(1\) in column \(e\), entry \(-1\) in column \(f\), and zeros elsewhere. Therefore, the row space contains a vector with support \(\{e,f\}\).} %take a path through the original edge. Choosing $e$ or $f$ gives two augmented rows whose difference is $\ee_e-\ee_f$. The row space therefore contains a vector supported on exactly these two labels. 
To prove that this support is a cocircuit, we
must \ma{rule out the existence of a vector in the row space whose support is \(\{e\}\) or \(\{f\}\). On the other hand, such} %exclude a nonzero row-space vector with support $\{e\}$ or $\{f\}$. Such 
a singleton cocircuit would make that element a coloop, which contradicts the matroidal
connectivity stated in \cref{prop:path-dictionary}. Thus, the support $\{e,f\}$ 
is minimal, hence a cocircuit \ma{of the augmented column matroid of the new network, saying that \(e\) and \(f\) form a series pair in that matroid.} %and a series pair. 
Finally,
under \eqref{eq:path-dictionary}, \ma{graphical contraction and deletion in the cycle matroid correspond, respectively, to deletion and contraction in the column matroid. Deleting one element of the parallel pair or contracting one element of the series pair yields the original augmented column matroid, with the surviving element identified with the original arc.} %graphic contraction corresponds to column-matroid deletion and graphic deletion corresponds to column-matroid contraction. The resulting reductions recover the original augmented matroid.
\end{proof}

\begin{table}[htbp]
\centering\small
\begin{tabular}{@{}p{.29\textwidth}p{.30\textwidth}p{.32\textwidth}@{}}
\toprule
Operation on the network & Incidence explanation & Operation on $N_{\HH}$\\
\midrule
Subdivide an arc into $e,f$ in series & Columns indexed by $e,f$ coincide & Parallel extension;
undo by deleting a copy\\[4pt]
Replace an arc by parallel $e,f$ & Two path rows differ only on $e,f$ &
Series extension; undo by contracting a member\\
\bottomrule
\end{tabular}
\caption{The series/parallel reversal: \ma{arc subdivision and parallel replacement corresponds to parallel and series extensions of \(N_{\mathcal H}\), respectively.} % concerns elementary edge operations. It does not identify arbitrary compound subnetworks with individual matroid elements.
}
\label{tab:sp-dictionary}
\end{table}

We examine two smallest examples. When $\HH$ consists of a two-edge path $ab$, we have
$B_\HH\triangleq[1\mid A_\HH]=[1\ 1\ 1]$ on labels $g,a,b$, hence $N_\HH=U_{1,3}$, where the two physical
series edges give \ma{identical columns in $B_\HH$, and are parallel elements in $N_\HH$}. %parallel columns. 
When $\mathcal H$ consists of two parallel one-edge paths $a$ and $b$, we have
\[
B_\HH=\begin{array}{c|rrr}&g&a&b\\\hline
 a&1&1&0\\ b&1&0&1\end{array}, %\qquad g=a+b.
\]
\ma{where every pair of columns is linearly independent, but the three columns are dependent. Thus \(N_{\mathcal H}=U_{2,3}\), in which every pair of elements is a cocircuit. In particular, \(a\) and \(b\) form a series pair.} %This is $U_{2,3}$, whose pairs are cocircuits: the physical parallel edges are in series in the column matroid, and their columns are independent. Both matroids are series-parallel, but their elementary operations have been exchanged.

\begin{corollary}[Recovery of the acyclic routing criterion]
\label{cor:routing-immunity}
Under the hypotheses of \cref{prop:path-dictionary}, the full directed
$s$-$t$ path family is immune if and only if $G$ is
two-terminal series-parallel.
\end{corollary}
\begin{proof}
\ma{By acyclicity, the vertices in $G$ admit a topological ordering \(v_1=s,\ldots,v_n=t\). Setting \(\pi(v_j)\triangleq j\) for \(j=1,\ldots,n\), we define positive weight $q_{(u,v)}\triangleq\pi(v)-\pi(u)>0$ for every arc \((u,v)\).} %Assign numbers $\pi(v)$ increasing along that ordering and set  on each arc. 
Along any $s$-$t$ path these
weights telescope to $\pi(t)-\pi(s)$. Thus all $s$-$t$ paths are minimizers
for the same positive weights $\boldsymbol q$, and the full path menu is trade-closed
by \cref{thm:positive-exposure}. If the system is immune, applying
\cref{thm:single} to this menu, \cref{prop:path-dictionary}
forces $G$ to be two-terminal series-parallel.

For sufficiency, \ma{we suppose that $G$ is two-terminal series-
parallel, and consider an arbitrary menu \(\mathcal H\) exposed by a strictly positive weight vector \(\boldsymbol q\). Let \(G_{\boldsymbol q}\) be the subnetwork of \(G\) formed by the vertices and arcs of the paths in \(\mathcal H\).} %let $\boldsymbol q>\boldsymbol 0$ expose a relevant menu $\HH$, and let $G_q$ be the union of the $s$-$t$ paths in $\HH$. 
We first show that $\HH$ contains
\emph{all} $s$-$t$ paths in $G_q$. Let $\ell(v)$ be the minimum
$\boldsymbol q$-weight of an $s$-$v$ path. Every prefix of a minimizing path is
itself minimizing. So each arc $(u,v)$ in $G_q$ satisfies
$q_{(u,v)}=\ell(v)-\ell(u)$. These equalities telescope along every
$s$-$t$ path in $G_q$, giving the same minimum weight
$\ell(t)-\ell(s)$. Every such path therefore belongs to $\HH$. \ma{Since \(G\) is two-terminal series-parallel and \(G_{\boldsymbol q}\) is the union of a nonempty family of its \(s\)-\(t\) paths, \(G_{\boldsymbol q}\) is also two-terminal series-parallel. Applying \cref{prop:path-dictionary}
to $G_q$, we deduce that $N_\HH$ is series-parallel. By \cref{thm:single}, the arbitrary choice of $\HH$ implies that the full directed $s$-$t$ path family of $G$ is immune.} %An $s$-$t$ path subnetwork of a two-terminal series-parallel network is again two-terminal series-parallel. This follows by induction on the construction: both factors of a series composition must remain, while empty factors of a parallel composition are discarded. Apply \cref{prop:path-dictionary} to $G_q$ and then \cref{thm:single} to all relevant menus.
\end{proof}

\Cref{cor:routing-immunity} recovers the classical criterion in the stated acyclic setting from
the general set-system theorem. \ma{However, the argument above does not establish \eqref{eq:path-dictionary} for the path-incidence matrix \(A_{\mathcal H}\) formed from all simple \(s\)–\(t\) paths in an arbitrary undirected network.} %It does not extend the unsigned path-incidence identity to arbitrary undirected networks.

\subsection{Immunity to the weak and strong paradoxes}
A \emph{chain-parallel network} in this paper means \ma{a series composition of bundles of parallel edges. Each \(s\)-\(t\) path uses exactly one edge from each bundle.} %a chain of parallel-edge bundles, without suppressing subdivisions of individual resources. A path chooses exactly one edge at each stage. 
This is a proper subclass of
two-terminal series-parallel networks. In the complete-path setting,
immunity to the strong paradox corresponds to two-terminal
series-parallel networks, while immunity to the weak paradox corresponds to chain-parallel networks. The first statement is the classical network
characterization \citep{milchtaich,chen}. The
second statement follows \ma{by combining the matroid-base characterization of
\citet[Theorem~3.1]{fujishige} and the characterization of matroid base families of paths proved below.} %  As explained after \cref{def:strong-braess}, excluding every resource-cost increase is a stronger requirement than excluding the cost increase of every population.

\ma{The full \(s\)-\(t\) path family is a matroid base family if and only if the network is chain-parallel. This can be verified directly. Let the arc set be partitioned into bundles of parallel arcs, and let an independent set contain at most one arc from each bundle. The bases of this partition matroid contain exactly one arc from each bundle and are therefore precisely the arc sets of the \(s\)-\(t\) paths.}
%For completeness, this consequence for path families can be checked directly. A chain of bundles has the bases of a partition matroid as its paths. 
Conversely, suppose the full directed
path family is a matroid base family. \ma{We fix an \(s\)–\(t\) path \(P\). For any arc \(e\notin P\), we choose an \(s\)-\(t\) path \(Q\) containing \(e\). By basis exchange, there is an arc \(f\in P\setminus Q\) such that \(Q'=(Q\setminus\{e\})\cup\{f\}\) is the arc set of another \(s\)–\(t\) path. Let \(D\) be the reduced vertex-arc incidence matrix as defined in the proof of Proposition 6.1; we write \(D_e\) and \(D_f\) for its columns indexed by \(e\) and \(f\). Both \(D\boldsymbol\chi^{Q'}\) and \(D\boldsymbol\chi^Q\) equal \(\boldsymbol b\). Their difference is \(D_f-D_e\), which implies \(D_e=D_f\). Each full incidence column has one entry equal to \(-1\) and one equal to \(1\); so its entries sum to zero. The entry in the deleted row is therefore determined by the retained entries. Since \(D_e=D_f\), the full incidence columns indexed by \(e\) and \(f\) coincide. It follows that \(e\) and \(f\) have the same tail and head, and therefore are parallel. Hence, every arc outside \(P\) is parallel to some arc of \(P\), implying that the graph is a chain-parallel network.}

\ma{The basis-exchange argument also applies to undirected networks. In an \(s\)-\(t\) path, \(s\) and \(t\) are the only vertices of odd degree. In a basis exchange, removing an edge \(e\) flips the degree parity at its two end-vertices, and inserting an edge \(f\) does the same at its end-vertices. Since the resulting path still has \(s\) and \(t\) as its only odd-degree vertices, \(e\) and \(f\) must join the same pair of vertices. As in the directed case, every edge outside \(P\) is therefore parallel to an edge of \(P\), so the network is chain-parallel.}

%A basis exchange replacing $e$ by $f$ preserves $Dx=b$, so $D_e=D_f$; the exchanged edges have the same ordered endpoints. Fix one path $P$. For any edge $e$ outside $P$, a path $Q$ containing it can exchange $e$ for an edge of $P$. Thus every edge is parallel to an edge of $P$, proving the chain-of-bundles form.The same argument applies to undirected paths by their mod-two boundary vectors: a one-edge exchange must preserve the unordered endpoints. This proves the specialization, rather than inferring its necessity merely from the partition-matroid example.

The chain case also has a more specific column description. If the
bundles are $C_1,\ldots,C_r$, the affine equations are
$\sum_{e\in C_j}x_e=1$. Thus $N_{\HH}^*$ has resource column $\ee_j$
for each $e\in C_j$ and demand column $-\one\in\R^r$. \ma{Deleting the demand column \(-\boldsymbol 1_r\) in the representation of $N_{\HH}^*$ leaves the matrix of resource columns. Its column matroid is a direct sum of rank-one matroids, one for each parallel-edge bundle.}
%Deleting $g$ from this dual representation leaves a direct sum of rank-one resource blocks, one for each bundle $C_j$. 
This is
a special form of \eqref{eq:path-dictionary}. %; it does not make the network and the augmented column matroid the same object.

\begin{example}[Series-parallel columns without matroid-base strategies]
\label{ex:network-two-paths}
Let the network consists of two internally disjoint $s$-$t$ paths, $ab$ and $cd$, in parallel.
The network is two-terminal series-parallel but not chain-parallel. Its
path family $\HH=\{ab,cd\}$ does not satisfy basis exchange axiom, yet the augmented column matroid $M(B_\HH)$ with
\[
B_{\HH}=\begin{array}{c|rrrrr}&g&a&b&c&d\\\hline
ab&1&1&1&0&0\\cd&1&0&0&1&1\end{array}
\]
has parallel pairs $a,b$ and $c,d$. Deleting $b,d$ from the matroid leaves
$U_{2,3}$; \ma{in its matrix representation, the column indexed by \(g\) is the sum of the columns indexed by \(a\) and \(c\).} %relation $g=a+c$. 
Hence $M(B_\HH)$ is
series-parallel. %Notice again that the \emph{series edges on each path} produce the \emph{parallel columns}.
\begin{figure}[htbp]
\centering
\begin{tikzpicture}[scale=.95]
\node[vertex,label=left:$s$](s)at(0,0){};
\node[vertex,label=above:$u$](u)at(1.5,.8){};
\node[vertex,label=below:$v$](v)at(1.5,-.8){};
\node[vertex,label=right:$t$](t)at(3,0){};
\draw[flow](s)--node[above left]{$a$}(u);
\draw[flow](u)--node[above right]{$b$}(t);
\draw[flow](s)--node[below left]{$c$}(v);
\draw[flow](v)--node[below right]{$d$}(t);
\node[align=center]at(7,.35){physical series pairs: $a,b$ and $c,d$\\[5pt]
matroidal parallel pairs: $a,b$ and $c,d$};
\node[align=center]at(7,-.6){$\boldsymbol{h}_a=\boldsymbol{h}_b,\quad \boldsymbol{h}_c=\boldsymbol{h}_d,\quad \boldsymbol{h}_g=\boldsymbol{h}_a+\boldsymbol{h}_c$};
\end{tikzpicture}
\caption{The same edge labels describe different dependencies in the network
and its augmented column matroid. Both the network and matriod are series-parallel, with
series and parallel relations interchanged by duality.}
\label{fig:path-dictionary}
\end{figure}
Initially, suppose that $c_a(x)=c_c(x)=x$, $c_b(x)=1$, and $c_d(x)=0$. At unit demand,
all equilibrium flow uses $cd$ and the population cost is $1$. Lowering
$c_b$ to zero splits the flow equally between the paths. The cost on
$a$ rises from $0$ to $1/2$, while the population cost falls to $1/2$.
This is a weak paradox but \ma{not} a strong one.
\end{example}

%\paragraph{Beyond paths.}
%\ma{By \eqref{eq:path-dictionary},} %The full-path identity makes 
%matroid $N_{\HH}$ is cographic, %meaning dual to a graph cycle matroid, 
%and hence binary. %, meaning representable over the two-element field $\mathbb F_2$.
%Accordingly, $U_{2,4}$ cannot be a minor there. For arbitrary set systems,
%$U_{2,4}$ is an additional possible obstruction; \cref{ex:U24-first}
%gives an explicit example. The network interpretation explains the classical
%obstruction, while the column criterion covers the non-graphical case.

\section{Examples, obstructions, and certificates}\label{sec:examples}
The examples presented in this section \ma{address three questions about the characterization:} %answer three questions left by the characterization:
why must \ma{the series-parallel condition be checked for every trade-closed menu rather than only for} %we test menus rather than only 
the full strategy family, how does a forbidden minor produce a paradox on the original resources,
and why do \ma{multiple} populations require a separate condition on demand labels? \ma{In this context, an \emph{obstruction} encompasses a forbidden minor, a menu and its augmented incidence column matroid that admits the minor, a marked block of a strategy system containing the minor, and non-parallel demand elements residing within the same marked block of a strategy system.}

\subsection{A full-family test is insufficient}
The \ma{examination} %quantifier 
over all trade-closed menus is necessary even for one
population. The next example passes the test on the full family, but a
positively exposed sub-menu reveals a negative basic point and hence
nonimmunity.

\begin{example}[A safe full affine hull with an unsafe sub-menu]\label{ex:submenus}
Let $\F=\{abcd,aef,bef,cef,def\}$ be the family of all strategies. 
The full affine hull \ma{of the incidence vectors of all strategies in \(\mathcal F\) is described by the following equations:} %has equations
\[
 x_e=x_f=t,\qquad x_a+x_b+x_c+x_d+3t=4.
\]
Its basic points are $4\ee_a,4\ee_b,4\ee_c,4\ee_d$
and the vector $\boldsymbol{x}$ with $x_e=x_f=4/3$ and all other coordinates zero.
They are all nonnegative. The full family is exposed, for example, by
weights $(1,1,1,1,3/2,3/2)$, which give each strategy weight 4. \ma{By \Cref{lem:bridge}, the augmented incidence column matroid \(N_{\mathcal F}=M(B_{\mathcal F})\) of the full strategy family is series-parallel.}
%The affine--matroid bridge therefore shows that the full augmented incidence matroid is series-parallel. This verifies the test for the full menu only; it does not yet establish immunity.

However, positive weight $\boldsymbol q=(2,3,2,2,3,4)^\top$ exposes exactly
$\HH=\{abcd,aef,cef,def\}$. The four strategies in $\HH$ have weight 9,
whereas the unique outsider $bef\in\mathcal F\setminus\HH$ has weight 10. Hence $\HH$ is trade-closed. \ma{The affine hull \(\operatorname{aff}\{\boldsymbol\chi^S:S\in\mathcal H\}\) contains the following point} %Its affine hull has the minimally supported point
\[
 -\tfrac12\boldsymbol \chi^{abcd}
 +\tfrac12\boldsymbol \chi^{aef}
 +\tfrac12\boldsymbol \chi^{cef}
 +\tfrac12\boldsymbol \chi^{def}
 =(0,-\tfrac12,0,0,\tfrac32,\tfrac32)^\top,
\]
where the vector's coordinates are indexed by $a,b,c,d,e,f$ in this order. Within this affine hull, the equations $x_a=x_c=x_d=0$ uniquely determine this point. \ma{Hence, it is an affine basic point. Nevertheless, its $b$-coordinate has a negative value \(-\tfrac12\). Thus \(\mathcal H\) fails the nonnegative basic-point criterion in \Cref{prop:single-analytic}, and} %its support cannot be reduced further. Its negative $b$-coordinate violates the basic-point criterion. Thus 
the strategy system is not immune. % even though its full-menu matroid is series-parallel. 
The sub-menu, rather than the full strategy
matrix, reveals the obstruction.
\end{example}
\subsection{\ma{Forbidden minors and paradox construction}%Unlabelled obstructions and their realization
}
We next illustrate the two forbidden minors with $M(K_4)$ and $U_{2,4}$ explicit strategy
families. \ma{For each strategy family, we show that the affine hull of its strategy-incidence vectors has an affine basic point with a negative coordinate. We then explain how to realize the corresponding strong Braess paradox on the original resources.} %After identifying their negative basic points, we explain how to realize the resulting paradoxes on the original resources.

\begin{example}[The Wheatstone menu]\label{ex:wheatstone-first}
Consider the strategy family $\mathcal W=\{ab,cd,ade\}$.
Its augmented incidence matroid is $M(K_4)$. To see this, let the edges \(12, 13, 23, 24, 14, 34\) of \(K_4\) be labeled as \(e, b, a, c, d, g\), respectively. \ma{Under this labeling, the four triangles of \(K_4\) have edge sets \(\{a,b,e\}\), \(\{c,d,e\}\), \(\{g,a,c\}\), and \(\{g,b,d\}\). These sets are exactly the three-element circuits of \(N_{\mathcal W}\). Both \(M(K_4)\) and \(N_{\mathcal W}\) have rank three, and neither has a circuit of size one or two. They have the same independent sets. Hence \(N_{\mathcal W}\cong M(K_4)\).}
%The four three-element circuits \[  \{a,b,e\},\quad\{c,d,e\},\quad\{g,a,c\},\quad\{g,b,d\}\] are exactly the three-element circuits of the corresponding incidence columns. Both matroids have rank three and no circuits of size one or two. Their dependent triples are exactly the four listed triples, and every set of four or more elements is dependent because the rank is three. Thus their independent sets coincide under the displayed labels, proving the identification. 

By \cref{cor:full-family}, the full strategy family \(\mathcal W\) is trade-closed. Indeed, with resource order
$(a,b,c,d,e)$, the positive weight vector $\boldsymbol q=(1,2,2,1,1)^\top$ gives equal strategy weight ${\boldsymbol  q}^\top\boldsymbol \chi^{ab}={\boldsymbol  q}^\top\boldsymbol \chi^{cd}={\boldsymbol  q}^\top\boldsymbol \chi^{ade}=3$.
%Every multiset of $m$ menu strategies therefore has total weight $3m$. A strict coordinatewise saving would decrease this total because $\boldsymbol q>\boldsymbol 0$, which is impossible. There are no outside strategies to check.
Nevertheless, the affine hull \(\operatorname{aff}\{\boldsymbol\chi^S:S\in\mathcal W\}\) contains the minimally supported point
\begin{equation}\label{eq:wheatstone-negative}
      \boldsymbol \chi^{ab}+\boldsymbol\chi^{cd}-\boldsymbol\chi^{ade}
       =(0,1,1,0,-1).
\end{equation}
%The coefficients in \eqref{eq:wheatstone-negative} sum to one, so the point is in the affine hull, although it is not a feasible load. 
To see it is
minimally supported, we write an affine point as
$\alpha\boldsymbol \chi^{ab}+\beta\boldsymbol \chi^{cd}+\gamma\boldsymbol \chi^{ade}$ with
$\alpha+\beta+\gamma=1$. The equations $x_a=x_d=0$ force
$(\alpha,\beta,\gamma)=(1,1,-1)$. Hence this is the only affine
point supported within $\{b,c,e\}$, and all three displayed
coordinates are nonzero. The negative $e$-coordinate will give a negative cost-response
coefficient. For suitable positive slopes, lowering the intercept of
$c_e$ raises the equilibrium population cost. %The construction below obtains such costs in the original strategy coordinates, and
See Appendix~\ref{app:numerical-witnesses} for numerical instances.
\end{example}
\begin{example}[A nonbinary \ma{forbidden minor}%obstruction
]\label{ex:U24-first}
The family of strategies
\begin{equation}\label{eq:four-strategies}
 \F=\{ad,bce,bde,cde\}.
\end{equation}
is positively exposed by weights $(2,1,1,1,1)^\top$, where
every strategy has weight 3. So the full menu $\F$ is trade-closed. \ma{The affine hull of \(\{\boldsymbol\chi^S:S\in\mathcal F\}\) is the solution set of the following} affine equations
\begin{equation}\label{eq:four-affine}
 x_a+x_e=1,\qquad x_a+x_b+x_c+x_d=2.
\end{equation}
The dual augmented matroid has columns
\[
[-\boldsymbol b\mid H]=
\begin{pmatrix}-1&1&0&0&0&1\\-2&1&1&1&1&0\end{pmatrix},
\]
where the columns are indexed by $g, a, b, c, d, e$ from left to right. \ma{The columns of \([-\boldsymbol b\mid H]\) indexed by \(g,a,b,e\) span a two-dimensional space, and every pair of these columns is linearly independent. Thus \(N_{\mathcal F}^{*}\setminus\{c,d\}=M([-\boldsymbol b\mid H])\setminus\{c,d\}\cong U_{2,4}\). Equivalently, contracting \(c\) and \(d\) in \(N_{\mathcal F}\) produces a \(U_{2,4}\) minor. The corresponding column deletion in \([-\boldsymbol b\mid H]\) amounts to setting \(x_c=x_d=0\) in \eqref{eq:four-affine}. Setting \(x_b=0\) as well uniquely determines the affine basic point \((2,0,0,0,-1)^\top\) in the original resource coordinates.}
%Delete $c,d$ in the dual of $M([-\boldsymbol b\mid H])$: the remaining four columns have rank two and every pair is independent, giving $U_{2,4}$. Equivalently, contract $c,d$ in the primal. Setting $x_b=x_c=x_d=0$ gives the negative basic point $(2,0,0,0,-1)$. This obstruction is possible for arbitrary set systems even though it is excluded by binary representability in the full-path setting. 
Appendix~\ref{app:numerical-witnesses} presents positive affine
costs for which lowering the intercept of $c_e$ raises the equilibrium
cost from $57$ to $115/2$.
\end{example}

\ma{\Cref{ex:U24-first} shows that the augmented incidence matroid of a general strategy family can have a \(U_{2,4}\) minor. In contrast, for the full path family in Proposition 6.1, the augmented path-incidence matroid is cographic and hence binary; so it has no \(U_{2,4}\) minor.}

\paragraph{From a minor to a \ma{cost} witness.}\label{sec:explicit-obstructions} \ma{The preceding examples show how forbidden minors certify a failure of immunity. We call a menu $\HH$ \emph{obstructed} if $N_\HH$ has a $U_{2,4}$ or $M(K_4)$ as a minor. The next proposition gives a general procedure for constructing resource costs that realize a strong Braess paradox in the original strategy family.}
%A matroid contraction is not an instruction to remove a physical resource. The witness must be constructed in the original strategy coordinates.

\begin{proposition}[A \ma{cost} witness \ma{for} %from
an obstructed menu]\label{prop:minor-witness}
\ma{Given the incidence matrix \(A_{\mathcal H}\) of a trade-closed menu \(\mathcal H\) whose augmented incidence matroid \(N_{\mathcal H}\) has a \(U_{2,4}\) or \(M(K_4)\) minor, a finite procedure constructs affine cost functions with positive rational slopes and intercepts for the resources of the original strategy family. The procedure also specifies how much to lower the intercept of a single resource cost function so that the equilibrium cost at unit demand strictly increases.} %Given the incidence table of a trade-closed menu with a $U_{2,4}$ or $M(K_4)$ minor, a finite procedure constructs rational positive affine costs and a one-intercept paradox at unit demand on the original family.
\end{proposition}
\begin{proof} \ma{Using \(A_{\mathcal H}\), the procedure computes a full-row-rank matrix \(H\) and a vector \(\boldsymbol b\), both with rational entries, such that \(\operatorname{aff}\{\boldsymbol\chi^S:S\in\mathcal H\}=\{\boldsymbol x:H\boldsymbol x=\boldsymbol b\}\).}
%Compute rational equations $H\boldsymbol x=b$ for the menu's affine hull.
\Cref{prop:automatic-connected,thm:standard-sp,lem:bridge} imply that \ma{there exists an affine basic point of \(\{\boldsymbol x:H\boldsymbol x=\boldsymbol b\}\) with a negative coordinate. By enumerating the bases of \(M(H)\), the procedure finds a basis \(J\) whose associated affine basic point has a negative coordinate.} %some basis has a negative basic coordinate; Enumeration finds such a point. 
\ma{Solving  \eqref{eq:single-recognition}, the procedure finds a rational vector \(\boldsymbol q\), which, by \Cref{prop:recognition}, positively exposes the menu \(\mathcal H\) in the original strategy family \(\mathcal F\). For \(\varepsilon>0\), let \(W_\varepsilon\) be the diagonal matrix whose entry indexed by \(e\) is \(1\) if \(e\in J\) and \(\varepsilon\) otherwise. By \cref{lem:cb}, \(k(W_\varepsilon,\boldsymbol b)\) converges to the affine basic point associated with \(J\) as \(\varepsilon\to0^+\). This point has a negative coordinate. So testing \(\varepsilon=\tfrac12,\tfrac14,\tfrac18,\ldots\) eventually yields a vector \(k(W_\varepsilon,\boldsymbol b)\) whose entry in the same coordinate is negative. Set \(\boldsymbol x^0\triangleq\frac{1}{|\mathcal H|}\sum_{S\in\mathcal H}\boldsymbol\chi^S\). For the value of \(\varepsilon\) found above, the procedure starts with \(W=W_\varepsilon\) and follows the construction in the proof of \cref{prop:single-analytic}, as described in \eqref{eq:affine-witness-costs}--\eqref{eq:single-cost-increase}.}

\ma{Let \(e_*\) be the resource index at which \(k(W_\varepsilon,\boldsymbol b)\) has the negative entry found above. The procedure tests \(W=2^mW_\varepsilon\) for \(m=0,1,2,\ldots\) and takes the first \(W\) for which all intercepts in \eqref{eq:affine-witness-costs} are positive. As \(m\to\infty\), these intercepts converge to the corresponding positive entries of \(\boldsymbol q\); so the search terminates. For each resource \(e\) unused by \(\mathcal H\), the procedure defines its cost by \(c_e(s)=s+q_e\) for all $s\ge0$. For the scaled matrix \(W=2^mW_\varepsilon\), we have \(k(W,\boldsymbol b)=k(W_\varepsilon,\boldsymbol b)\). In particular, \(k(W,\boldsymbol b)_{e_*}<0\).}

\ma{Fix this \(W\) and let \(G=\operatorname{conv}\{\boldsymbol\chi^S:S\in\mathcal H\}\). For each trial value \(\delta=\tfrac12,\tfrac14,\tfrac18,\ldots\), the procedure lowers the intercept of \(c_{e_*}\) by \(\delta\) to obtain \(\bar c\), and form the candidate load \(\boldsymbol x^\delta=\boldsymbol x^0+\delta\boldsymbol v\), where \(\boldsymbol v\) is computed from \(W\) and \(e_*\) by \eqref{eq:witness-xp}. When evaluating costs of strategies in \(\mathcal F\), the procedure extends \(\boldsymbol x^0\) and \(\boldsymbol x^\delta\) by zero on resources unused by strategies in \(\mathcal H\). It accepts the first \(\delta\) for which the reduced intercept is positive, \(\boldsymbol x^\delta\in\operatorname{ri}G\), and every strategy in \(\mathcal F\setminus\mathcal H\) has strictly higher cost than every strategy in \(\mathcal H\) under \(\bar c\) at \(\boldsymbol x^\delta\).
The intercept and cost inequalities can be checked by exact rational arithmetic, and membership in \(\operatorname{ri}G\) can be checked by linear programming. Since \(\boldsymbol x^0\in\operatorname{ri}G\) and \(H\boldsymbol v=\boldsymbol 0\), the candidate load remains in \(\operatorname{ri}G\) for all sufficiently small \(\delta>0\). The intercept and cost inequalities are strict at \(\delta=0\) and persist by continuity. It follows that the search terminates.}

\ma{For the accepted \(\delta\), the equilibrium argument in the proof of \cref{prop:single-analytic} applies. By \eqref{eq:single-cost-increase}, \(\lambda(\bar c,1)-\lambda(c,1)=-\delta k(W,\boldsymbol b)_{e_*}>0\). The procedure therefore returns cost functions with positive rational slopes and intercepts for which lowering the intercept of \(c_{e_*}\) strictly increases the equilibrium cost at unit demand.}
%Solve \eqref{eq:single-recognition} for a rational exposing vector $\boldsymbol q$. In \cref{lem:cb} put weight one on the chosen basis and weight $\varepsilon>0$ elsewhere. Successively decreasing rational $\varepsilon$ eventually makes the corresponding response coordinate negative. Use the barycenter for $x^0$ and apply the construction in \cref{prop:single-analytic}, equations \eqref{eq:affine-witness-costs}--\eqref{eq:single-cost-increase}. Scaling $W$ and decreasing a rational $\delta$ eventually preserve positive intercepts, relative-interior feasibility, and every outside gap. All conditions are strict and rationally checkable, so the search terminates.
\end{proof}

\paragraph{\ma{Many strategy families as witnesses for the same paradox type.}%One obstruction, many strategy realizations.
}
The forbidden minors classify dependencies rather than a unique physical layout.
For example, given a congestion game instance and a resource $e$ used by some strategy, we replace $e$ in every strategy using it by two
resources $e_1$ and $e_2$ that must be used together. Their loads agree at every
feasible flow. We assign them costs $\theta c_e$ and $(1-\theta)c_e$,
with $0<\theta<1$. \ma{This preserves every strategy cost and hence the equilibrium cost. If lowering the intercept of \(c_e\) by \(\delta>0\) strictly increases the equilibrium cost, we may choose \(\theta\) so that \(\delta/c_e(0)<\theta<1\), which is possible because \(c_e(0)-\delta>0\). Reducing the intercept of \(c_{e_1}=\theta c_e\) by \(\delta\) then gives the positive intercept \(\theta c_e(0)-\delta\). The combined cost of \(e_1\) and \(e_2\) at their common load \(s\) becomes \(c_e(s)-\delta\). Thus the original equilibrium-cost increase is preserved, showing the same paradox.} %every equilibrium comparison. If an affine witness lowers the intercept of $e$, the same decrease can be made in just $e_1$ after choosing it sufficiently small to keep that intercept positive. 
In the augmented incidence matrix,
this replacement corresponds to duplicating a resource column, which is a parallel extension.

Likewise, \ma{given a unit-demand example for a single population in which lowering a resource cost strictly increases the equilibrium cost, we add a new resource \(r\) to every strategy and give it the same positive affine cost function \(c_r\) before and after this reduction. The equilibrium costs before and after the reduction each increase by \(c_r(1)\). It follows that the equilibrium cost after the reduction is still strictly higher than the equilibrium cost before it. The paradox is preserved.} %for a one-population witness at fixed unit demand, adjoining a resource to every strategy adds the same constant $c_r(1)$ to every strategy cost in both instances. It preserves the paradox and 
This corresponds to adding a
column \ma{identical to} the demand column indexed by $g$ in the associated augmented incidence matrix. 

\ma{Repeating the above two types of extensions produces strategy families with arbitrarily many resources. Each admits the same type of strong Braess paradox with strictly positive affine resource costs. Additional strategies can also be accommodated, provided that their inclusion preserves the menu’s strictly positive exposure gap. Without a strictly positive exposure gap, adding strategies may change the equilibria; consequently, the original cost functions and intercept reduction may no longer yield a strong Braess paradox.}
%\ma{The proof of \cref{prop:single-analytic} uses this gap to ensure that, after a sufficiently small intercept reduction, every strategy outside the menu remains strictly more costly than the menu strategies.}
%The above two types of extensions of game instances give arbitrarily many larger realizations of one obstruction while keeping costs strictly positive and affine. Additional strategies can also be accommodated when the obstructed menu retains a strictly positive exposure gap, precisely the condition used in \cref{prop:single-analytic}. Without that gap, the menu and its witness cannot simply be assumed to survive.

%\ma{The forbidden-minor hypothesis of \Cref{prop:minor-witness} guarantees that the enumeration finds an affine basic point with a negative coordinate. The procedure then uses this point to construct cost functions for the resources of the original strategy family and to identify the resource whose cost intercept is to be lowered, showing a paradox.} %The procedure separates detection from realization. The forbidden minor certifies that the menu's associated matroid is not series-parallel; This negative coordinate then helps to identify a resource and slopes that produce a paradox. These costs are assigned to the original resources, even if the minor no longer contains the demand element $g$. For either matroid, $U_{2,4}$ or $M(K_4)$, the beta invariant is two. Its coordinate arrangement therefore has a second bounded chamber, and some vertex of that chamber has a negative coordinate, as in \cref{lem:bridge}.

\paragraph{The geometry behind a negative coefficient.} \ma{We first consider \(U_{2,4}\). By normalizing the three resource columns and the demand vector to lie on a common affine line, we can see why the representation of the demand vector in at least one resource basis has a negative coefficient. Suppose that the matrix \([-\boldsymbol b\mid\boldsymbol h_1\ \boldsymbol h_2\ \boldsymbol h_3]\) represents \(U_{2,4}\), where \(-\boldsymbol b\) is the demand column and \(\boldsymbol h_1,\boldsymbol h_2,\boldsymbol h_3\) are resource columns. Assume that there exists a vector \(\boldsymbol\ell\) such that \(\boldsymbol\ell^\top\boldsymbol h_j>0\) for \(j=1,2,3\), and that there exist strictly positive coefficients \(u_1^+,u_2^+,u_3^+\) such that \(\boldsymbol b=\sum_{j=1}^{3}u_j^+\boldsymbol h_j\).}
%The rank-two obstruction has an elementary picture. Suppose a dual representation $[-b\mid h_1\ h_2\ h_3]$ is $U_{2,4}$, the resource columns lie in an open halfspace $\ell h_j>0$, and $b=\sum_j u_j^+h_j$ with $u_j^+>0$. 
We normalize the resource columns
and demand direction to the affine line $\boldsymbol\ell^\top\boldsymbol v=1$ as follows:
\[
\boldsymbol v_j\triangleq\frac{\boldsymbol h_j}{\boldsymbol\ell^\top\boldsymbol h_j},\;\text{ for }j=1,2,3\qquad\text{and}\qquad\boldsymbol u\triangleq\frac{b}{\boldsymbol\ell^\top\boldsymbol  b}.
\]
The normalized point $\boldsymbol u$ is a strictly positive convex combination
of the $\boldsymbol v_j$. Indeed, its coefficient at $\boldsymbol v_j$ is
$u_j^+\boldsymbol \ell^\top\boldsymbol  h_j/(\boldsymbol \ell^\top\boldsymbol  b)>0$, and these coefficients sum to one because
\(\boldsymbol b=\sum_{j=1}^{3}u_j^+\boldsymbol h_j\).
After ordering the three distinct points as $\boldsymbol v_1<\boldsymbol v_2<\boldsymbol v_3$ on this line,
$\boldsymbol u$ lies strictly between $\boldsymbol v_1$ and $\boldsymbol v_3$. It is distinct from $\boldsymbol v_2$
because every pair of columns of $U_{2,4}$ is independent. If $\boldsymbol u<\boldsymbol v_2$,
then $\boldsymbol u$ lies outside the segment $[\boldsymbol v_2,\boldsymbol v_3]$; if $\boldsymbol u>\boldsymbol v_2$, it lies
outside $[\boldsymbol v_1,\boldsymbol v_2]$. In either case, its affine coordinates on one
resource pair include a negative coefficient. The positive normalizations
do not change that negativity. Therefore, a resource basis gives a negative basic
coordinate for $\boldsymbol b$. For example, \ma{suppose that the points \(\boldsymbol v_1,\boldsymbol v_2,\boldsymbol v_3,\boldsymbol u\) have affine coordinates \(0,1,2,\tfrac12\), respectively, on this line. Then} %with line coordinates
%$v_1=0$, $v_2=1$, $v_3=2$, and $u=1/2$,
$\boldsymbol u=\tfrac32\boldsymbol  v_2-\tfrac12\boldsymbol  v_3$.
Although the point $\boldsymbol u$ is a positive mixture of all resources, it cannot be expressed
nonnegatively on every independent resource pair. For $M(K_4)$, \cref{fig:chambers} \ma{shows a second bounded chamber with a vertex having a negative coordinate.} %gives the analogous obstruction as a second bounded chamber. 

Algebraically, the characteristic polynomials \ma{of the two matroids are \(\psi_{U_{2,4}}(t)=(t-1)(t-3)\) and \(\psi_{M(K_4)}(t)=(t-1)(t-2)(t-3)\).}
%are $(t-1)(t-3)$ for $U_{2,4}$ and $(t-1)(t-2)(t-3)$ for $M(K_4)$; 
Substituting their derivatives at $t=1$
into \eqref{eq:bounded-beta} gives \ma{\(\beta(U_{2,4})=\beta(M(K_4))=2\). The coordinate arrangements therefore have a second bounded chamber,
and some vertex of that chamber has a negative coordinate, as in
\cref{lem:bridge}.}

\ma{These geometric arguments show how affine basic points with negative coordinates arise in the \(U_{2,4}\) and \(M(K_4)\) cases. A forbidden minor of \(N_{\mathcal H}\) may omit the demand element \(g\). Even when \(g\) is retained, the corresponding demand vector in the minor’s dual representation need not be a strictly positive linear combination of the resource columns of that representation.} %These pictures explain the two core obstructions. A minor of a larger menu need not retain the distinguished demand element or its positive representation, so the general realization argument is \cref{prop:minor-witness}, which finds the negative basis in the original affine equations.

\subsection{The independent obstruction from demand labels}
\ma{For multiple populations, the series-parallel condition in \Cref{thm:multi}(ii) alone does not ensure immunity. Condition (i) additionally requires the demand elements in each marked block to form a single parallel class. In the next example, condition (ii) holds while condition (i) fails. We explicitly compute the resulting increase in one population’s equilibrium cost.} %For \ma{multiple} populations, series-parallel structure alone leaves a second possible obstruction: demand directions in the same resource block may fail to be parallel. The next example isolates this failure and computes the resulting increase in one population's equilibrium cost.

\begin{example}[Incompatible labels and an \ma{explicit} %exact 
paradox]
\label{ex:marked-first}\label{ex:marked-costs}
Given an instance with three populations and $\F_1=\{ab,c\}$, $\F_2=\{a\}$, $\F_3=\{b\}$,
%\begin{equation}\label{eq:marked-example-families}
 %      \F_1=\{ab,c\},\qquad
  %     \F_2=\{a\},\qquad
 %      \F_3=\{b\},
%\end{equation}
the full menu tuple $\boldsymbol{\HH}=(\F_i)_{i=1}^3$ is jointly trade-closed. Indeed, let resources $a,b,c$ be assigned weights $1,1,2$, respectively. Every strategy of population~1 then has
weight 2, while the sole strategies of populations~2 and~3 have
weight 1. \ma{Thus the full menu tuple is jointly positively exposed and hence jointly trade-closed by \Cref{lem:trade-multi}.} %A population-preserving coordinatewise weak saving must therefore preserve the weighted total; since all weights are positive, every resource count is preserved. There are no outside strategies.
The marked incidence matrix is
\[
 B_{\boldsymbol\HH}=\begin{array}{c|rrr|rrr}
 &g_1&g_2&g_3&a&b&c\\\hline
 (1,ab)&1&0&0&1&1&0\\
 (1,c) &1&0&0&0&0&1\\
 (2,a) &0&1&0&1&0&0\\
 (3,b) &0&0&1&0&1&0
 \end{array}.
\]
The dual of $M(B)$ is the following marked matroid
\[
 M=M\!\left(
 \begin{array}{rrr|rrr}
 -1&-1&0&1&0&1\\
 -1&0&-1&0&1&1
 \end{array}\right).
\]
After deleting the demand elements, \ma{we see that the resource columns represent the connected matroid \(U_{2,3}\). Thus \(\{a,b,c\}\) is the only resource block. In the marked matroid $M$, the demand elements \(g_1,g_2,g_3\) are parallel to the resource elements \(c,a,b\), respectively.} %the three resource columns form $U_{2,3}$'s representation, hence one connected resource block. In that block,
%\[
  %     g_1\parallel c,\qquad
  %     g_2\parallel a,\qquad
  %     g_3\parallel b.
%\]
The \emph{whole} matroid $M$ is series-parallel: starting with
the resource circuit $U_{2,3}$ and adding the three parallel demand copies.
However, the demand labels lie in three different parallel classes of
the same resource block, violating condition~(i) of \cref{thm:multi}.
\ma{The matroid \(M\) has no minor isomorphic to \(U_{2,4}\) or \(M(K_4)\)} %An unlabelled excluded-minor test passes 
because it is series-parallel. The calculation below shows that the
additional parallel-label condition is needed for immunity.

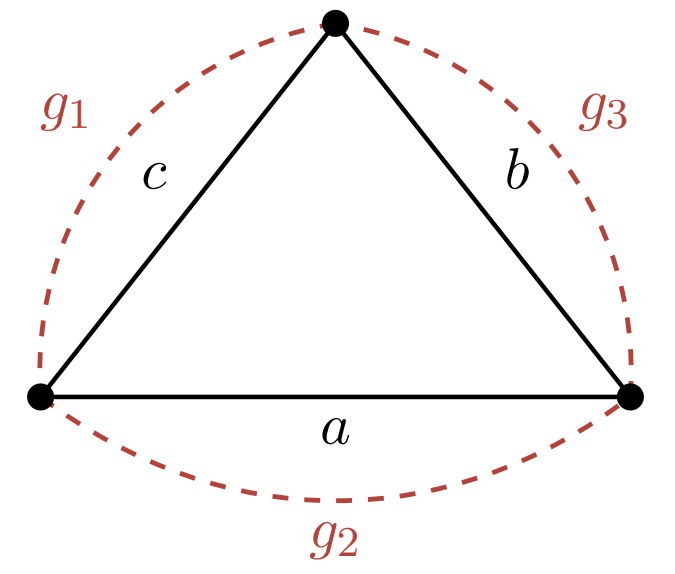
\begin{figure}[htbp]
\centering
\begin{tikzpicture}[scale=1.3]
\coordinate(A)at(0,0);\coordinate(B)at(3,0);
\coordinate(C)at(1.5,1.9);
\draw[resource](A)--node[below]{$a$}(B);
\draw[resource](B)--node[above right]{$b$}(C);
\draw[resource](C)--node[above left]{$c$}(A);
\draw[bad,dashed,line width=.9pt](A)to[bend right=37]node[below]{$g_2$}(B);
\draw[bad,dashed,line width=.9pt](B)to[bend right=42]node[above right]{$g_3$}(C);
\draw[bad,dashed,line width=.9pt](C)to[bend right=42]node[above left]{$g_1$}(A);
\foreach \v in {A,B,C}{\node[vertex]at(\v){};}
\end{tikzpicture}
\caption{Graphic picture of the marked matroid $M$ in
\cref{ex:marked-first}. Solid edges are resources and dashed edges are demand
labels. Every demand edge is individually parallel to a resource edge, so the
whole %unlabelled 
structure is series-parallel. The problem is that the three
demand labels lie in different parallel classes of the same resource block.}
\label{fig:marked}
\end{figure}
Suppose that all three populations have demand 1. Let resource costs be given by $c_a(s)=s+2$, $c_b(s)=s+2$ and $c_c(s)=s+13/2$.
%\begin{equation}\label{eq:marked-costs} c_a(s)=s+2,\qquad c_b(s)=s+2,\qquad c_c(s)=s+13/2.\end{equation}
\ma{Let \(r\in[0,1]\) be population 1’s flow on strategy \(ab\), so its flow on strategy \(c\) is \(1-r\). Populations 2 and 3 each have unit demand and use their sole strategies \(a\) and \(b\), respectively. Thus the resource loads, in the order \((a,b,c)\), are \((1+r,1+r,1-r)\).} %If population~1 sends fraction $r$ on $ab$, the forced populations already contribute one unit to $a$ and $b$, so the total resource loads are $(1+r,1+r,1-r)$. 
The costs of population~1's strategies $ab$ and $c$ are
$6+2r$ and $15/2-r$, respectively. Equating these two costs, we have $r=1/2$, which is strictly
between zero and one. It follows that both strategies $ab$ and $c$ are used in equilibrium. The equilibrium costs of populations are $\lambda_1(c,\boldsymbol{1})=7$, $\lambda_2(c,\boldsymbol{1})=\lambda_3(c,\boldsymbol{1})=7/2$.
Reducing only the intercept of $c_a$ by $\varepsilon\in(0,3/2)$, we have reduced costs $\bar c$, under which the new tie equation gives
$r=1/2+\varepsilon/3$. The upper bound 3/2 on $\varepsilon$ guarantees that this
fraction $r$ is strictly between zero and one and that the reduced intercept
is positive. So, the same strategies remain used, and the equilibrium costs become $\lambda_1(\bar c,\boldsymbol{1})=7-\frac{\varepsilon}{3}$,  $\lambda_2(\bar c,\boldsymbol{1})=\frac72-\frac{2\varepsilon}{3}$ and $\lambda_3(\bar c,\boldsymbol{1})=\frac72+\frac{\varepsilon}{3}$.
%\begin{equation}\label{eq:marked-increase}
% (\bar\lambda_1,\bar\lambda_2,\bar\lambda_3)
% =\left(7-\frac{\varepsilon}{3},\,
%        \frac72-\frac{2\varepsilon}{3},\,
%        \frac72+\frac{\varepsilon}{3}\right).
%\end{equation}
Population~3 pays more (even though its only available strategy is
unchanged), where improving $a$ shifts population~1 toward $ab$, raising the
load and cost on $b$. This shows a strong Braess paradox.

\ma{The matrix \(H\) and the demand vectors \(\boldsymbol b_1,\boldsymbol b_2,\boldsymbol b_3\) are given in} %The joint affine data are those in 
\eqref{eq:marked-preview}. \ma{With unit demand for each population, \(\boldsymbol b=\boldsymbol b_1+\boldsymbol b_2+\boldsymbol b_3=(2,2)^\top\). The affine space \(\{\boldsymbol x:H\boldsymbol x=\boldsymbol b\}\) has exactly two distinct affine basic points, \((2,2,0)^\top\) and \((0,0,2)^\top\), both of which are coordinatewise nonnegative. However, the affine space \(\{\boldsymbol z:H\boldsymbol z=\boldsymbol b_2\}\) has the affine basic point \((0,-1,1)^\top\), whose \(b\)-coordinate is negative. The nonnegativity condition in \Cref{prop:multi-analytic} fails for population 2.} %, so the resource matroid is connected. The aggregate vector $b=(2,2)\T$ has only the nonnegative basic points $(2,2,0)$ and $(0,0,2)$. Thus the aggregate single-population test passes. But $Hz=b_2$ has the basic point $(0,-1,1)$, so the labelled criterion fails. 
As shown in \cref{fig:marked}, the whole marked matroid is
series-parallel and contains neither excluded minor. The obstruction
is the failure of demand parallelism, not an unlabelled forbidden minor.
\end{example}
The aggregate direction passes every basis nonnegativity test,
but the individual direction $\boldsymbol b_2$ does not. The point $(0,-1,1)^\top$
is not asserted to be a feasible load for population~2; it is a basic
response in the joint exchange space. This is precisely the distinction
illustrated by the three-column example preceding \cref{lem:ray}.

\subsection{Matroid bases, harmless overlap, and shared factors}
The characterization in \Cref{thm:multi} also explains why some familiar forms of resource
overlap are harmless. We first recover immunity for matroid base
systems. Then we give a construction in which \ma{each population forms its strategies by choosing one strategy from each of several immune strategy families on disjoint resource sets. Different populations may use some of the same families.} %populations choose independently from shared immune factors.

\begin{corollary}[Matroid bases under arbitrary fixed sharing]
\label{cor:matroid-sharing}
Suppose that, for every population $i\in I$, the inclusion-minimal members of
$\F_i$ form the complete base family of a matroid of positive rank on $E$. Then
$(\F_i)_{i\in I}$ is immune, for every fixed resource-sharing pattern.
\end{corollary}
\begin{proof}
Fix any jointly positively exposed menu tuple $(\HH_i)_{i\in I}$, with common exposing weights
$\boldsymbol q>\boldsymbol 0$. %A strategy strictly containing another available strategy cannot minimize $\boldsymbol q$, so
Each $\HH_i$ consists exactly of the minimum-weight bases of
population $i$'s matroid. By \cref{lem:min-weight-bases}, these are the
complete bases of a matroid, denoted as $M_i^q$. We restrict the attention to the union $E'$ of the
resources occurring in the menus $\HH_i$, $i\in I$. The set $E'$ is nonempty under
our standing assumption that every available strategy uses a resource.

For a matroid base family, its exchange space is generated by
$\ee_e-\ee_f$ \ma{over all pairs of resources \(e,f\) belonging to the same nonloop connected component of the matroid (see Appendix~\ref{app:matroid-comparisons} for details). We construct an undirected graph with vertex set \(E'\) by joining two distinct vertices \(e,f\in E'\) with an edge if they belong to the same nonloop connected component of \(M_i^q\) for some population \(i\). Let} %for elements in the same nonloop matroid component; the circuit-to-base-exchange argument in \cref{app:matroid-comparisons} establishes this for arbitrary matroids. On $E'$, join two resources when they belong to one such component of any $M_i^q$, and let
$C_1,\ldots,C_m$ be the components of this graph, including isolated
resources (if any). The joint exchange space \(L=\sum_{i\in I}L_i\) defined in \eqref{eq:joint-affine-data} is therefore given by
%Summing the exchange spaces gives
\[
 L=\left\{\boldsymbol v\in\R^{E'}:
             \sum_{e\in C_j}v_e=0\text{ for all }j=1,\ldots,m\right\}.
\]
Indeed, \ma{each generating vector \(\boldsymbol e_e-\boldsymbol e_f\) has zero coordinate sum on every \(C_j\), since \(e\) and \(f\) belong to the same component of the auxiliary graph. Conversely, summing differences of unit vectors along a path in the graph shows that \(\boldsymbol e_e-\boldsymbol e_f\in L\) whenever \(e,f\) lie in the same \(C_j\). For a fixed \(f\in C_j\), every vector \(\boldsymbol v\) supported on \(C_j\) with zero coordinate sum satisfies \(\boldsymbol v=\sum_{e\in C_j\setminus\{f\}}v_e(\boldsymbol e_e-\boldsymbol e_f)\), and hence belongs to \(L\).} %each exchange preserves these sums; conversely, differences of unit vectors along a graph path telescope, generating every zero-sum direction within a component.

\ma{Let matrix \(H\) with rows indexed by \(j=1,\ldots,m\) and columns indexed by \(e\in E'\) be defined by setting \(H_{je}=1\) if \(e\in C_j\) and \(H_{je}=0\) otherwise. So,} %Choose $H$ with row $j$ equal to the indicator of $C_j$. For 
for $e\in C_j$,
its column in $H$ is $\boldsymbol h_e=\ee_j$. \ma{It is clear that $\ker H=L$. For each population \(i\in I\), the demand vector \(\boldsymbol b_i\) is given by} % Each population direction is
\[
 \boldsymbol b_i=H\boldsymbol \chi^{S_i}
     =\bigl(|S_i\cap C_1|,\ldots,|S_i\cap C_m|\bigr)\T\ge0,
 \qquad \text{for every }S_i\in\HH_i.
\]
\ma{For each population \(i\) and each component \(C_j\), the value of \(|S_i\cap C_j|\) is the same for all \(S_i\in\mathcal H_i\),} %These counts are independent of $S_i$ 
because $\ker H=L$. \ma{The resource blocks are therefore \(C_1,\ldots,C_m\), and the restriction of \(M(H)\) to each \(C_j\) has rank one. Therefore, each marked block has a matrix representation in which all resource columns and retained demand columns are nonzero scalar multiples of a single vector. Clearly,}
%The resource blocks are therefore the rank-one sets $C_j$. In each marked block, all retained demand columns and all resource columns are nonzero multiples of one vector. There is 
Each marked block has at least a retained demand label
because the block contains a resource used by some population.
Hence, the marked block is a rank-one series-parallel matroid and all
its nonloop demand labels are parallel. Both conditions of
\cref{thm:multi} hold for every tuple, which imply immunity.
\end{proof}

\Cref{cor:matroid-sharing} recovers immunity to the strong Braess paradox for matroid
base systems, a consequence of \citet[Theorem~3.1]{fujishige}, from
our marked-block characterization. %Their theorem also excludes increases in equilibrium resource costs. 
The calculation above explains why
arbitrary overlap of matroid base systems satisfies the parallel-label
condition. \ma{Since every marked block has rank one, all its retained demand elements belong to a single parallel class.} %every resource block has rank one.

\begin{example}[Partial resource overlap can be harmless]\label{ex:singleton-overlap}
Let $\F_1=\{a,b\}$ and $\F_2=\{b,c\}$ be strategy families of populations 1 and 2, respectively. %, where each displayed letter denotes a singleton strategy. 
The two populations overlap only on resource $b$, yet the
system is immune by \cref{cor:matroid-sharing}. Both families are rank-one
matroid bases. On the full joint menu $(\F_1,\F_2)$,
$L=\spanop\{\ee_a-\ee_b,\ee_b-\ee_c\}$, for which one may take
$H=(1,1,1)$ and $\boldsymbol b_1=\boldsymbol b_2=1$. The common resource block has compatible
demand directions even though the physical strategy sets differ.
\end{example}

\ma{Parallel demand elements need not be represented by identical columns.} %Parallel labels need not have identical magnitudes. 
If population~1 chooses
one resource from $\{a,b,c\}$ and population~2 chooses two, both strategy
families are matroid bases and the system is immune by
\cref{cor:matroid-sharing}. On the full joint menu,
$H=(1,1,1)$, $\boldsymbol b_1=1$, and $\boldsymbol b_2=2$. \ma{Since the demand column for population 2 is twice that for population 1, the two demand elements are parallel.} %: the demand columns are parallel, not equal. 
The general distinction between strategy matroids and augmented
incidence matroids is discussed in Appendix~\ref{app:matroid-comparisons}.

\ma{The following proposition constructs immune systems by combining immune strategy families on disjoint resource sets.} %The following sufficient construction gives a direct factor interpretation of another class of immune systems.

\begin{proposition}[Shared independent factors]\label{prop:network}
Let $E_1,\ldots,E_m$ be disjoint resource sets, and let
$\mathcal Q_r\subseteq2^{E_r}\setminus\{\varnothing\}$ be a nonempty immune
single-population family for every $r=1,\ldots,m$.  \ma{For each population \(i\in I\), let \(J_i\subseteq\{1,\ldots,m\}\) be a nonempty index set, which along with $\mathcal Q_r$'s defines the strategy family of  population $i$ as $\F_i\triangleq \{\bigcup_{r\in J_i}S_r:S_r\in\mathcal Q_r
                  \text{ for every }r\in J_i \}$} %Suppose each population $i$ has a nonempty set $J_i\subseteq\{1,\ldots,m\}$ of factors and the complete product family \[\F_i=\left\{\bigcup_{r\in J_i}S_r:S_r\in\mathcal Q_r                  \text{ for every }r\in J_i\right\}.\]
Then $(\F_i)_{i\in I}$ is immune.
\end{proposition}
\begin{proof} \ma{By the definition of \(\mathcal F_i\), replacing any \(S_r\) by another strategy from \(\mathcal Q_r\) yields another strategy in \(\mathcal F_i\). Since the total strategy cost is the sum of the costs of the selected \(S_r\), a strategy minimizes cost exactly when each \(S_r\) minimizes cost in \(\mathcal Q_r\). All populations \(i\) with \(r\in J_i\) choose from the same strategy family \(\mathcal Q_r\) on \(E_r\). Their total demand is \(d^{(r)}\triangleq\sum_{i:r\in J_i}d_i\). At an equilibrium of the original system, we group the strategies of populations \(i\) with \(r\in J_i\) according to their intersection with \(E_r\), and take the total flow assigned to the strategies in each group. The total flow assigned to the strategies in \(\mathcal Q_r\) equals \(d^{(r)}\). Since every strategy carrying positive flow minimizes cost in \(\mathcal Q_r\), this flow is a Wardrop equilibrium for the single-population system with demand \(d^{(r)}\) and resource costs \(c|_{E_r}\).}

%Because all combinations of factor strategies are available, a population can change its choice in one factor while keeping all other choices fixed. Its full strategy is therefore minimizing exactly when its choice is minimizing in every factor. In factor $r$, all participating populations have the same local family $\mathcal Q_r$, and their demands sum to $D_r=\sum_{i:r\in J_i}d_i$. A global equilibrium consequently induces an equilibrium of this single-population factor problem.

Conversely, \ma{for each \(r\) with \(d^{(r)}>0\), we choose an equilibrium of the corresponding single-population system. For each  \(S\in\mathcal Q_r\), let \(p_r(S)\) denote the fraction of its total flow assigned to \(S\). For each population \(i\) and each of its strategies \(\bigcup_{r\in J_i}S_r\), where \(S_r\in\mathcal Q_r\) for all \(r\in J_i\), we assign a flow of \(d_i\prod_{r\in J_i}p_r(S_r)\) to that strategy. Consider arbitrary \(r\) with \(d^{(r)}>0\) and \(S_r\in\mathcal Q_r\). For each population \(i\) with \(r\in J_i\), the total flow assigned to strategies \(S\in\mathcal F_i\) satisfying \(S\cap E_r=S_r\) is \(d_i p_r(S_r)\). Summing over these populations $i$ with $r\in J_i$, we have \(d^{(r)}p_r(S_r)\), which is exactly the flow assigned to \(S_r\) in the selected single-population equilibrium. In this way, the constructed strategy flow reproduces the selected equilibrium flow on every \(\mathcal Q_r\).} %choose an equilibrium distribution in each used factor, and let $p_r(S)$ be its proportion on $S\in\mathcal Q_r$. Assign population $i$ flow $d_i\prod_{r\in J_i}p_r(S_r)$ to the full strategy $\bigcup_{r\in J_i}S_r$. Summing over the other factor choices gives marginal flow $d_i p_r(S_r)$ in factor $r$, and aggregation over its participating populations gives $D_rp_r(S_r)$. Thus these product flows realize all the chosen factor equilibria simultaneously. 
It follows that

\[
\lambda_i(c,\boldsymbol d)=\sum_{r\in J_i}\lambda_r(c|_{E_r},d^{(r)})\quad\text{for all }i\in I.
\]
\ma{Under a resource-cost reduction and demands \(\bar d_i\le d_i\) for every population \(i\in I\), each single-population system on \(E_r\) also has reduced resource costs and total demand \(\bar d^{(r)}=\sum_{i:r\in J_i}\bar d_i\le d^{(r)}\). By the assumed immunity of each \(\mathcal Q_r\), the equilibrium cost of the single-population system on \(E_r\) does not increase under these reductions. Summing these costs over \(r\in J_i\), we conclude that the equilibrium cost of population \(i\) does not increase either for each $i\in I$, proving the immunity of  $(\F_i)_{i\in I}$.} %A resource-cost improvement and componentwise demand reduction improve each factor's data. Its immunity makes every summand nonincreasing, proving the result. Unused factors may be omitted.
\end{proof}

\ma{Suppose each \(\mathcal Q_r\) is the full family of source-sink paths in a two-terminal series-parallel network. The immunity of each \(\mathcal Q_r\)} %For complete two-terminal series-parallel path factors, the required local immunity 
follows from \cref{cor:routing-immunity} in the directed acyclic
case, or from \citet{milchtaich} in the undirected case. \ma{\Cref{prop:network} gives a sufficient condition for immunity when each population’s strategy family has the product form defined above on disjoint resource sets.} %This product construction supplies a sufficient physical factor decomposition; the general characterization uses the resource dependencies of each menu.

\subsection{Recognition and short certificates}
\ma{\Cref{thm:multi} implies a finite procedure for deciding immunity when the strategy families are explicitly listed. Enumerate all nonempty menu tuples and retain those for which \eqref{eq:recognition} is feasible. %By Corollary 3.6, these are exactly the jointly trade-closed menu tuples. 
The system is immune if and only if, for every retained tuple, each associated marked block is series-parallel and all its demand elements belong to a single parallel class.} %For explicitly listed strategy families, enumerate all nonempty menu tuples, retain those satisfying \eqref{eq:recognition}, and test the marked blocks of \cref{thm:multi}. 
For a specified tuple, rational linear algebra
computes a dual representation and its resource components in polynomial
time. A pair of nonzero parallel columns has rank one. Let \(F\) denote the ground set of the marked block under consideration. A series pair $e,f$ is
detected whenever $e$ and $f$ satisfy $ r(F\setminus e)=r(F\setminus f)=r(F)$ and $r(F\setminus\{e,f\})=r(F)-1$. Repeatedly delete one element of a parallel pair or contract one element of a series pair in each marked block. The block is series-parallel if and only if this process reduces it to \(U_{1,2}\). \ma{If the remaining matroid has more than two elements and contains neither a parallel pair nor a series pair, the original marked block is not series-parallel.}
%Successive parallel deletions and series contractions test whether a connected block is series-parallel: such a block reduces to $U_{1,2}$; if the reductions stop earlier, the test fails. 

Enumeration over all menu
tuples may still be exponential. To certify that a system is not immune,
however, it suffices to exhibit one \ma{jointly trade-closed menu tuple and an affine basic point with a negative coordinate.} %relevant tuple and one negative basic coordinate.

\begin{proposition}[Polynomial certificates for nonimmunity]\label{prop:coNP}
When all strategies are explicitly listed by their $0$-$1$ incidence
vectors, nonimmunity has a certificate of polynomial binary encoding
length, which \ma{is} verifiable in polynomial time. Hence deciding immunity
belongs to $\mathrm{coNP}$ under this input model.
\end{proposition}
\begin{proof}
A certificate consists of nonempty menu tuple $(\HH_j)_{j\in I}$, %a rational solution $\boldsymbol q$ of \eqref{eq:recognition}, 
a population index $i\in I$, \ma{a set \(J\) of resource indices appearing in these menus, and an index \(e\in J\).} %a resource $e$, and a column basis $J$ for the joint equation matrix $H$ such that $(H_J^{-1}b_i)_e<0$ on that basis. The basis coordinates are indexed by $J$, so the certificate requires $e\in J$.

\ma{The verification algorithm (verifier) first decides whether \eqref{eq:recognition} is feasible, in polynomial time by \Cref{cor:joint-recognition}. If it is infeasible, the algorithm rejects the certificate. Otherwise, the verifier constructs $L$ as the span of all vectors \(\boldsymbol\chi^S-\boldsymbol\chi^T\), taken over all populations \(j\in I\) and all pairs of strategies \(S,T\in\mathcal H_j\). It then uses a \emph{fixed} row-reduction procedure over the rationals to compute a full-row-rank matrix \(H\) satisfying} %from the within-menu incidence differences, chooses $H$ by a fixed rational row-reduction procedure with 
$\ker H=L$. \ma{It chooses a reference strategy \(S_i^0\in\mathcal H_i\) and set \(\boldsymbol b_i=H\boldsymbol\chi^{S_i^0}\). The verifier accepts the certificate only if it verifies that \(H_J\) is nonsingular and \((H_J^{-1}\boldsymbol b_i)_e<0\), using exact rational arithmetic.}
%and sets $b_i=H\chi^{S_i^0}$. It checks the exposure inequalities, the nonsingularity of $H_J$, and the negative coordinate exactly.
By \cref{lem:trade-multi,prop:multi-analytic}, every accepted certificate
proves nonimmunity, and every nonimmune system \ma{admits a certificate that the verifier accepts.} %has such menus and a basis.

It remains to bound the encoding lengths. The menus are subsets of the
explicit input. %The feasible system \eqref{eq:recognition} has polynomially many variables and constraints with coefficients in $\{-1,0,1\}$.  Standard rational-polyhedral bounds provide a feasible $\boldsymbol q$ of polynomial bit length \citep[Chapter~10]{schrijverLP}. 
The matrix of exchange differences has entries in $\{-1,0,1\}$. Rational elimination and determinant bounds
therefore give polynomial bit lengths for the computed $H$, $\boldsymbol b_i$, and
\ma{\((H_J^{-1}\boldsymbol b_i)_e\)}. %the basis coefficients. 
The remaining certificate data are just indices.
All checks use polynomial-time exact rational arithmetic.
\end{proof}

The certificate records the obstructed menu and \ma{the indices needed to recover an affine basic point with a negative coordinate.} % a negative basic response;
The witness construction then supplies numerical costs. \ma{For explicitly listed strategy families, nonimmunity admits certificates of polynomial encoding length that can be verified in polynomial time.} %The bound concerns verification of a negative answer for explicitly listed strategies.
Polynomial-time recognition and $\mathrm{coNP}$-hardness of the entire
problem remain open.

\section{Conclusion}
For finite nonatomic games with \ma{fixed strategy families on a common resource set and common continuous nondecreasing resource cost functions,} %fixed sharing and common continuous nondecreasing costs,
we show that immunity is determined by the incidence structure of
positively exposed complete menus. \ma{For a single population, immunity holds if and only if the augmented incidence matrix of every positively exposed menu represents a series-parallel matroid.} %Series-parallel augmented columns characterize the single-population case. 
For \ma{multiple} populations, the ray
lemma adds compatibility of the individual demand directions within each
resource block. Whenever a system is not immune, the proof constructs a strong
Braess paradox with unit demands and strictly positive affine costs,
by lowering the intercept of one resource cost. The same criterion recovers matroid
base systems under arbitrary sharing parterns and complete acyclic series-parallel
path systems.

\ma{Our characterizations require that the matroid conditions hold for every positively exposed menu or jointly positively exposed menu tuple. This} %The menu dependence of the characterization 
raises two further questions:
can a smaller sufficient collection of menus be found efficiently, and
can the marked-block conditions be assembled into a constructive global
description of immune systems? \Cref{ex:submenus} shows why testing only
the full family is insufficient. Resolving these questions would clarify
the recognition complexity beyond the present certificate bound.
Player-specific costs, atomic users, and elastic demand provide further
directions in which the equilibrium response argument would need to be
extended.

\appendix

\section{From local derivatives to general continuous costs}\label{app:regularity}
\ma{In this section, the main proofs analyze changes in population equilibrium costs on intervals where the minimum-cost strategy menus and the active affine pieces of the resource cost functions are fixed. The analysis assumes that the resource cost functions are piecewise affine, take strictly positive values, and have strictly positive slopes on every affine piece.} %The main proofs calculate how equilibrium population costs change while the complete menus and the affine pieces of the cost functions are fixed. That calculation  assumes positive costs and slopes. 
The main
theorems, however, allow continuous nondecreasing costs, including flat
pieces and nondifferentiable points. Menus may change, and equilibrium
loads need not be unique. This section justifies the passage from the
local calculation to that general setting.

We proceed in three steps.  First, we prove that equilibrium costs are continuous
under uniform changes of the resource-cost functions and convergent changes
of demand, even when loads are not unique. This is consistent with classical
sensitivity theory \citep{hall,patriksson,josefssonpatriksson}.
Second, \ma{for strictly positive piecewise affine resource cost functions with strictly positive slopes, we show that the equilibrium load vector and population equilibrium costs are piecewise rational along the interpolation path, with finitely many pieces. If a population’s equilibrium cost has a nonpositive derivative on each smooth interval, continuity at the interval endpoints implies that the cost is nonincreasing along the entire interpolation path.} %for positive piecewise-affine costs with positive slopes, the equilibrium follows finitely many rational branches. The derivative inequalities therefore join across their endpoints. 
Third, we approximate
any ordered pair of continuous nondecreasing resource cost functions \ma{uniformly on a bounded interval containing all feasible loads by ordered pairs of such piecewise affine functions. Continuity of equilibrium costs then implies that the inequality between the initial and final equilibrium costs proved for these approximations also holds for the original cost functions.} %by such an ordered pair of piecewise-affine functions and pass the endpoint inequality to the limit. 
These regularity arguments do not use the matroid
characterization.

\subsection{Continuity of equilibrium costs}
For a demand vector $\boldsymbol d$, let $P(\boldsymbol d)$ denote the polytope of feasible aggregate loads.  Recall that Wardrop loads are precisely the minimizers of the
Beckmann potential, or equivalently the solutions $\boldsymbol x$ of the variational
inequality
\begin{equation}\label{eq:vi-regularity}
 c(\boldsymbol x)\T(\boldsymbol y-\boldsymbol x)\ge 0\qquad\text{for all }\boldsymbol y\in P(\boldsymbol d).
\end{equation}
The variational-inequality argument in \cref{sec:model} shows that all
equilibria have the same resource-cost vector, even if their loads differ.
We include the corresponding perturbation argument below.

\begin{lemma}[Continuity of equilibrium population costs]\label{lem:cost-continuity} \ma{Let \((\boldsymbol d^{(n)})_{n\ge1}\) be a sequence of strictly positive demand vectors converging to a strictly positive demand vector \(\boldsymbol d\).}
%Let $d^n\to d$ with positive demands, and 
Suppose that $c^{(n)}$ converges
uniformly to $c$ on a \ma{common bounded} interval containing all feasible resource loads.  \ma{For each \(n\), let \(\boldsymbol x^{(n)}\) be an equilibrium load vector for the instance \((c^{(n)},\boldsymbol d^{(n)})\). Then every accumulation point of \((\boldsymbol x^{(n)})_{n\ge1}\) is an equilibrium load vector for \((c,\boldsymbol d)\).} %If $x^n$ is an equilibrium for $(c^{(n)},\boldsymbol d^{(n)})$, then every accumulation point of $(x^n)$ is an equilibrium for $(c,\boldsymbol d)$.  
Moreover, \ma{for every population \(i\in I\), \(\lambda_i(c^{(n)},\boldsymbol d^{(n)})\to\lambda_i(c,\boldsymbol d)\) as \(n\to\infty\). If every resource cost function \(c_e\), \(e\in E\), is strictly increasing, the instance \((c,\boldsymbol d)\) has a unique equilibrium load vector \(\boldsymbol x\), and \(\boldsymbol x^{(n)}\to\boldsymbol x\) as \(n\to\infty\).} %every population equilibrium cost converges to its equilibrium cost in the limiting instance.  If all limiting resource costs $c_e$ are strictly increasing, the aggregate equilibrium load is unique and $x^n\to x$.
\end{lemma}
\begin{proof}

Because $\boldsymbol d^{(n)}\to \boldsymbol  d$ as $n\to\infty$, the total demands are bounded, and so are all
feasible resource loads. \ma{Let \((\boldsymbol x^{(n_j)})_{j\ge1}\) be any convergent subsequence of \((\boldsymbol x^{(n)})_{n\ge1}\), and denote its limit by \(\boldsymbol x\). We first show that \(\boldsymbol x\in P(\boldsymbol d)\). Since \(\boldsymbol x^{(n_j)}\in P(\boldsymbol d^{(n_j)})\), we may choose vectors \(\boldsymbol x_i^{(n_j)}\in P_i\), $i\in I$ such that \(\boldsymbol x^{(n_j)}=\sum_{i\in I}d_i^{(n_j)}\boldsymbol x_i^{(n_j)}\).} %Take a convergent subsequence $x^{n_j}\to x$. To see that its limit is feasible, decompose $x^{n_j}=\sum_i d_i^{n_j}x_i^{n_j}$ with $x_i^{n_j}\in P_i$.
The finite product $\prod_{i\in I}P_i$ is compact. By compactness, \ma{there is a subsequence \((n_{j_l})_{l\ge1}\) of \((n_j)_{j\ge1}\) such that \(\boldsymbol x_i^{(n_{j_l})}\to\boldsymbol x_i\in P_i\) for every \(i\in I\).} %so after taking a further subsequence these component vectors converge to $x_i\in P_i$.
It follows that $\boldsymbol x=\sum_{i\in I} d_i\boldsymbol x_i\in P(\boldsymbol d)$.

\ma{For an arbitrary \(\boldsymbol y\in P(\boldsymbol d)\), we choose vectors \(\boldsymbol y_i\in P_i\), \(i\in I\), such that \(\boldsymbol y=\sum_{i\in I}d_i\boldsymbol y_i\). For each \(l\ge1\), we define \(\boldsymbol y^{(n_{j_l})}\triangleq\sum_{i\in I}d_i^{(n_{j_l})}\boldsymbol y_i\). Then \(\boldsymbol y^{(n_{j_l})}\in P(\boldsymbol d^{(n_{j_l})})\) and \(\boldsymbol y^{(n_{j_l})}\to\boldsymbol y\).} 
%Fix any comparison load $\boldsymbol y\in P(\boldsymbol d)$. Write $y=\sum_i d_iy_i$ with $y_i\in P_i$ and use the same component vectors to define $y^{n_j}=\sum_i d_i^{n_j}y_i$. Then $y^{n_j}\in P(d^{n_j})$ and $y^{n_j}\to y$. Uniform convergence of the costs and convergence of the loads give $c^{n_j}(x^{n_j})\to c(\boldsymbol x)$. 
\ma{For each \(l\ge1\), the equilibrium condition for \(\boldsymbol x^{(n_{j_l})}\) and the feasibility of \(\boldsymbol y^{(n_{j_l})}\) give $c^{(n_{j_l})}\!(\boldsymbol x^{(n_{j_l})})^\top
(\boldsymbol y^{(n_{j_l})}-\boldsymbol x^{(n_{j_l})})\ge0$.
Uniform convergence of the cost functions, continuity of \(c\), and \(\boldsymbol x^{(n_{j_l})}\to\boldsymbol x\) imply that \(c^{(n_{j_l})}(\boldsymbol x^{(n_{j_l})})\to c(\boldsymbol x)\). Letting \(l\to\infty\) and using \(\boldsymbol y^{(n_{j_l})}\to\boldsymbol y\), we obtain \(c(\boldsymbol x)^\top(\boldsymbol y-\boldsymbol x)\ge0\). Since \(\boldsymbol y\in P(\boldsymbol d)\) is arbitrary and \(\boldsymbol x\in P(\boldsymbol d)\), the vector \(\boldsymbol x\) is an equilibrium load vector for \((c,\boldsymbol d)\).} %We may therefore pass to the limit in each equilibrium variational inequality to obtain \[ c(\boldsymbol x)\T(\boldsymbol y-\boldsymbol x)\ge0,\] so $x$ is an equilibrium of the limit instance.

\ma{Let \(\boldsymbol x,\hat{\boldsymbol x}\in P(\boldsymbol d)\) be two equilibrium load vectors for the same instance \((c,\boldsymbol d)\). Applying (A.1) to each vector, using the other as the comparison load, and adding the two inequalities, we have $\sum_{e\in E}
(c_e(x_e)-c_e(\hat x_e))(x_e-\hat x_e)\le0$.
Each term in this sum is nonnegative because the resource cost function \(c_e\) is nondecreasing. It follows that} %For a fixed continuous nondecreasing cost vector, the variational inequality implies uniqueness of the equilibrium resource-cost vector on any two solutions: applying the inequality in both directions yields $(c(\boldsymbol x)-c(\hat x))\T(x-\hat x)\le0$, whereas monotonicity gives the reverse inequality term by term.  Hence 
every term is zero. When two resource loads differ, this
forces equality of their cost values; when they agree, equality is
immediate. Thus, all equilibria have the same resource-cost vector. \ma{Let \(\boldsymbol p\) denote the unique equilibrium resource cost vector for \((c,\boldsymbol d)\). Along any convergent subsequence \((\boldsymbol x^{(n_{j_l})})_{l\ge1}\) of the equilibrium load sequence, uniform convergence of the cost functions and continuity of \(c\) imply that \(c^{(n_{j_l})}(\boldsymbol x^{(n_{j_l})})\to\boldsymbol p\). Since \(\mathcal F_i\) is finite, the equilibrium cost of population \(i\) along this subsequence converges to \(\min_{S\in\mathcal F_i}\sum_{e\in S}p_e=\lambda_i(c,\boldsymbol d)\).}
%Along any convergent load subsequence, uniform convergence gives this same limiting vector. Each population cost is the minimum of finitely many sums of resource costs, so it has the corresponding unique limit.

\ma{For each $n\ge1$, let \(\boldsymbol p^{(n)}=c^{(n)}(\boldsymbol x^{(n)})\). Suppose that \(\boldsymbol p^{(n)}\) does not converge to \(\boldsymbol p\). Then there exist \(\varepsilon>0\) and a subsequence \((n_j)_{j\ge1}\) such that \(\|\boldsymbol p^{(n_j)}-\boldsymbol p\|\ge\varepsilon\) for every \(j\). Since \((\boldsymbol x^{(n_j)})_{j\ge1}\) is bounded, it has a convergent subsequence \((\boldsymbol x^{(n_{j_l})})_{l\ge1}\). The preceding argument implies that \(\boldsymbol p^{(n_{j_l})}\to\boldsymbol p\), contradicting the lower bound $\varepsilon$. Thus, when $n\to\infty$, we have \(\boldsymbol p^{(n)}\to\boldsymbol p\), and consequently \(\lambda_i(c^{(n)},\boldsymbol d^{(n)})\to\lambda_i(c,\boldsymbol d)\) for every \(i\in I\).}

\ma{If each \(c_e\), $e\in E$, is strictly increasing, then the equality \(c(\boldsymbol x)=c(\hat{\boldsymbol x})\) implies \(\boldsymbol x=\hat{\boldsymbol x}\). Hence the instance \((c,\boldsymbol d)\) has a unique equilibrium load vector \(\boldsymbol x\). Every accumulation point of the bounded sequence \((\boldsymbol x^{(n)})_{n\ge1}\) must equal \(\boldsymbol x\), giving \(\boldsymbol x^{(n)}\to\boldsymbol x\) as $n\to\infty$.}
%If the full cost sequence did not converge, a subsequence bounded away from this limit would have a convergent load subsequence, contradicting what we have proved. Under strict increase, equal resource-cost values also force equal loads, giving uniqueness and convergence of $x^n$.
\end{proof}

\subsection{Strictly increasing piecewise-affine costs}\label{apx:piecewise}
\ma{Suppose first that the resource cost functions \(c_e\) and \(\bar c_e\), \(e\in E\), are piecewise affine, with all breakpoints contained in a common finite set. Each function takes strictly positive values, and each affine piece has a strictly positive slope. Assume that \(\bar c_e(s)\le c_e(s)\) for every \(e\in E\) and \(s\ge0\). Let \(\boldsymbol d\) and \(\bar{\boldsymbol d}\) be two strictly positive demand vectors. For \(t\in[0,1]\), we define \(c^{(t)}\triangleq(1-t)c+t\bar c\) and \(\boldsymbol d^{(t)}\triangleq(1-t)\boldsymbol d+t\bar{\boldsymbol d}\). For each \(t\in[0,1]\), the interpolated resource cost functions \(c_e^{(t)}\), $e\in E$, are strictly increasing, so the instance \((c^{(t)},\boldsymbol d^{(t)})\) has a unique equilibrium load vector \(\boldsymbol x(t)\). \cref{lem:cost-continuity} implies that the map \(t\mapsto\boldsymbol x(t)\) is continuous. For each population $i\in I$, we set \(\lambda_i(t)\triangleq\lambda_i(c^{(t)},\boldsymbol d^{(t)})\).}
%Suppose first that $c$ and $\bar c$ are piecewise affine on a common finite grid, have strictly positive values and slopes, and satisfy $\bar c\le c$.  Interpolate both costs and the two demand vectors linearly in $t\in[0,1]$.  Strict increase makes the equilibrium aggregate load $x(t)$ unique, and \cref{lem:cost-continuity} makes it continuous.

A combinatorial-affine state records the complete \ma{minimizing} menu of each
population, the set $E'$ of their used resources, and \ma{the affine piece of each resource cost function that applies at its current load. Since each strategy family is finite and each resource cost function has finitely many affine pieces, there are} %the affine segment containing each resource load. There are finitely many strategies and segments, hence 
only finitely many states. Consider an arbitrary state \ma{with the complete minimizing menu \(\mathcal H_i\) of each population \(i\) and used resource set \(E'\). We restrict the load and resource cost vectors to \(E'\), and construct \(H\) and \(B_{\mathrm{dem}}\) from these menus as in \eqref{eq:joint-affine-data} and \eqref{eq:dual-H}. For the fixed state, let \(D(t)\) be the diagonal matrix of the slopes of the selected affine pieces, and let \(\boldsymbol\gamma(t)\) be the vector of their intercepts. The resource cost vector \(\boldsymbol p(t)=(c_e^{(t)}(x(t)_e))_{e\in E'}\) then satisfies \(\boldsymbol p(t)=D(t)\boldsymbol x(t)+\boldsymbol\gamma(t)\). All diagonal entries of \(D(t)\) are strictly positive. So we may define \(W(t)\triangleq D(t)^{-1}\).}
%On its resource coordinates, use the joint equation matrix $H$ and demand-direction matrix $B_{\mathrm{dem}}$ from \cref{eq:joint-affine-data}. The cost vector has the form \[ p=D(t)x+\gamma(t),\qquad D(t)>0,  \qquad W(t)=D(t)^{-1}.\]
The slopes and intercepts \ma{of the selected affine pieces depend affinely on \(t\), since \(c^{(t)}=(1-t)c+t\bar c\). If the equilibrium has the fixed state, then, for each \(i\in I\), all strategies in \(\mathcal H_i\) have the same cost under \(\boldsymbol p(t)\). Thus \((\boldsymbol\chi^S-\boldsymbol\chi^T)^\top\boldsymbol p(t)=0\) for all \(S,T\in\mathcal H_i\). Since these differences span \(L\), we have \(\boldsymbol p(t)\in L^\perp=\operatorname{row}(H)\). Suppose each population \(i\) allocates its demand \(d_i^{(t)}\) among strategies in \(\mathcal H_i\), producing aggregate load \(\boldsymbol x(t)\). Since \(H\boldsymbol\chi^S=\boldsymbol b_i\) for every \(S\in\mathcal H_i\), we have \(H\boldsymbol x(t)=\sum_{i\in I}d_i^{(t)}\boldsymbol b_i=B_{\mathrm{dem}}\boldsymbol d^{(t)}\).} 
%are affine in $t$ because the two endpoint cost functions are interpolated linearly. Ties within each complete menu make $p$ orthogonal to all joint exchange directions, so $p\in\operatorname{row}(H)$. Feasibility on these menus gives $Hx=B_{\mathrm{dem}}\boldsymbol d^{(t)}$. 
To solve these equations, we write $\boldsymbol p(t)=H\T \boldsymbol y$.
Then $\boldsymbol x(t)=W(t)H\T \boldsymbol y-W(t)\gamma(t)$, and
\[
 HW(t)H\T \boldsymbol y=B_{\mathrm{dem}}\boldsymbol d^{(t)}+HW(t)\gamma(t).
\]
\ma{Since \(HW(t)H^\top\) is positive definite, the preceding system has a unique solution \(\boldsymbol y\). Substituting this solution into \(\boldsymbol x(t)=W(t)H^\top\boldsymbol y-W(t)\boldsymbol\gamma(t)\) gives the following formula for the equilibrium load whenever the fixed state occurs.}
%Inverting this positive definite matrix and substituting back gives the candidate branch
\begin{equation}\label{eq:rational-branch}
\boldsymbol  x(t)=W(t)H\T(HW(t)H\T)^{-1}
           \bigl(B_{\mathrm{dem}}\boldsymbol d^{(t)}+HW(t)\gamma(t)\bigr)-W(t)\gamma(t)
\end{equation}
\ma{The formula is well defined and finite for every \(t\in[0,1]\). It specifies the coordinates of \(\boldsymbol x(t)\) indexed by \(E'\). For every \(e\in E\setminus E'\), we set \(x(t)_e=0\). Thus, on any interval where the state remains fixed, every coordinate of \(\boldsymbol x(t)\)}
%on $E'$, with zero coordinates outside $E'$. Here $E'$ is nonempty under the standing assumption that all strategies are nonempty. Every coordinate of the candidate branch 
is a rational function of $t$,
that is, a quotient of polynomials in $t$. \ma{The entries of \(D(t)\), \(\boldsymbol\gamma(t)\), and \(\boldsymbol d^{(t)}\) are affine functions of \(t\) with real coefficients. The additions, multiplications, and matrix inversions in (A.2) therefore yield quotients of polynomials in \(t\) with real coefficients.} %This does not require rational numerical cost data: the polynomial coefficients may be real. Indeed, the entries of $W(t)$ are reciprocal affine functions, and matrix inversion uses cofactors divided by a determinant. The resulting branch has no pole on $[0,1]$: all diagonal entries of $D(t)$ are positive, and full row rank of $H$ makes $HW(t)H\T$ positive definite throughout.

\begin{lemma}[\ma{Piecewise rationality on a finite partition}%Finite-branch regularity
]\label{lem:finite-branches}
\ma{The maps \(t\mapsto\boldsymbol x(t)\) and \(t\mapsto\lambda_i(t)\), \(i\in I\), are continuous on \([0,1]\). Moreover, there exists a finite partition $0=t_0<t_1<\cdots<t_m=1$
such that, on each open interval \((t_{j-1},t_j)\), every coordinate \(x(t)_e\), \(e\in E\), and every cost \(\lambda_i(t)\), \(i\in I\), can be expressed as a quotient of polynomials in \(t\) with real coefficients and a denominator that is nonzero throughout that interval. In particular, these functions are differentiable on \((0,1)\) except possibly at \(t_1,\ldots,t_{m-1}\).}
%For strictly increasing piecewise-affine costs on a common finite grid, the equilibrium path $x(t)$ and every equilibrium population cost $\lambda_i(t)$ are continuously pieced together from finitely many rational functions.  After merging identical branches, only finitely many switching times remain. Consequently they are differentiable except at finitely many values of $t$.
\end{lemma}

\begin{proof}
\ma{\Cref{lem:cost-continuity} implies that \(\boldsymbol x(t)\) and every \(\lambda_i(t)\) are continuous on \([0,1]\). For each state \(\sigma\), let \(\boldsymbol x_\sigma(t)\) be the vector defined by \eqref{eq:rational-branch}, assigning zero load to resources unused by its menus. At every \(t\), the equilibrium load \(\boldsymbol x(t)\) equals \(\boldsymbol x_\sigma(t)\) for at least one state \(\sigma\).}

\ma{We group together states whose expressions \(\boldsymbol x_\sigma(t)\) agree for every \(t\in[0,1]\), and retain one expression from each group. For two distinct retained expressions \(\boldsymbol x_\sigma(t)\) and \(\boldsymbol x_\tau(t)\), we choose \(e\in E\) such that \(x_\sigma(t)_e-x_\tau(t)_e\) is not identically zero. The numerator of this rational function  \(x_\sigma(t)_e-x_\tau(t)_e\) is a nonzero polynomial and therefore has only finitely many zeros. Hence, there are only finitely many values \(t\in[0,1]\) for which \(\boldsymbol x_\sigma(t)=\boldsymbol x_\tau(t)\).  Partition \([0,1]\) at all these values. On each resulting open interval, the retained expressions have pairwise distinct values. By continuity of the map \(t\mapsto\boldsymbol x(t)\) and each retained map \(t\mapsto\boldsymbol x_\sigma(t)\), the index of the expression equal to \(\boldsymbol x(t)\) is locally constant on the interval. Since the interval is connected, this index is constant throughout the interval. Thus \(\boldsymbol x(t)\) agrees there with one fixed expression \(\boldsymbol x_\sigma(t)\).}

\ma{The preceding partition of interval $[0,1]$ fixes the expression for \(\boldsymbol x(t)\), but several states may share that expression on the same subinterval. We further refine the partition so that each resource cost is evaluated using a fixed affine piece on every open interval. On each subinterval obtained by the above partition, we subdivide further at the solutions of \(x(t)_e=a\), for every resource \(e\) and every cost breakpoint \(a\), unless the equality holds throughout the interval. Since \(x(t)_e\) is rational there, only finitely many points are added. On each refined open interval, every resource load lies in one fixed affine piece or remains at a breakpoint. In the former case, \(c_e^{(t)}(x(t)_e)\) is rational in \(t\), since \(x(t)_e\) is rational and the selected affine formula has coefficients that are affine in \(t\). In the latter case, its value is \((1-t)c_e(a)+t\bar c_e(a)\), which is affine in \(t\) and hence rational.}

\ma{Consequently, every strategy cost \(\sum_{e\in S}c_e^{(t)}(x(t)_e)\), \(S\in\mathcal F_i\), is rational on each subinterval. Two strategy costs of the same population either agree throughout the subinterval or only finitely many values of \(t\) within that subinterval. We finally subdivide at these values for pairs that do not agree throughout. On every final open subinterval, the ordering of strategy costs is fixed. It follows that each population has a fixed complete minimizing menu and \(\lambda_i(t)\) equals the cost of one fixed strategy. All load coordinates and population equilibrium costs are therefore rational, and hence differentiable, on the open subintervals of the resulting finite partition.}
%
%At each value of $t$, the equilibrium belongs to a state and hence agrees with one of the finitely many candidate branches. Merge branches that are identical as vector functions. Any remaining pair can intersect only finitely often: at least one coordinate of their difference is a nonzero rational function, whose umerator has finitely many roots. Since the equilibrium load is continuous, it can switch between distinct branches only where they intersect. Thus only finitely many branch switches are possible.
%
%We also subdivide at changes of cost segments and minimizing menus. Along a fixed branch, a load reaches a grid point or a strategy gap becomes zero at a root of a rational function. There are finitely many such roots unless the equality holds identically on an interval. A persistent equality adds no switching times: retain the complete tied menu on that interval, and handle a persistent cost kink as explained below. On every resulting open interval, the relevant cost formulas and menus are fixed, so all strategy costs, including $\lambda_i(t)$, are rational functions. This gives the claimed finite subdivision.
\end{proof}

On an open interval of the subdivision specified in the above proof, \ma{the complete minimizing menus \((\mathcal H_i)_{i\in I}\) and the resource set \(E'\) are fixed. The associated matrix \(H\) and demand vectors \(\boldsymbol b_i\), constructed as in \Cref{sec:joint}, can therefore be chosen independently of \(t\). For every population \(i\in I\) and every strategy \(S\in\mathcal H_i\), we have \((\boldsymbol\chi^S)^\top\boldsymbol p(t)=\lambda_i(t)\) throughout the interval. Differentiating with respect to \(t\) gives \((\boldsymbol\chi^S)^\top\boldsymbol p'(t)=\lambda_i'(t)\).}
%retain the fixed complete menus and their equation data $H,b_i$ on $E'$. Resources outside $E'$ have zero load. For a minimizing strategy $a$ of population $i$, the gap $\boldsymbol a^\top p(t)-\lambda_i(t)$ is zero; differentiating gives $\boldsymbol a^\top \boldsymbol p'=\lambda_i'$. 
For every \(i\in I\) and every pair \(S,T\in\mathcal H_i\), subtracting the derivative equalities for \(S\) and \(T\) gives \((\boldsymbol\chi^S-\boldsymbol\chi^T)^\top\boldsymbol p'(t)=0\). Since these differences span \(L=\ker H\), we obtain \(\boldsymbol p'(t)\perp\ker H\). \ma{The equilibrium load \(\boldsymbol x(t)\) satisfies \(H\boldsymbol x(t)=\sum_{i\in I}d_i^{(t)}\boldsymbol b_i\) throughout the interval. Since \(H\) and the vectors \(\boldsymbol b_i\) are independent of \(t\), differentiating gives \(H\boldsymbol x'(t)=\sum_{i\in I}(\bar d_i-d_i)\boldsymbol b_i\). Moreover, differentiating \(\boldsymbol p(t)=D(t)\boldsymbol x(t)+\boldsymbol\gamma(t)\) gives \(\boldsymbol p'(t)=D(t)\boldsymbol x'(t)+D'(t)\boldsymbol x(t)+\boldsymbol\gamma'(t)\). Combining this identity with \(\boldsymbol p'(t)\perp\ker H\) and \(H\boldsymbol x'(t)=\sum_{i\in I}(\bar d_i-d_i)\boldsymbol b_i\) gives the linear system used to derive \eqref{eq:single-sensitivity} and \eqref{eq:multi-sensitivity}.}
%All feasible loads on these menus satisfy $Hx=\sum_i b_id_i(t)$, and $H,b_i$ are fixed on the interval. Thus \[ Hx'=\sum_i b_id_i'.\] Together with $\boldsymbol p'=D\boldsymbol x'+\boldsymbol h$, where $h=\partial_t   c^{(t)}(x)$ holds the load fixed, these are precisely the equations used in \eqref{eq:single-sensitivity} and \eqref{eq:multi-sensitivity}.

\ma{For each resource \(e\), let \(a\) be a load value at which the affine formula for \(c_e\) or \(\bar c_e\) changes. The subdivision constructed above ensures that, on each open subinterval, for each resource $e$, either \(x(t)_e\ne a\) throughout the subinterval or \(x(t)_e=a\) throughout it. In the latter case, \(x'(t)_e=0\) and $p(t)_e=c_e^{(t)}(a)=(1-t)c_e(a)+t\bar c_e(a)$, giving \(p'(t)_e=\bar c_e(a)-c_e(a)\). For every \(t\), the affine expressions for \(c_e^{(t)}\) on the two sides of \(a\) have the same value at \(a\), because \(c_e^{(t)}\) is continuous in the load variable. Differentiating either affine expression with respect to \(t\), with the load fixed at \(a\), therefore gives the same value. Thus, \(D(t)_{ee}\) and \(\gamma(t)_e\) may be taken as the slope and intercept, respectively, of the affine expression for \(c_e^{(t)}\) on either side of \(a\), using the same side throughout the interval. The following derivative identity remains valid with either choice:
\[
\boldsymbol p'(t)
=D(t)\boldsymbol x'(t)
+D'(t)\boldsymbol x(t)
+\boldsymbol\gamma'(t).
\]
If the derivative calculation gives \(\lambda_i'(t)\le0\) on every open subinterval, then continuity at the finitely many subdivision points implies that \(\lambda_i\) is nonincreasing on \([0,1]\). In particular, \(\lambda_i(1)\le\lambda_i(0)\).}
%A persistent cost kink needs a separate observation. If a resource load reaches a grid point only at an isolated time, that time is already a subdivision point. If it stays at the grid point on an interval,  then $x_e'=0$ there. Both adjacent affine pieces have the same value at the grid point for every $t$, so their partial time derivatives there also agree. Either adjacent positive slope therefore gives the same identity $p_e'=D_{ee}x_e'+h_e=h_e$. This justifies the derivative calculation even on such intervals. The local inequalities make each population cost nonincreasing on every open interval. Continuity joins the inequalities at the finitely many subdivision points, yielding the comparison between $t=0$ and $t=1$.

\subsection{Order-preserving approximation}\label{apx:order}
\ma{We now use approximation to extend the equilibrium-cost comparison to arbitrary continuous nondecreasing resource-cost functions. }
%We finally remove strict increase, piecewise affinity, and positivity.  
Let $c$ and $\bar c$ be arbitrary continuous nondecreasing resource costs satisfying
$\bar c_e(s)\le c_e(s)$ for all $e\in E$ and $s\ge0$, and let the improved demand vector $\bar{\boldsymbol{d}}$ be coordiantewise no larger
than the original demand $\boldsymbol{d}$.  Let positive integer $M$ be chosen such that
$M>\sum_{i\in I} d_i$, which guarantees that every feasible resource load along the linear demand
interpolation lies in $[0,M]$.

\ma{To approximate the cost functions in the load variable \(s\), we choose a sequence of partitions of \([0,M]\) whose maximum subinterval length tends to zero. For each partition and each resource \(e\), we construct piecewise-affine interpolants of \(s\mapsto c_e(s)\) and \(s\mapsto\bar c_e(s)\). On each subinterval \([a,b]\) of the partition, the value of the linear interpolant of \(c_e\) at any load in  \([a,b]\) is a convex combination of \(c_e(a)\) and \(c_e(b)\). The same holds for \(\bar c_e\).}
%Use common grids on $[0,M]$ whose maximum interval length tends to zero, and interpolate both $c$ and $\bar c$ linearly. On each grid interval, an interpolated value is a convex combination of its endpoint values.
The inequalities $\bar c_e(s)\le c_e(s)$ therefore preserve the order
between the interpolants. \ma{Since \(c_e\) and \(\bar c_e\) are nondecreasing, their linear interpolants have nonnegative slopes on every subinterval of the partition and are therefore nondecreasing on \([0,M]\).}
%monotonicity of the endpoint values preserves nondecreasingness. 
By continuity on the compact interval, we obtain uniform
convergence of these interpolants to the original functions.

\ma{We index the partitions by \(n=1,2,\ldots\), and choose positive numbers \(\varepsilon_n\to0^+\) as \(n\to\infty\). For the \(n\)th partition and each resource \(e\), we increase the interpolated values of \(c_e\) and \(\bar c_e\) at each load \(s\in[0,M]\) by \(\varepsilon_n(1+s)\). The resulting functions have strictly positive values and slopes, while preserving the order between the two interpolants.}
%Add the same term $\varepsilon_n(1+s)$ to both interpolants, with $\varepsilon_n\downarrow0$. This gives positive cost values and slopes without changing their order. 
The added term  \(\varepsilon_n(1+s)\)  tends uniformly to
zero on $[0,M]$. \ma{For each \(n\) and each \(e\in E\), we denote the resulting piecewise-affine approximations of \(c_e\) and \(\bar c_e\) on \([0,M]\) by \(c_e^{(n)}\) and \(\bar c_e^{(n)}\), respectively. Extend \(c_e^{(n)}\) and \(\bar c_e^{(n)}\) from their respective values at \(M\) using a common positive slope. The inequality \(\bar c_e^{(n)}(M)\le c_e^{(n)}(M)\) then implies \(\bar c_e^{(n)}(s)\le c_e^{(n)}(s)\) for every \(s\ge M\).}
%Extend the two functions beyond $M$ with the same positive slope; their endpoint order then persists on $[M,\infty)$.
Because all feasible loads are smaller than $M$, this extension does not affect any comparison for
equilibria.

\ma{For each \(n\), applying the argument in \Cref{apx:piecewise} to the resource-cost functions \(c^{(n)}\) and \(\bar c^{(n)}\), we have \(\lambda_i(\bar c^{(n)},\bar{\boldsymbol d})
\le
\lambda_i(c^{(n)},\boldsymbol d)\) for every \(i\in I\). }
%For each approximating pair, the finite-branch argument gives $\lambda_i(\bar c^{\,n},\bar d)\le\lambda_i(c^{(n)},d)$ for every population. 
By \cref{lem:cost-continuity}, \ma{as \(n\to\infty\), we have \(\lambda_i(c^{(n)},\boldsymbol d)\to\lambda_i(c,\boldsymbol d)\) and \(\lambda_i(\bar c^{(n)},\bar{\boldsymbol d})
\to\lambda_i(\bar c,\bar{\boldsymbol d})\) for all $i\in I$. It follows that}
%the costs on both sides converge to their respective equilibrium costs for the original pair. Passing to the limit yields
$\lambda_i(\bar{c},\bar{\boldsymbol d})\le\lambda_i(c,\boldsymbol d)$ for all $i\in I$.
This completes the extension of
\cref{prop:single-analytic,prop:multi-analytic} to the continuous
nondecreasing cost functions, as stated in \cref{sec:model}. 
%, including flat and zero costs, nonunique loads, and changes of complete menus.

\section{Supplementary matroid comparisons}\label{app:matroid-comparisons}
\subsection{The two \ma{forbidden minors}%unlabelled obstructions
}
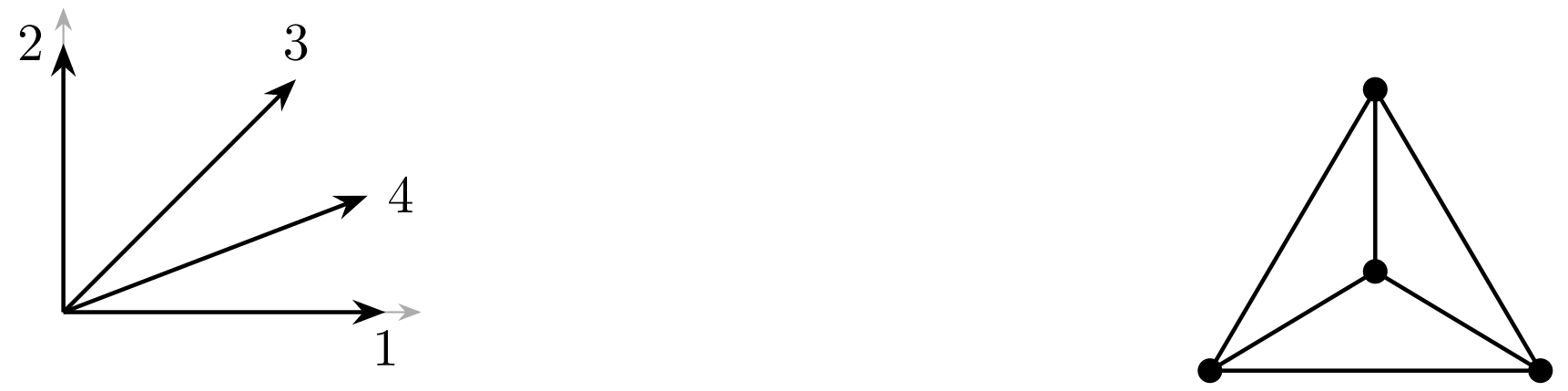
\begin{figure}[htbp]
\centering
\begin{subfigure}{.46\textwidth}\centering
\begin{tikzpicture}[scale=1.3]
\draw[gray!65,-{Stealth}](0,0)--(2,0);
\draw[gray!65,-{Stealth}](0,0)--(0,1.7);
\draw[resource,-{Stealth}](0,0)--(1.8,0)node[below]{$1$};
\draw[resource,-{Stealth}](0,0)--(0,1.5)node[left]{$2$};
\draw[resource,-{Stealth}](0,0)--(1.3,1.3)node[above]{$3$};
\draw[resource,-{Stealth}](0,0)--(1.7,.65)node[right]{$4$};
\end{tikzpicture}
\caption{$U_{2,4}$: four distinct directions in rank two.}
\end{subfigure}\hfill
\begin{subfigure}{.46\textwidth}\centering
\begin{tikzpicture}[scale=1.2]
\coordinate(A)at(0,0);\coordinate(B)at(2,0);
\coordinate(C)at(1,1.7);\coordinate(D)at(1,.6);
\draw[resource](A)--(B)--(C)--cycle;
\draw[resource](D)--(A)(D)--(B)(D)--(C);
\foreach \v in {A,B,C,D}{\node[vertex]at(\v){};}
\end{tikzpicture}
\caption{$M(K_4)$: the six edge dependencies of $K_4$.}
\end{subfigure}
\caption{The two excluded minors in \Cref{thm:single} for immunity of single population. The augmented incidence column matroid of an obstructed
 menu contains at least one of them as a minor. \ma{Revealing such a minor may require deletion,
contraction, and row operations.} %The minor need not occur as a literal submatrix of the original incidence matrix: deletion, contraction, and row operations may be needed to reveal it.
}
\label{fig:obstructions}
\end{figure}

The excluded minors in \cref{thm:single} have different representability properties:
\begin{itemize}
\item $U_{2,4}$, the rank-two uniform matroid on four elements. Every pair
is independent and every triple is a circuit. It is self-dual, so its
cocircuits also have size three. The matroid therefore has neither a parallel pair nor a
series pair. It has real representation 
$\left(\begin{smallmatrix}1&0&1&1\\0&1&1&2\end{smallmatrix}\right)$.
A matroid is \emph{binary} if it has a column representation over
the two-element field $\mathbb F_2$. This matroid is not binary, because
a two-dimensional vector space over $\mathbb F_2$ has
only three nonzero directions.
\item $M(K_4)$, the cycle matroid of the complete graph on four vertices $K_4$.
Its rank is three and its six elements are edges of $K_4$. Independent sets
are forests and circuits are cycles of $K_4$. The graph is simple and every
nontrivial edge cut has at least three edges, so this matroid also has no
parallel or series pair. It is graphic and binary, yet not series-parallel.
\end{itemize}
Both matroids are connected irreducible obstructions, illustrated in
\cref{fig:obstructions}. The presence of either matroid suffices to violate the condition on the right-hand side of
\eqref{eq:sp-excluded-minors}; \ma{Violating the condition does not require the presence of both minors.} %the condition does not require exclusion of both.
Excluding only $U_{2,4}$ still permits $M(K_4)$, while excluding only
$M(K_4)$ still permits $U_{2,4}$.

\subsection{Strategy bases and incidence columns}\label{apx:differentmatriod}
If the strategies in $\HH$ are the bases of a matroid $M_0$, then $A_{\HH}$ may be
called a \emph{base-element incidence matrix}, whose rows enumerate bases. \ma{The column matroid \(M(A_{\mathcal H})\) need not coincide with \(M_0\).}
%This need not mean that its column matroid equals $M_0$. 
For example, taking  $M_0=U_{2,4}$, its six bases give
\begin{equation}\label{eq:base-incidence-distinction}
 A_{\HH}=\begin{pmatrix}
 1&1&0&0\\1&0&1&0\\1&0&0&1\\
 0&1&1&0\\0&1&0&1\\0&0&1&1
 \end{pmatrix},\qquad
 M(A_{\HH})=U_{4,4},\qquad M([\one\mid A_{\HH}])=U_{4,5}.
\end{equation}
The strategy matroid $M_0$ has rank two, but the four resource
columns of $A_{\HH}$ are linearly independent over $\R$. Hence its
column matroid is $U_{4,4}$, not $M_0$. \ma{In \(B_{\mathcal H}=[\boldsymbol 1\mid A_{\mathcal H}]\), twice the all-ones column equals the sum of the four resource columns. Up to scaling, this is the only nontrivial linear relation among the five columns, and all five coefficients are nonzero.}
%Adding $g$ gives the unique dependence $2g=a+b+c+d$, with every coefficient nonzero. 
Every proper
subset of those five columns is therefore independent. Hence, the augmented
column matroid is the five-element circuit $U_{4,5}$. This matroid is
series-parallel even though the strategy matroid $U_{2,4}$ is not.

In fact, for the full base family of any positive-rank matroid $M_0$,
the augmented incidence matroid is connected and series-parallel after removing
unused resources. To see it, let $C_j$'s
be nonloop components of $M_0$ and write $r_j\triangleq r_{M_0}(C_j)>0$. \ma{After omitting loop coordinates, the affine hull of the base polytope of \(M_0\) consists of all vectors \(\boldsymbol x\) satisfying \(\sum_{e\in C_j}x_e=r_j\) for every \(j\) \citep{schrijver}. Indeed,}
%Its base polytope has affine hull $\sum_{e\in C_j}x_e=r_j$ for every $j$ \citep{schrijver}. To verify this description, 
every base has exactly $r_j$ elements
in $C_j$. %, so its incidence vector satisfies the displayed equations.
Conversely, suppose that $e$  and $f$ lie on a common circuit $C$. We extend
$C\setminus\{f\}$ to a base $B$. Adding $f$ to $B$ creates exactly on circuit, which must be
the circuit $C$. Removing $e$ therefore gives another base $B\cup\{f\}\setminus\{e\}$. \ma{Replacing \(e\) by \(f\) in the basis \(B\) changes the incidence vector by the \(\boldsymbol e_f-\boldsymbol e_e\).}
%The difference of these base vectors is $\ee_f-\ee_e$. 
Since any two distinct elements of a connected component lie in a common circuit, these differences  \ma{between incidence vectors of bases therefore span the subspace of vectors \(\boldsymbol v\) with \(v_e=0\) for all $e\notin C_j$ and \(\sum_{e\in C_j}v_e=0\).}
%generate the full zero-sum subspace on $C_j$. Thus base-vector differences generate every direction permitted by the displayed equations, proving the claimed affine hull.

The dual of the augmented incidence matroid therefore has resource columns
$\ee_j$ for $e\in C_j$ and demand column $-(r_1,\ldots,r_m)\T$. \ma{Retaining one resource column for each $C_j$ gives a matrix with resource columns}
%Deleting repeated resource columns leaves
$\ee_1,\ldots,\ee_m$ and the demand column $-(r_1,\ldots,r_m)\T$. The resource columns
are independent, and their unique dependence with the demand column
uses every coefficient $r_j>0$. Thus these $m+1$ elements form a
single circuit, obtainable from $U_{1,2}$ by series extensions.
Restoring the parallel resource copies and then dualizing preserves
series-parallel structure, proving the assertion.
Note that here we do not need any representability assumption on the strategy matroid $M_0$. %is needed.

\begin{lemma}[Minimum-weight matroid bases]\label{lem:min-weight-bases}
Let $M_0$ be a matroid on a finite set $E$ and let $\boldsymbol  q\in\R^E$ assigns each $e\in E$ a weight $q_e$. The
family of all minimum-weight bases of $M_0$ is itself the complete
base family of a matroid on $E$, possibly with additional loops.
\end{lemma}
\begin{proof}
The claim is immediate if $E=\varnothing$. Otherwise, let
$\omega_1<\cdots<\omega_p$ be the distinct resource weights. We set $A_0\triangleq\varnothing$ and
$A_j\triangleq\{e:q_e\le\omega_j\}$ for $j=1,\ldots,p$. 
For an element $e$ with $q_e=\omega_s$, the differences
$\omega_{j+1}-\omega_j$ for $j\ge s$ telescope to
$\omega_p-\omega_s$. \ma{Let \(B\) be any basis of \(M_0\). Summing this identity over \(e\in B\), and using \(|B|=r_{M_0}(E)\), we obtain}
%Summing this identity over a base gives
\[
 {\boldsymbol  q}^\top\boldsymbol  \chi^B
 =\omega_p r_{M_0}(E)
   -\sum_{j=1}^{p-1}(\omega_{j+1}-\omega_j)|B\cap A_j|.
\]
\ma{For every basis \(B\) of \(M_0\) and every \(j\), we have \(|B\cap A_j|\le r_{M_0}(A_j)\). Processing the elements in nondecreasing order of weight, the matroid greedy algorithm constructs a basis for which equality holds for every \(j\). Since \(\omega_{j+1}-\omega_j>0\) for every \(j<p\), a basis $B$ satisfying \(|B\cap A_j|=r_{M_0}(A_j)\) for every \(j\) minimizes the total weight \(\boldsymbol q^\top\boldsymbol\chi^B\).}
%Every base has $|B\cap A_j|\le r_{M_0}(A_j)$, and the matroid greedy algorithm attains all these bounds simultaneously. Each count is subtracted with a positive coefficient in the displayed formula. Its simultaneous upper bounds therefore give the minimum possible weight. If any bound with $j<p$ is not attained, the weight is strictly larger; the bound for $j=p$ holds for every base. Consequently $B$ is minimum-weight exactly when $|B\cap A_j|=r_{M_0}(A_j)$ for all $j$.

\ma{The minimum-weight bases of \(M_0\) are exactly the bases of the direct sum $M_0^q\triangleq\bigoplus_{j=1}^p (M_0|_{A_j})/A_{j-1}$. For each \(j\), the summand \((M_0|A_j)/A_{j-1}\) is obtained by restricting \(M_0\) to \(A_j\) and then contracting \(A_{j-1}\). Its ground set \(A_j\setminus A_{j-1}\) consists of the elements with weight \(\omega_j\). For each \(j=1,\ldots,p\), let us choose a basis \(B_j\) of \((M_0|A_j)/A_{j-1}\). Suppose inductively that \(B_1\cup\cdots\cup B_{j-1}\) is a basis of \(M_0|A_{j-1}\). The basis-extension property of contraction then implies that \(B_1\cup\cdots\cup B_j\) is a basis of \(M_0|A_j\). By induction, \(B=\bigcup_{j=1}^{p}B_j\) is a basis of \(M_0\) satisfying \(|B\cap A_j|=r_{M_0}(A_j)\) for every \(j=1,\ldots,p\). Conversely, let \(B\) be a basis of \(M_0\) satisfying \(|B\cap A_j|=r_{M_0}(A_j)\) for every \(j=1,\ldots,p\). Then \(B\cap A_j\) is a basis of \(M_0|A_j\) for every \(j\). Consequently, \(B_j=B\cap(A_j\setminus A_{j-1})\) is a basis of \((M_0|A_j)/A_{j-1}\).} %These are precisely the bases of the direct sum of the layer minors. The  summands have disjoint ground sets $A_j\setminus A_{j-1}$. Suppose the choices in the first $j-1$ layers form a basis on $A_{j-1}$. Choosing a basis of the contracted $j$-th layer completes it to a basis on $A_j$. Inducting over the layers gives a base with all the prefix-rank equalities above. Conversely, a base satisfying those equalities restricts to a basis on every prefix $A_j$; its choices in layer $j$ must therefore be a basis of that contracted layer.
This proves both directions of the direct-sum description. The greedy
and contraction facts are standard matroid properties
\citep{oxley,schrijver}.
\end{proof}

\subsection{\ma{Series-parallel matrices and incidence column matroids}%Literal series-parallel matrices
}
\ma{The series-parallel property of the augmented incidence column matroid \(N_{\mathcal H}\) should be distinguished from Walter’s notion of a \emph{series-parallel matrix} \cite{walter}.}
%This construction must also be distinguished from the \emph{series-parallel matrix} of \citet{walter}. 
For a binary matrix $Q$, \ma{Walter’s definition permits deletion of a zero row or column, a row or column containing exactly one $1$, or a duplicate row or column.}
%that literal matrix property permits reductions by deleting a zero row or column, a row or column with a single one, or a duplicate row or column.
In the associated fundamental-matrix convention, the binary matroid associated with $Q$ is the column matroid of  $[\,\mathbb I\mid Q\,]$ over $\mathbb F_2$, where $\mathbb I$ denotes the identity matrix. \ma{The rows of \(Q\) are indexed by the elements of the basis corresponding to the identity matrix \(\mathbb I\). By contrast, the rows of \(A_{\mathcal H}\) are indexed by the bases of \(M_0\). In the matrix \(A_{\mathcal H}\) displayed in \eqref{eq:base-incidence-distinction}, every row contains exactly two \(1\)s and every column contains exactly three \(1\)s. No two rows are identical, and no two columns are identical. Thus none of the deletions permitted by Walter’s definition can be applied to \(A_{\mathcal H}\). However, the augmented incidence column matroid \(N_{\mathcal H}\) is isomorphic to \(U_{4,5}\) and is series-parallel. Consequently, Walter’s algorithm therefore cannot be used directly on \(A_{\mathcal H}\) to determine whether \(N_{\mathcal H}\) is series-parallel, which servers as a criterion of  immunity.}
%The row labels index a chosen basis, not all bases of a matroid. Our object is instead $M_{\R}([\one\mid A_{\HH}])$. In \eqref{eq:base-incidence-distinction}, every row has two ones and every column has three, with no duplicates: the literal incidence matrix has no first series-parallel matrix reduction, although its augmented \emph{column matroid} is series-parallel. Consequently Walter's matrix recognition algorithm cannot simply be applied to the raw strategy table as a test for our theorem.

\section{Numerical realizations of the two obstruction mechanisms}
\label{app:numerical-witnesses}

This section presents \ma{explicit strong Braess paradoxes for the strategy families in \cref{ex:wheatstone-first,ex:U24-first}.}
%concrete instances of the two algebraic obstructions. 
Each instance is checked by listing feasible strategy
flows and verifying that all strategies carrying flow have minimum
cost. We then vary the slopes of the resource costs to exhibit families of distinct
paradoxes on the same strategy systems. The examples illustrate the
general witness construction of \cref{prop:minor-witness}. %; they do not restrict the range of systems to which that construction applies.

\subsection{The $M(K_4)$ obstruction}
We go back to the Wheatstone menu $\{ab,cd,ade\}$ in
\cref{ex:wheatstone-first}. It is the path family of the network in
\cref{fig:wheatstone}. Given the unit population demand, we set
\[
 c_a(s)=c_d(s)=s\qquad \text{and}\qquad c_b(s)=c_c(s)=1.
\]
Originally, under $c_e(s)=1$, half of the equilibrium flow uses $ab$, and half uses $cd$,
yielding the equilibrium population cost of $3/2$. After reducing $c_e$ to zero, we see that sending all
flow along $ade$ is now an equilibrium, since all three paths have cost
two at the resulting load. The common cost has increased from $3/2$
to $2$. The originally unappealing cross-arc $(u,v)$ becomes attractive\ma{, increasing the loads on \(a\) and \(d\) from \(1/2\) to \(1\).} %and loads both congestible resources. 
The negative $e$-coordinate of the basic point in
\eqref{eq:wheatstone-negative} identifies a resource whose improvement
can produce the paradox. Not that the actual equilibrium loads remain
nonnegative; it is the signed affine basic point that reveals the
structural obstruction. This familiar example uses constant and
zero costs for simplicity; \cref{prop:single-analytic} gives strictly
positive affine witnesses for the same structural failure.

\begin{figure}[htbp]
\centering
\begin{tikzpicture}[scale=1.05]
\node[vertex,label=left:$s$](S)at(0,0){};
\node[vertex,label=above:$u$](U)at(2,1.2){};
\node[vertex,label=below:$v$](V)at(2,-1.2){};
\node[vertex,label=right:$t$](T)at(4,0){};
\draw[flow](S)--node[above left]{$a:\ x$}(U);
\draw[flow](U)--node[above right]{$b:\ 1$}(T);
\draw[flow](S)--node[below left]{$c:\ 1$}(V);
\draw[flow](V)--node[below right]{$d:\ x$}(T);
\draw[bad,line width=1pt,-{Stealth[length=2mm]}](U)--node[right]{$e:\ 1\rightarrow0$}(V);
\node[anchor=west,align=left]at(5,.5){Before: $f_{ab}=f_{cd}=1/2$, $\lambda_1(c,1)=3/2$.};
\node[anchor=west,align=left]at(5,-.3){After: $f_{ade}=1$, $\lambda_1(\bar c,1)=2$.};
\end{tikzpicture}
\caption{The Wheatstone routing example. The physical network shown here
and the $K_4$ representation of the associated augmented incidence column matroid are
different graphs. \Cref{thm:single} concerns the latter's column dependencies.}
\label{fig:wheatstone}
\end{figure}

\subsection{The $U_{2,4}$ obstruction}
For the family displayed in \eqref{eq:four-strategies}, \cref{ex:U24-first}
establishes positive exposure, the affine equations
\eqref{eq:four-affine}, and provides the negative basic point
$(2,0,0,0,-1)$. The following unit-demand witness realizes that
obstruction with positive slopes and intercepts for every resource cost:
\begin{equation}\label{eq:four-costs}c_a(s)=4s+37,\;\;\;c_b(s)=24s+7,\;\;\;c_c(s)=24s+7,\;\;\; c_d(s)=24s+1,\;\;\;c_e(s)=4s+16.\end{equation}
%\begin{equation}\label{eq:four-costs}
%\begin{aligned}
 %c_a(s)&=4s+37,&c_b(s)&=24s+7,&c_c(s)&=24s+7,\\
% c_d(s)&=24s+1,&c_e(s)&=4s+16.
%\end{aligned}
%\end{equation}
Only the intercept of $c_e$ is lowered, from $16$ to $14$. The explicit
equilibria are listed in the following table, with strategy order $(ad,bce,bde,cde)$
and resource order $(a,b,c,d,e)$:
\begin{center}
\small
\begin{tabular}{@{}lll@{}}
\toprule
 & Before & After\\
\midrule
Strategy flow & $(1/4,1/4,1/4,1/4)$ & $(1/8,5/24,1/3,1/3)$\\
Resource costs & $(38,19,19,19,19)$ & $(75/2,20,20,20,35/2)$\\
Population cost & $57$ & $115/2$\\
\bottomrule
\end{tabular}
\end{center}
\ma{At the aggregate loads induced by the original flow allocation, strategy \(ad\) has cost \(38+19=57\), while each of \(bce\), \(bde\), and \(cde\) has cost \(19+19+19=57\). After the intercept reduction, the corresponding flow allocation gives cost \(75/2+20=115/2\) for \(ad\) and \(20+20+35/2=115/2\) for each of the other three strategies. Each allocation is feasible for unit demand, since its entries are positive and sum to one. All four strategies have equal cost in each instance, so both allocations are Wardrop equilibria for their respective instances.}
%To check the table, sum resource costs along each of the four strategies. Before the change, $38+19=57$ on $ad$ and $19+19+19=57$ on each other strategy. Afterwards, $75/2+20=115/2$ on $ad$ and $20+20+35/2=115/2$ on each other strategy. The displayed flows are positive and sum to one, so both rows describe Wardrop equilibria.
Lowering the intercept
therefore increases the population's equilibrium cost by $1/2$. Unlike the
Wheatstone example, this witness need not come from a path system.

\subsection{Parameter families rather than a unique paradox}
For the Wheatstone coordinates $(a,b,c,d,e)$, an affine representation is given by
\[
 H=\begin{pmatrix}1&0&1&0&0\\0&1&0&1&0\\-1&1&0&0&1\end{pmatrix}
 \qquad \text{and}\qquad\boldsymbol b=(1,1,0)\T.
\]
\ma{Let \(\alpha,\beta,\eta>0\), and define the diagonal matrix of resource-cost slopes by} %Choose positive slopes
$D=\operatorname{diag}(\alpha,\beta,\beta,\alpha,\eta)$. Recalling the relation \(W=D^{-1}\) introduced at the beginning of \Cref{sec:basic} and the definition of \(k(W,\boldsymbol b)\) in \Cref{lem:cb}, we write \(\boldsymbol k\triangleq k(W,\boldsymbol b)\). Then \(H\boldsymbol k=\boldsymbol b\) and \(D\boldsymbol k\in\operatorname{row}(H)\).
%Here $W=D^{-1}$ and $k=k(W,\boldsymbol b)$ satisfies $Hk=b$ and $Dk\in\operatorname{row}(H)$. 
Equivalently, \(\boldsymbol k\) is the unique minimizer of \(\tfrac12\boldsymbol z^\top D\boldsymbol z\) subject to \(H\boldsymbol z=\boldsymbol b\). \ma{Indeed, for any \(\boldsymbol z\) satisfying \(H\boldsymbol z=\boldsymbol b\), we have \(H(\boldsymbol z-\boldsymbol k)=\boldsymbol 0\). Since \(D\boldsymbol k\perp\ker H\), it follows that \(\boldsymbol k^\top D(\boldsymbol z-k)=0\). Hence $\frac12\boldsymbol z^\top D\boldsymbol z-\frac12\boldsymbol k^\top D\boldsymbol k
=\frac12(\boldsymbol z-\boldsymbol k)^\top D(\boldsymbol z-\boldsymbol k)\ge0$. Since \(D\) is positive definite, equality holds only when \(\boldsymbol z=k\). Thus \(\boldsymbol k\) is the unique minimizer.} %: the displayed response satisfies the constraint, and $Dk$ is orthogonal to every feasible direction.
By symmetry and uniqueness of the minimizer, \(k_a=k_d\) and \(k_b=k_c\). These common values are written as \(u\) and \(v\), respectively. The equations $H\boldsymbol k=\boldsymbol b$ imply $u+v=1$ and $k_e=u-v$. \ma{Let \(\boldsymbol w=(1,-1,-1,1,2)^\top\). Since \(H\boldsymbol w=\boldsymbol0\) and \(D\boldsymbol k\in\operatorname{row}(H)\), we have $0=\boldsymbol w^\top D\boldsymbol k=\alpha k_a-\beta k_b-\beta k_c+\alpha k_d+2\eta k_e=2(\alpha u-\beta v+\eta k_e)$, giving \(\alpha u-\beta v+\eta k_e=0\). Eliminating \(u\) and \(v\) from \(u+v=1\), \(k_e=u-v\), and \(\alpha u-\beta v+\eta k_e=0\), we obtain}
%Orthogonality to the kernel direction $(1,-1,-1,1,2)$ gives $\alpha u-\beta v+\eta k_e=0$. Solving these equations yields
\begin{equation*}%\label{eq:wheat-parameter}
 k_e=\frac{\beta-\alpha}{\alpha+\beta+2\eta}.
\end{equation*}
If $\alpha>\beta>0$ and $\eta>0$, then $k_e<0$.
By \eqref{eq:single-cost-increase}, a sufficiently small decrease in
resource $e$'s intercept increases the equilibrium population cost. \ma{To construct affine resource-cost functions with strictly positive intercepts, we} %For a fully explicit positive-intercept initialization 
take
$\boldsymbol x^0\triangleq(2/3,1/3,1/3,2/3,1/3)^\top$ and
$\boldsymbol q\triangleq Q(1,2,2,1,1)^\top$ with $Q>\max\{\alpha,\beta,\eta\}$, and set
$\boldsymbol\gamma\triangleq q-D\boldsymbol x^0$. \ma{At the load vector \(\boldsymbol x^0\), all three strategies have cost \(3Q\), and every cost intercept \(\gamma_{e'}\), $e'\in E$, is strictly positive.} 
%All three strategies cost $3Q$ and every intercept is positive. 
For sufficiently small $\delta>0$, lowering $\gamma_e$ by $\delta$ raises
the common equilibrium cost by
$\delta(\alpha-\beta)/(\alpha+\beta+2\eta)$.

For the strategy family in \Cref{ex:U24-first}, we use %For \eqref{eq:four-strategies}, use
\[
 H=\begin{pmatrix}1&0&0&0&1\\1&1&1&1&0\end{pmatrix},
 \qquad b=(1,2)\T,
 \qquad D=\operatorname{diag}(\alpha,\beta,\beta,\beta,\eta).
\]
\ma{Permuting the coordinates indexed by \(b,c,d\) leaves both the objective and the constraints unchanged. By uniqueness of the minimizer, we have \(k_b=k_c=k_d\). Let this common value be denoted by \(v\).}
%The three equally weighted coordinates satisfy $k_b=k_c=k_d=v$. Writing
With $u\triangleq k_a$, we deduce from the equations $H\boldsymbol k=\boldsymbol b$ that
$u+k_e=1$ and $u+3v=2$. Orthogonality of $D\boldsymbol k$ to the
kernel direction $(1,-1/3,-1/3,-1/3,-1)^T$ implies
$\alpha u-\beta v-\eta k_e=0$. Eliminating $u$ and $v$, we derive
\begin{equation*}%\label{eq:u24-parameter}
 k_e=\frac{3\alpha-\beta}{3\alpha+\beta+3\eta}.
\end{equation*}
For any $\beta>3\alpha>0$ and $\eta>0$, we take
$\boldsymbol x^0\triangleq(1/4,1/2,1/2,3/4,3/4)^\top$, $\boldsymbol q\triangleq Q(2,1,1,1,1)^\top$ with number
$Q>\max\{\alpha/8,3\beta/4,3\eta/4\}$, and $\boldsymbol\gamma\triangleq\boldsymbol q-D\boldsymbol x^0$. \ma{At the load vector \(\boldsymbol x^0\), all four strategies have cost \(3Q\), and every cost intercept is strictly positive.}
%The four strategies tie at $3Q$ with positive intercepts.
A small intercept reduction $\delta$ on $e$ increases their equilibrium
cost by $\delta(\beta-3\alpha)/(3\alpha+\beta+3\eta)$. \ma{Taking \((\alpha,\beta,\eta,Q)=(4,24,4,19)\) gives the cost functions in \eqref{eq:four-costs}. In this example, the intercept of \(c_e\) is reduced by \(\delta=2\). The table shows that all four strategies carry positive equilibrium flow before and after the reduction, so the complete minimizing menu remains \(\{ad,bce,bde,cde\}\).}
%The exact numerical instance above corresponds to $(\alpha,\beta,\eta,Q)=(4,24,4,19)$; its larger perturbation $\delta=2$ remains within the same complete menu, as its displayed positive flows verify directly.

The inequalities on $\alpha,\beta,\eta$ specify when these
particular affine instances yield a paradox after improving $e$.
They do not characterize immunity of the strategy systems. The
forbidden-minor criterion concerns all admissible costs and demands. A forbidden minor guarantees that some choice of parameters produces a paradox,
while other choices on the same system can still be harmless.

\section{Detailed proofs of auxiliary lemmas}\label{app:auxiliary-proofs}
This section provides detailed proofs of auxiliary results stated and used in the main text.  
%The main text states each auxiliary result where it is used and explains its role in the proof. This appendix supplies the longer elementary derivations without interrupting the route to the immunity criteria.

\subsection{Strict separation modulo a subspace}\label{app:separation-proof}
\ma{We prove the separation result used in the trade-closure argument by orthogonal projection. When the origin lies outside the convex hull of the projected vectors, strict separation gives the required linear functional. }
%The trade-closure argument needs a separation statement that ignores within-menu exchange directions. We prove it by projecting away those directions and examining whether the remaining vectors have a convex combination equal to zero.

\begin{proof}[Proof of \Cref{lem:strict-alternative}] \ma{The proof proceeds in four steps. In the trade-closure application, \(L\) is spanned by differences between incidence vectors of strategies in the same menu. Every weight vector \(\boldsymbol q\in L^\perp\) has zero inner product with every vector in \(L\).}

\emph{Step 1:} %remove the directions that the weights must ignore.
\ma{We project the given vectors orthogonally onto \(L^\perp\), removing their components in \(L\) while preserving their inner products with every weight vector in \(L^\perp\).}
Let $\pi$ be orthogonal projection onto $L^\perp$. For $j=1,\ldots,r$, we write $\boldsymbol y_j\triangleq\pi(\boldsymbol v_j)$.
Thus $\boldsymbol v_j-\boldsymbol y_j\in L$ and ${\boldsymbol  q}^\top \boldsymbol v_j={\boldsymbol  q}^\top\boldsymbol y_j$ for every $\boldsymbol  q\in L^\perp$. \ma{Let \(\eta_1,\ldots,\eta_r\) be nonnegative coefficients, not all zero. Since \(\pi\) is linear, \(\ker\pi=L\), and \(\boldsymbol y_j=\pi(\boldsymbol v_j)\), we have
\[
\sum_{j=1}^{r}\eta_j\boldsymbol v_j\in L
\quad\Longleftrightarrow\quad
\pi\!\left(\sum_{j=1}^{r}\eta_j\boldsymbol v_j\right)=\boldsymbol0
\quad\Longleftrightarrow\quad
\sum_{j=1}^{r}\eta_j\boldsymbol y_j=\boldsymbol0.
\] Let}  \(K=\operatorname{conv}\{\boldsymbol y_1,\ldots,\boldsymbol y_r\}\), a nonempty compact convex subset of \(L^\perp\). \ma{By definition, \(\boldsymbol0\in K\) if and only if a convex combination of the projected vectors \(\boldsymbol y_1,\ldots,\boldsymbol y_r\) equals \(\boldsymbol0\).}
%Since $\ker\pi=L$, we also have \[ \sum_j\eta_jv_j\in L  \quad\Longleftrightarrow\quad \sum_j\eta_jy_j=0. \]
% Thus $0\in K$ asks whether a normalized nonnegative combination of the projected vectors cancels completely. 
The condition is equivalent to
alternative (ii) of the lemma, because any nonzero nonnegative coefficients can be
divided by their positive sum.

\emph{Step 2:} We derive (ii) or (i) by distinguishing between two cases depending whether $\boldsymbol0$ belongs to $K$ or not.  If $\boldsymbol0\in K$, then we may write $\boldsymbol0=\sum_{j=1}^r\theta_j\boldsymbol y_j$ with $\theta_j\ge0$ for all $j=1,\ldots,r$ and
$\sum_{j=1}^r\theta_j=1$. Taking $\boldsymbol\eta\triangleq\boldsymbol\theta$, we derive (ii), \ma{where \(\sum_{j=1}^r\theta_j=1\) ensures that the coefficients \(\theta_j\), $j=1,\ldots,r$, are not all zero.} %; the normalization ensures that the coefficients are not all zero.
It remains to consider the case where $\boldsymbol0\not\in K$.
By compactness, we can take a point $\boldsymbol p\in K$ of minimum Euclidean norm. The case hypotheses implies $\boldsymbol p\ne\boldsymbol 0$. For any $\boldsymbol y\in K$ and any $t\in[0,1]$, we have $\boldsymbol p+t(\boldsymbol y-\boldsymbol p)\in K$. Expanding its squared norm, we obtain
\[
 \|\boldsymbol p+t(\boldsymbol y-\boldsymbol p)\|^2
 =\|\boldsymbol p\|^2+2t\,\boldsymbol p\T(\boldsymbol y-\boldsymbol p)+t^2\|\boldsymbol y-\boldsymbol p\|^2.
\] If $\boldsymbol p\T(\boldsymbol y-\boldsymbol p)<0$, a sufficiently small positive $t$ decreases the
norm, contrary to minimality. Hence $\boldsymbol p\T \boldsymbol y\ge\|\boldsymbol p\|^2>0$ for every
$\boldsymbol y\in K$. Setting $\boldsymbol q\triangleq \boldsymbol p\in L^\perp$, we see that
${\boldsymbol  q}^\top \boldsymbol v_j={\boldsymbol  q}^\top \boldsymbol y_j\ge\|\boldsymbol p\|^2>0$ for every $j=1,\ldots,r$, proving (i).
%This gives a concrete proof of the strict separation used here.

\emph{Step 3:} We prove that the two alternatives are mutually exclusive.
If both (i) and (ii) held, then orthogonality and positivity would give $ 0={\boldsymbol  q}^\top(\sum_{j=1}^r\eta_j\boldsymbol v_j)
   =\sum_{j=1}^r\eta_j\boldsymbol q\T \boldsymbol v_j>0$,  a contradiction. The cases $\boldsymbol 0\in K$ and $\boldsymbol 0\notin K$ exhaust all
possibilities, proving that exactly one alternative holds.

\emph{Step 4: } \ma{We obtain rational coefficients by solving a linear system. Assume that every \(\boldsymbol v_j\) has rational coordinates and that \(L\) admits a basis of rational vectors. Choose such a basis and let \(C\) be the matrix whose columns are the chosen basis vectors.}
Rational data means that each $v_j$ has rational coordinates and $L$
has a rational basis, say the columns of $C$. \ma{Suppose that nonnegative real coefficients \(\eta_1,\ldots,\eta_r\) satisfying alternative (ii). Dividing these coefficients by their positive sum, we may assume that \(\sum_{j=1}^{r}\eta_j=1\).}
%Normalize a solution of (b) to have $\sum_j\eta_j=1$. 
Since \(\sum_j\eta_j\boldsymbol v_j\in L\) and the columns of \(C\) form a basis of \(L\), there exists a coefficient vector \(\boldsymbol\gamma\) such that $\sum_{j=1}^{r}\eta_j\boldsymbol v_j=C\boldsymbol\gamma$. \ma{We now treat both \(\boldsymbol\eta=(\eta_1,\ldots,\eta_r)\) and \(\boldsymbol\gamma\) as unknowns and seek a rational solution of the following system
\[
\begin{cases}
\displaystyle\sum_{j=1}^{r}\eta_j\boldsymbol v_j=C\boldsymbol\gamma,\\[2pt]
\displaystyle\sum_{j=1}^{r}\eta_j=1,\\
\eta_j\ge0\quad\text{for all }j=1,\ldots,r.
\end{cases}
\]
 When seeking a rational solution of this system, we impose \(\eta_j=0\) at every index where the chosen real solution has a zero coefficient. Gaussian elimination gives a parametrization of all solutions to the equality constraints. In this parametrization, every component of \(\boldsymbol\eta\) and \(\boldsymbol\gamma\) is an affine function of free real parameters with rational coefficients.}
%Fix to zero the coordinates of $\eta$ that vanish in this solution. Gaussian elimination expresses the remaining rational equation system as a rational particular solution plus rational nullspace directions with free real parameters. 
We approximate these parameters by rational
ones sufficiently closely to preserve every remaining strict inequality
$\eta_j>0$. This leads to rational $\boldsymbol\eta$ and $\boldsymbol\gamma$, as required.
This is the finite-dimensional Gordan--Stiemke alternative in the form
needed by the trade argument.
\end{proof}

\subsection{Positive series-parallel elimination}\label{app:elimination-proof}
This subsection provides the reduction argument that turns series-parallel
column structure into nonnegative \ma{affine basic} %minimal-support 
points, completing
the establishment of the bridge between affine geometry and matroid structure.

\begin{proof}[Proof of \Cref{lem:bridge}(iii)$\Rightarrow$(i)]
%\emph{A direct elimination proof of (iii)$\Rightarrow$(i).}
This implication closes the equivalence cycle; recall that the chamber argument in the main text
has proved (i)$\Rightarrow$(ii), but not its converse. 
%The elimination proof also makes the sign role of positivity transparent. 
We prove by induction on the number $n=|E|$ of resources, retaining the
demand element $g$ as permitted by the reduction rule in
\cref{sec:matroid-primer}. The inductive reduction must preserve three properties. \ma{Each reduced affine space must contain a strictly positive point and satisfy an affine equation with strictly positive coefficients and a strictly positive right-hand side. We must also show that if every affine basic point of the reduced space is coordinatewise nonnegative, then every affine basic point of the original space is coordinatewise nonnegative. }:
%a strictly positive point, a positive constant equation, and a correspondence that lifts nonnegative minimal-support points to nonnegative points. 
\ma{If \(n=1\), the equation \(q_1x_1=\rho\) determines the unique point of \(U\), with coordinate \(x_1=\rho/q_1>0\). For \(n\ge2\), by (iii), we may choose a series or parallel pair $(e,f)$ in \(N(U)\). We distinguish four cases according to the type of pair and whether it contains the demand element \(g\).}
%If $n=1$, the positive equation fixes the only point at $\rho/q_1>0$. For $n\ge2$, a series or parallel pair has one of the following four forms.

\emph{Case 1: } \ma{$\{e.f\}\subseteq E$.  and \(e\) and \(f\) are parallel elements of \(N(U)\). It follows that there exists a nonzero scalar \(\kappa\) such that \(x_e=\kappa x_f\) for every \(\boldsymbol x\in U\). Evaluating this identity at a strictly positive point of \(U\) shows that \(\kappa>0\). Deleting coordinate \(e\) from each point of \(U\), we obtain the reduced affine space \(U'\). Accordingly, by substituting \(x_e=\kappa x_f\) into \(\boldsymbol q^\top\boldsymbol x=\rho\), we replace the coefficient \(q_f\) by \(q_f+\kappa q_e>0\). Every point of \(U'\) extends uniquely to a point of \(U\) by restoring \(x_e=\kappa x_f\). Since \(x_e=0\) exactly when \(x_f=0\), a point of \(U\) has inclusion-minimal support precisely when the corresponding point of \(U'\) has inclusion-minimal support.}
%If $e,f$ are parallel, $x_e=\kappa x_f$ throughout $U$ with $\kappa>0$, as witnessed by a positive point. Delete $e$, replacing $q_f$ by $q_f+\kappa q_e$. Minimal-support points correspond under deletion and restoration: $x_e=0$ exactly when $x_f=0$, and restoring $x_e=\kappa x_f$ preserves inclusions between supports. 
Thus, a basic point
in the reduced space  corresponds to a basic point in the
original space. Since $\kappa>0$, nonnegativity is preserved, and
induction applies.

\emph{Case 2:} $e\in E$ is parallel $g$. \ma{It follows that the coordinate \(x_e\) has a fixed value \(\kappa>0\) throughout \(U\). We delete coordinate \(e\) and denote the resulting affine space by \(U'\). A strictly positive point \(\boldsymbol x^+\in U\) projects to a strictly positive point of \(U'\), and every point of \(U'\) satisfies \(
\sum_{j\ne e}q_jx_j
=\rho-q_e\kappa
=\sum_{j\ne e}q_jx_j^+>0\).
Since \(n\ge2\), $\sum_{j\ne e}q_jx_j^+$ has at least one term, and every term is strictly positive.}
%Here $x_e=\kappa>0$ is constant. Delete it. The remaining coordinates inherit a positive point. Their constant equation has right-hand side $\rho-q_e\kappa>0$, since at the positive point this difference equals $\sum_{j\ne e}q_jx_j^+$ and at least one resource remains. 
Since $x_e>0$ for all $\boldsymbol x\in U$, minimality of support is determined by the other coordinates. \ma{By induction, every affine basic point of \(U'\) is coordinatewise nonnegative. Reinserting the fixed coordinate \(x_e=\kappa>0\) leaves all other coordinates unchanged. We deduce that every affine basic point of \(U\) is also coordinatewise nonnegative.}
%Restoring this fixed positive coordinate preserves the assertion.

\emph{Case 3:} $\{e,f\}\subseteq E$, and $e$ and $f$ are in series. \ma{In the matrix representation of \(N(U)\) with row space \(V=\{(t,\boldsymbol x^\top)^\top:H\boldsymbol x=t\boldsymbol b\}\), the cocircuit \(\{e,f\}\) is the support of a vector \((0,\boldsymbol v)\). Thus \(H\boldsymbol v=\boldsymbol0\), and only \(v_e\) and \(v_f\) are nonzero. For any \(\boldsymbol x\in U\), we also have \(\boldsymbol x+\boldsymbol v\in U\), implying \(\boldsymbol q^\top\boldsymbol v
=\boldsymbol q^\top(\boldsymbol x+\boldsymbol v)-\boldsymbol q^\top\boldsymbol x
=\rho-\rho=0\).
Consequently, \(q_ev_e+q_fv_f=0\). Since \(q_e,q_f>0\), the nonzero coordinates \(v_e\) and \(v_f\) have opposite signs. Replacing \(\boldsymbol v\) by \(-\boldsymbol v\) if necessary, we may assume that \(v_e>0\) and \(v_f<0\). Setting \(\beta=v_e>0\) and \(\alpha=-v_f>0\), we obtain \(\boldsymbol v=\beta\boldsymbol e_e-\alpha\boldsymbol e_f\). Suppose that linear map \(T:\mathbb R^E\to\mathbb R^{E\setminus\{e\}}\)  maps $\boldsymbol x$ to the image $T\boldsymbol x$ defined by
\[
(T\boldsymbol x)_f\triangleq\alpha x_e+\beta x_f
\qquad\text{and}\qquad
(T\boldsymbol x)_j\triangleq x_j
\quad\text{for all}j\in E\setminus\{e,f\}.
\] 
Let \(U'=T(U)\triangleq\{T\boldsymbol x:\boldsymbol x\in U\)\} denote the image of \(U\) under \(T\).  A strictly positive point \(\boldsymbol x^+\in U\) maps to a strictly positive point of \(U'\), since \(\alpha x_e^++\beta x_f^+>0\) and the other coordinates are unchanged. Since \(q_e/\alpha=q_f/\beta\), every point $\boldsymbol{x}$ of \(U'\) satisfies \(\sum_{j\notin\{e,f\}}q_jx_j+\frac{q_e}{\alpha}y=\rho\).
All coefficients and the right-hand side of this equation are strictly positive. The matroid \(N(U')\) is obtained from \(N(U)\) by contracting \(e\). Indeed, for any \(\boldsymbol x\in U\), set $\widehat{\boldsymbol x}
\triangleq\boldsymbol x-\frac{x_e}{\beta}\boldsymbol v$. 
Since \(H\boldsymbol v=\boldsymbol0\), we have \(\widehat{\boldsymbol x}\in U\). Moreover, $\widehat x_e=0$, $\beta\widehat x_f=\alpha x_e+\beta x_f$,
and all coordinates outside \(\{e,f\}\) are unchanged. Thus \(U'\) is obtained from \(U\cap\{x_e=0\}\) by deleting coordinate \(e\) and multiplying coordinate \(f\) by \(\beta\). The homogenization of \(U\cap\{x_e=0\}\) is \(V\cap\{x_e=0\}\). A matrix representing \(N(U)/e\) is obtained by taking the transposes of a basis of this subspace as rows and deleting column \(e\). Multiplying column \(f\) of this matrix by \(\beta\) gives a representation of \(N(U')\). Since nonzero column scaling preserves the represented matroid, \(N(U')=N(U)/e\).}
%A two-element cocircuit is the support of a row-space vector whose $g$-coordinate is zero. By the homogeneous equations, its resource part is a nonzero direction in $\ker H$ supported on $e,f$. The positive equation gives ${\boldsymbol  q}^\top v=0$, so the two nonzero coordinates must have opposite signs. After scaling and, if necessary, reversing the direction, write it as $v=\beta\ee_e-\alpha\ee_f$ with $\alpha,\beta>0$. Replace the two coordinates by $y=\alpha x_e+\beta x_f$. The image $U'$ has a positive point and a positive constant equation because $q_e/\alpha=q_f/\beta>0$. Its matroid is $N(U)/e$, up to scaling the remaining column. Indeed, represented column contraction restricts the row space to vectors whose $e$-coordinate is zero and then deletes that coordinate. Here this is the homogenized slice $x_e=0$. Every fiber of the map to $y$ meets that slice by movement along $v$, and on it $y=\beta x_f$. Thus the image has the contracted representation with a positive scaling at $f$. 

\ma{Taking an arbitrary affine basic point \(\boldsymbol z\in U\), we prove it is nonnegative. If both \(z_e\) and \(z_f\) were nonzero, then \(\boldsymbol z-(z_e/\beta)\boldsymbol v\in U\) would have support properly contained in that of \(\boldsymbol z\). Thus at most one of \(z_e\) and \(z_f\) is nonzero. Let \(\boldsymbol z'\triangleq T\boldsymbol z\) denote $\boldsymbol z$'s image in $U'$. Suppose that some \(\boldsymbol w'\in U'\) has support properly contained in that of \(\boldsymbol z'\). Construct \(\boldsymbol w\) by \(w_j\triangleq w'_j\) for \(j\notin\{e,f\}\) and
\[
(w_e,w_f)\triangleq
\begin{cases}
(0,w'_f/\beta),&z_e=0,\\
(w'_f/\alpha,0),&z_e\ne0\quad(\text{which enforces }z_f=0).
\end{cases}
\]
For any \(\boldsymbol x\in U\) with \(T\boldsymbol x=\boldsymbol w'\), these two constructions give \(\boldsymbol{w}=\boldsymbol x-(x_e/\beta)\boldsymbol v\) when $x_e=0$, and \(\boldsymbol{w}=\boldsymbol x+(x_f/\alpha)\boldsymbol v\) otherwise. Hence \(\boldsymbol w\in U\). In either case, its support is properly contained in that of \(\boldsymbol z\), a contradiction. Therefore, \(\boldsymbol z'\) is an affine basic point of \(U'\) and is coordinatewise nonnegative by induction. Applying the same reconstruction formulas to \(\boldsymbol z'\) recovers \(\boldsymbol z\). Since \(\alpha,\beta>0\), every coordinate of \(\boldsymbol z\) is nonnegative.}
%A minimal-support point of $U$ cannot have both $e,f$ nonzero, since movement along $v$ would reduce its support. Its image is minimally supported in $U'$; otherwise a smaller image could be lifted keeping the same zero side of the pair. By induction the image is nonnegative. A minimal-support lift has either $x_e=y/\alpha$, $x_f=0$, or $x_e=0$, $x_f=y/\beta$. Its remaining coordinates are unchanged, so the lift is nonnegative.

\emph{Case 4:} $e\in E$ and $g$ are in series. \ma{Since \(\{g,e\}\) is a cocircuit of \(N(U)\), the corresponding row space \(V\) contains a vector supported exactly on \(\{g,e\}\). Dividing this vector by its nonzero \(g\)-coordinate, the resulting vector has \(g\)-coordinate \(1\) and resource subvector \(\boldsymbol u\), with \(u_j=0\) for \(j\ne e\). The equation defining \(V\) gives \(H\boldsymbol u=\boldsymbol b\), yielding \(\boldsymbol u\in U\). Consequently, \(q_eu_e=\boldsymbol q^\top\boldsymbol u=\rho\), and hence \(\boldsymbol u=(\rho/q_e)\boldsymbol e_e\). Contracting \(e\) in \(N(U)\) corresponds to restricting \(U\) to \(x_e=0\) and then deleting coordinate \(e\). To obtain a strictly positive point in the reduced space, we construct a point of \(U\) whose \(e\)-coordinate is zero and whose other coordinates are strictly positive. We begin with a strictly positive point \(\boldsymbol x^+\in U\). Since \(n\ge2\) and \(q_jx_j^+>0\) for every \(j\in E\), we have $0<q_ex_e^+
<\sum_{j\in E}q_jx_j^+
=\rho$.}
%The corresponding row-space vector, normalized at $g$, gives $u=(\rho/q_e)\ee_e\in U$. Contract $e$, which corresponds to taking  $x_e=0$ and deleting this coordinate. To obtain a positive point on this slice, choose $x^+\in U$ with all coordinates positive. Since $n\ge2$, the other resources contribute positively to ${\boldsymbol  q}^\top \boldsymbol x^+=\rho$, so $0<q_ex_e^+<\rho$. 
Setting
\[
 \theta\triangleq q_e x_e^+/\rho\in(0,1)\qquad\text{and}\qquad
 \boldsymbol w\triangleq(\boldsymbol x^+-\theta\boldsymbol  u)/(1-\theta),
\]
we have $\boldsymbol w\in U$ with $w_e=0$, and all its other coordinates positive. Indeed, the coefficients of $\boldsymbol x^+$ and $\boldsymbol u$  in the above definition of $\boldsymbol w$ sum to one, so $\boldsymbol w$ is still in the
affine space $U$; its $e$-coordinate cancels by the definition of $\theta$.
For $j\ne e$, $w_j=x_j^+/(1-\theta)>0$. \ma{Let \(U'\) be obtained from \(U\cap\{x_e=0\}\) by deleting coordinate \(e\). Then every \(\boldsymbol x'\in U'\) satisfies $\sum_{j\ne e}q_jx'_j=\rho$
with strictly positive coefficients and right-hand side.}

\ma{Taking any affine basic point \(\boldsymbol z\in U\), we show that it is nonnegative. If \(z_e\ne0\), then \(\operatorname{supp}(\boldsymbol u)=\{e\}\subseteq\operatorname{supp}(\boldsymbol z)\). Inclusion-minimality forces \(\operatorname{supp}(\boldsymbol z)=\{e\}\), and \(\boldsymbol q^\top\boldsymbol z=\rho\) gives \(\boldsymbol z=\boldsymbol u\ge\boldsymbol0\). If \(z_e=0\), let \(\boldsymbol z'\in U'\) be obtained from $\boldsymbol z$ by deleting coordinate \(e\). A point of \(U'\) with support properly contained in that of \(\boldsymbol z'\) would, after inserting zero at coordinate \(e\), contradict the support minimality of \(\boldsymbol z\). Thus \(\boldsymbol z'\) is an affine basic point of \(U'\) and is coordinatewise nonnegative by induction. Since \(z_e=0\), \(\boldsymbol z\) is also coordinatewise nonnegative, which completes the proof.}
%The contracted affine space still satisfies the required positive equation. A minimal-support point with $x_e\ne0$ must have support $\{e\}$ and equal $u$. Every other minimal-support point is covered by induction on the contracted space. This finishes (iii)$\Rightarrow$(i).
\end{proof}

\subsection{Positive averaging of basic solutions}\label{app:averaging-proof}
\ma{This subsection gives a direct proof of \Cref{lem:cb}, which is used in the proofs of \Cref{prop:single-analytic,prop:multi-analytic}. We show that the coefficients \(\omega_J(W)\) in \eqref{eq:cb} are strictly positive and sum to one, prove the averaging identity coordinate by coordinate, and establish the equivalence \eqref{eq:k-basic-equivalence} by a limiting argument.}
%The analytic criterion uses the averaging identity to transfer signs from basic points to equilibrium responses. We verify that its determinant weights form a convex combination and recover each coordinate of the response vector.

\begin{proof}[Proof of \Cref{lem:cb}] 
We begin with a normalization of the weights.
Applying Cauchy-Binet formula to $HW^{1/2}$ and $W^{1/2}H\T$, we obtain
\[
 \det(HWH\T)=\sum_{J\text{ a basis}}
               \det(H_J)^2\prod_{j\in J}w_j.
\]
\ma{Each summand contains the factor \(\det(H_J)^2\), where \(J\) is a set of \(m=\operatorname{rank}H\) column indices and \(H_J\) is the \(m\times m\) submatrix formed by those columns of \(H\).}
%Each summand uses a square submatrix with $m=\rank H$ columns. A nonbasis has determinant zero and contributes nothing.
For every index set \(J\) of a column basis of \(H\), the matrix \(H_J\) is nonsingular. Since all diagonal entries \(w_j\) of \(W\) are strictly positive, each summand \(\det(H_J)^2\prod_{j\in J}w_j\) is strictly positive. \ma{Dividing both sides of the identity by the strictly positive number \(\det(HWH^\top)\) gives
\[
1
=\sum_{J\text{ a basis}}
\frac{\det(H_J)^2\prod_{j\in J}w_j}{\det(HWH^\top)}
=\sum_{J\text{ a basis}}\omega_J(W),
\]where the last equation is by definition in \eqref{eq:cb}.}
%Dividing by the positive determinant on the left gives the coefficients $\omega_J$ in \eqref{eq:cb}, with total sum one.

\emph{Step 1: Proof of the averaging identity \eqref{eq:cb}.} %identify the coordinate at resource $e$.}
We perturb only the column of $H$ with index $e$. Let \(s\in\mathbb R\), and define \(H(s)\) by replacing \(\boldsymbol h_e\) with \(\boldsymbol h_e+s\boldsymbol b\), leaving all other columns unchanged. Equivalently, \(H(s)\triangleq H+s\boldsymbol b\boldsymbol e_e^\top\). Accordingly, $R$ is perturbed to
$R(s)\triangleq H(s)WH(s)\T$. Then
\[
 R'(0)=w_e(\boldsymbol b\boldsymbol h_e\T+\boldsymbol h_e\boldsymbol b\T).
\]
Applying Jacobi’s formula to \(R(s)\) at \(s=0\) gives 
$\frac{d}{ds}\det R(s)|_{s=0}
=\det(R)\operatorname{tr}(R^{-1}R'(0))$,
where \(R=R(0)\). % and $\operatorname{tr}$ is the sum of the diagonal entries.
Since $R$ is symmetric, by substitution, we derive
\[
 \left.\frac{d}{ds}\det R(s)\right|_{s=0}
 =2\det(R)w_e\boldsymbol h_e\T R^{-1}\boldsymbol b
 =2\det(R)k_e(W,\boldsymbol b).
\]
For a basis $J$ not containing $e$, \ma{\(H(s)_J=H_J\) for every \(s\). Hence $\left.\frac{d}{ds}\det(H(s)_J)\right|_{s=0}
=0 =\det(H_J)z_e^J$, 
where \(z_e^J=0\) by the definition of the affine basic point in \eqref{eq:basic}.}
%$H(s)_J$ is unchanged and its derivative is zero, matching $z_e^J=0$. 
For a basis containing $e$,
multilinearity of the determinant, or the column form of Cramer's rule,
gives
\[
 \left.\frac{d}{ds}\det H(s)_J\right|_{s=0}
   =\det(H_J)(H_J^{-1}b)_e=\det(H_J)z_e^J.
\]
Consequently, \ma{applying the chain rule at \(s=0\), we have $\left.\frac{d}{ds}\bigl[\det(H(s)_J)\bigr]^2\right|_{s=0}
=2\det(H_J)^2z_e^J$. For the perturbed matrix, we use the Cauchy–Binet expansion of \(\det R(s)\) over all \(m\)-element sets \(J\) of column indices. If \(\det(H_J)=0\), then $\left.\frac{d}{ds}\bigl[\det(H(s)_J)\bigr]^2\right|_{s=0}
=
2\det(H_J)
\left.\frac{d}{ds}\det(H(s)_J)\right|_{s=0}
=0$.
Thus, only index sets of column bases of the original matrix \(H\) contribute to the derivative at \(s=0\). Differentiating the Cauchy–Binet expansion of \(\det R(s)\) term by term at \(s=0\), we obtain} 
%the derivative of its squared determinant is $2\det(H_J)^2z_e^J$. A nonbasis can make no first-order contribution to the squared determinant at $s=0$, since its determinant is zero there.  Differentiating the finite Cauchy--Binet sum therefore gives
\[
 2\det(R)k_e(W,b)
 =2\sum_{J\text{ a basis}}\det(H_J)^2
                    \Bigl(\prod_{j\in J}w_j\Bigr)z_e^J.
\]
Division by $2\det(R)$ proves the coordinate identity, and hence
\eqref{eq:cb}.

\emph{Step 2: Proof of the nonnegativity equivalence \eqref{eq:k-basic-equivalence}.} %concentrate on any prescribed basis.} 
\ma{Fix an index set \(J\) for a column basis of \(H\). For \(\varepsilon>0\), we define \(W_\varepsilon\triangleq\operatorname{diag}(w_j:j\in E)\), where \(w_j=1\) for \(j\in J\) and \(w_j=\varepsilon\) for \(j\notin J\). Let \(K\ne J\) be the index set of another column basis of \(H\). Since \(|K|=|J|\), we have \(K\setminus J\ne\varnothing\). The summand indexed by \(K\) therefore satisfies
\[
\det(H_K)^2\prod_{j\in K}w_j
=\det(H_K)^2\varepsilon^{|K\setminus J|}
\longrightarrow0
\qquad(\varepsilon\to0^+).
\]
The summand indexed by \(J\) equals \(\det(H_J)^2>0\) and is independent of \(\varepsilon\).}
%Fix a basis $J$, with weights one on $J$ and $\varepsilon>0$ elsewhere. Since every other basis contains at least one off-$J$ column, since all bases have the same size. Its determinant term tends to zero, whereas the term for $J$ is the positive constant $\det(H_J)^2$. Thus
It follows that \ma{as \(\varepsilon\to0^+\), \(\omega_J(W_\varepsilon)\to1\) and \(\omega_K(W_\varepsilon)\to0\) for every index set \(K\ne J\) of a column basis of \(H\).} 
%$\omega_J\to1$ and all other weights tend to zero. 
By the averaging
identity \ma{\eqref{eq:cb}, we deduce that \(k(W_\varepsilon,\boldsymbol b)\to\boldsymbol z^J\) as \(\varepsilon\to0^+\). }

\ma{On the one hand, suppose \(k(W,\boldsymbol b)\ge\boldsymbol0\) for every diagonal matrix \(W\) with strictly positive diagonal entries. Passing to the limit in \(k(W_\varepsilon,\boldsymbol b)\ge\boldsymbol0\) as \(\varepsilon\to0^+\) gives \(\boldsymbol z^J\ge\boldsymbol0\) for every index set \(J\) of a column basis of \(H\). On the other hand,} %$k(W,\boldsymbol b)\to \boldsymbol z^J$. If every basic point is nonnegative, all its weighted averages are nonnegative. Conversely, if every response for positive $W$ is nonnegative, these limits force every $\boldsymbol z^J$ to be nonnegative. 
\ma{if every affine basic point is coordinatewise nonnegative, then the averaging identity \eqref{eq:cb} gives \(k(W,\boldsymbol b)\ge\boldsymbol0\) for every diagonal matrix \(W\) with strictly positive diagonal entries.} This
proves \eqref{eq:k-basic-equivalence}.
\end{proof}

\ma{The argument above shows that every affine basic point can be approached by \(k(W,\boldsymbol b)\). This does not imply that every limit obtained by varying the positive diagonal matrix \(W\) is an affine basic point. For example, let \(H=(1,1,1)\), \(\boldsymbol b=1\), and \(W_\varepsilon=\operatorname{diag}(1,1,\varepsilon)\). Then
\[
k(W_\varepsilon,\boldsymbol b)
=\frac{(1,1,\varepsilon)^\top}{2+\varepsilon}
\longrightarrow (1/2,1/2,0)^\top
\qquad\text{as }\varepsilon\to0^+.
\]
The limit is not basic, since \((1,0,0)^\top\) is another solution of \(H\boldsymbol x=\boldsymbol b\) whose support is a proper subset of the support of the limit.}
%Other limits can be nonbasic: for $H=(1,1,1)$, $b=1$, and $W_\varepsilon=\operatorname{diag}(1,1,\varepsilon)$, \[ k(W_\varepsilon,b)  =\frac{(1,1,\varepsilon)\T}{2+\varepsilon} \longrightarrow (1/2,1/2,0)\T,\] whose support is not minimal.

\section*{Use of generative AI}
Generative AI tools assisted with the exploration of structural ideas and possible characterizations, examples, and proof arguments through iterative discussions with the authors. These tools also assisted with proof checking and the organization and revision checking of the manuscript. The authors take full responsibility for the correctness and originality of the mathematical results, the verification of the proofs and references, and the final presentation of the work.

\begingroup
\raggedright
\bibliographystyle{plainnat}
\bibliography{references}

@article{wardrop,
  author = {Wardrop, John G.},
  title = {Some Theoretical Aspects of Road Traffic Research},
  journal = {Proceedings of the Institution of Civil Engineers},
  volume = {1},
  number = {3},
  pages = {325--362},
  year = {1952},
  doi = {10.1680/ipeds.1952.11259}
}

@book{beckmann,
  author = {Beckmann, Martin and McGuire, C. B. and Winsten, C. B.},
  title = {Studies in the Economics of Transportation},
  publisher = {Yale University Press},
  address = {New Haven, CT},
  year = {1956}
}

@article{braess,
  author = {Braess, Dietrich},
  title = {\"Uber ein Paradoxon aus der Verkehrsplanung},
  journal = {Unternehmensforschung},
  volume = {12},
  pages = {258--268},
  year = {1968}
}

@article{fujishige,
  author = {Fujishige, Satoru and Goemans, Michel X. and Harks, Tobias and Peis, Britta and Zenklusen, Rico},
  title = {Matroids Are Immune to {Braess}' Paradox},
  journal = {Mathematics of Operations Research},
  volume = {42},
  number = {3},
  pages = {745--761},
  year = {2017},
  doi = {10.1287/moor.2016.0825}
}

@article{milchtaich,
  author = {Milchtaich, Igal},
  title = {Network Topology and the Efficiency of Equilibrium},
  journal = {Games and Economic Behavior},
  volume = {57},
  number = {2},
  pages = {321--346},
  year = {2006},
  doi = {10.1016/j.geb.2005.09.005}
}

@article{chen,
  author = {Chen, Xujin and Diao, Zhuo and Hu, Xiaodong},
  title = {Network Characterizations for Excluding {Braess}'s Paradox},
  journal = {Theory of Computing Systems},
  volume = {59},
  number = {4},
  pages = {747--780},
  year = {2016},
  doi = {10.1007/s00224-016-9710-4}
}

@article{cenciarelli,
  author = {Cenciarelli, Pietro and Gorla, Daniele and Salvo, Ivano},
  title = {A Polynomial-Time Algorithm for Detecting the Possibility of {Braess} Paradox in Directed Graphs},
  journal = {Algorithmica},
  volume = {81},
  number = {4},
  pages = {1535--1560},
  year = {2019},
  doi = {10.1007/s00453-018-0486-6}
}

@article{cominetti,
  author = {Cominetti, Roberto and Dose, Valerio and Scarsini, Marco},
  title = {Monotonicity of Equilibria in Nonatomic Congestion Games},
  journal = {European Journal of Operational Research},
  volume = {316},
  number = {2},
  pages = {754--766},
  year = {2024},
  doi = {10.1016/j.ejor.2024.01.050}
}

@article{verbreecherukuri,
  author = {Verbree, Jasper and Cherukuri, Ashish},
  title = {Wardrop Equilibrium and {Braess}'s Paradox for Varying Demand},
  journal = {arXiv preprint arXiv:2310.04256},
  year = {2023},
  url = {https://arxiv.org/abs/2310.04256}
}

@article{cominettiphase,
  author = {Cominetti, Roberto and Dose, Valerio and Scarsini, Marco},
  title = {Phase Transitions of the Price-of-Anarchy Function in Multi-Commodity Routing Games},
  journal = {Transportation Research Part B: Methodological},
  volume = {182},
  pages = {102922},
  year = {2024},
  doi = {10.1016/j.trb.2024.102922}
}

@article{duffin,
  author = {Duffin, R. J.},
  title = {Topology of Series-Parallel Networks},
  journal = {Journal of Mathematical Analysis and Applications},
  volume = {10},
  number = {2},
  pages = {303--318},
  year = {1965},
  doi = {10.1016/0022-247X(65)90125-3}
}

@article{brylawski,
  author = {Brylawski, Thomas H.},
  title = {A Combinatorial Model for Series-Parallel Networks},
  journal = {Transactions of the American Mathematical Society},
  volume = {154},
  pages = {1--22},
  year = {1971},
  doi = {10.1090/S0002-9947-1971-0288039-7}
}

@book{oxley,
  author = {Oxley, James},
  title = {Matroid Theory},
  edition = {2},
  publisher = {Oxford University Press},
  address = {Oxford},
  year = {2011},
  doi = {10.1093/acprof:oso/9780198566946.001.0001}
}

@article{zaslavsky,
  author = {Zaslavsky, Thomas},
  title = {Facing Up to Arrangements: Face-Count Formulas for Partitions of Space by Hyperplanes},
  journal = {Memoirs of the American Mathematical Society},
  volume = {1},
  number = {154},
  year = {1975}
}

@book{schrijver,
  author = {Schrijver, Alexander},
  title = {Combinatorial Optimization: Polyhedra and Efficiency},
  series = {Algorithms and Combinatorics},
  volume = {24},
  publisher = {Springer},
  address = {Berlin},
  year = {2003}
}

@article{walter,
  author = {Walter, Matthias},
  title = {Recognizing Series-Parallel Matrices in Linear Time},
  journal = {INFORMS Journal on Computing},
  volume = {35},
  number = {6},
  pages = {1404--1418},
  year = {2023},
  doi = {10.1287/ijoc.2021.0233}
}

@article{hall,
  author = {Hall, Michael A.},
  title = {Properties of the Equilibrium State in Transportation Networks},
  journal = {Transportation Science},
  volume = {12},
  number = {3},
  pages = {208--216},
  year = {1978},
  doi = {10.1287/trsc.12.3.208}
}

@article{patriksson,
  author = {Patriksson, Michael},
  title = {Sensitivity Analysis of Traffic Equilibria},
  journal = {Transportation Science},
  volume = {38},
  number = {3},
  pages = {258--281},
  year = {2004},
  doi = {10.1287/trsc.1030.0043}
}

@article{josefssonpatriksson,
  author = {Josefsson, Magnus and Patriksson, Michael},
  title = {Sensitivity Analysis of Separable Traffic Equilibrium Equilibria with Application to Bilevel Optimization in Network Design},
  journal = {Transportation Research Part B: Methodological},
  volume = {41},
  number = {1},
  pages = {4--31},
  year = {2007},
  doi = {10.1016/j.trb.2005.12.004}
}

@book{schrijverLP,
  author = {Schrijver, Alexander},
  title = {Theory of Linear and Integer Programming},
  publisher = {John Wiley \& Sons},
  year = {1986}
}

@article{oxleybeta,
  author = {Oxley, James G.},
  title = {On {Crapo}'s Beta Invariant for Matroids},
  journal = {Studies in Applied Mathematics},
  volume = {66},
  number = {3},
  pages = {267--277},
  year = {1982},
  doi = {10.1002/sapm1982663267}
}

@article{derezinskiwarmuth,
 author = {Derezi{\'n}ski, Micha{\l} and Warmuth, Manfred K.},
 title = {Reverse Iterative Volume Sampling for Linear Regression},
 journal = {Journal of Machine Learning Research},
 volume = {19}, number = {23}, pages = {1--39}, year = {2018},
 url = {https://jmlr.org/papers/v19/17-781.html}
}

@incollection{stanley,
 author = {Stanley, Richard P.},
 title = {An Introduction to Hyperplane Arrangements},
 booktitle = {Geometric Combinatorics},
 editor = {Miller, Ezra and Reiner, Victor and Sturmfels, Bernd},
 series = {IAS/Park City Mathematics Series}, volume = {13},
 pages = {389--496}, publisher = {American Mathematical Society},
 address = {Providence, RI}, year = {2007},
 url = {https://math.mit.edu/~rstan/arrangements/arr.html}
}

@inproceedings{matsubayashi,
 author = {Matsubayashi, Akira and Saito, Yushi},
 title = {A Faster Algorithm for Recognizing Directed Graphs Invulnerable to {Braess}'s Paradox},
 booktitle = {23rd Symposium on Algorithmic Approaches for Transportation Modelling, Optimization, and Systems (ATMOS 2023)},
 series = {Open Access Series in Informatics}, volume = {115},
 pages = {12:1--12:19}, year = {2023}, publisher = {Schloss Dagstuhl--Leibniz-Zentrum f{\"u}r Informatik},
 doi = {10.4230/OASIcs.ATMOS.2023.12}
}
\endgroup
\end{document}